\documentclass[preprint,3p,12pt]{elsarticle}

\usepackage{mathrsfs}
\usepackage{amsmath}
\usepackage{stmaryrd}
\usepackage{bbding}
\usepackage{dcolumn}
\usepackage{graphicx}
\usepackage{amsfonts}
\usepackage{amssymb}
\usepackage{psfrag}
\usepackage{wrapfig}
\usepackage{subfigure}
\usepackage{makeidx}
\usepackage{bm}
\usepackage{epsf}
\usepackage{epsfig}
\usepackage{setspace}
\usepackage{graphicx}
\usepackage{epstopdf}
\usepackage{psfrag}
\usepackage{subfigure}
\usepackage{color}
\usepackage{comment}
\usepackage{algorithm}
\usepackage{algorithmic}
\newcommand \td {\mathrm{~d}}

\journal{}

\begin{document}

\begin{frontmatter}



\title{High-order compact gas-kinetic scheme in arbitrary Lagrangian-Eulerian formulation}
\author[HKUST1]{Yue Zhang}
\ead{yzhangnl@connect.ust.hk}	
\author[HKUST1,HKUST2,HKUST3]{Kun Xu\corref{cor1}}
\ead{makxu@ust.hk}

\address[HKUST1]{Department of Mathematics, Hong Kong University of Science and Technology, Clear Water Bay, Kowloon, Hong Kong}
\address[HKUST2]{Department of Mechanical and Aerospace Engineering, Hong Kong University of Science and Technology, Clear Water Bay, Kowloon, Hong Kong}
\address[HKUST3]{Shenzhen Research Institute, Hong Kong University of Science and Technology, Shenzhen, China}
\cortext[cor1]{Corresponding author}

\begin{abstract}
This study proposes an extension of the high-order compact gas-kinetic scheme (CGKS) to compressible flow simulation in an arbitrary Lagrangian-Eulerian (ALE) formulation in unstructured mesh. The ALE method is achieved by subdividing arbitrary mesh into tetrahedrons and integrating flux function in a local coordinate system at the cell interface to ensure geometric conservation law. The scheme incorporates a compact reconstruction with third-order accuracy for updating both cell-averaged conservative flow variables and their gradients.
HWENO-type nonlinear reconstruction and gradient compression factors are adopted to improve the accuracy and robustness of the scheme.
A multi-stage multi-derivative (MSMD) time-stepping method is also implemented to achieve high-order time accuracy with fewer middle stages. The scheme is used to study problems involving moving boundaries. The numerical experiments demonstrate the effectiveness of the scheme in capturing the accurate solutions of both low-speed smooth flow and highly compressible ones with strong shock waves.
\end{abstract}


\end{frontmatter}


\section{Introduction}
There are two common frameworks for describing fluid motion in computational fluid dynamics: the Eulerian and the Lagrangian systems.
The Eulerian system describes flow on a fixed grid, while in the Lagrangian system, the grid moves with the fluid velocity. While the Eulerian system is more straightforward to implement, it has difficulties in handling moving boundaries and introducing dissipation when dealing with contact discontinuities. The Lagrangian system can resolve contact discontinuities well, such as material interfaces or free surfaces, but may lead to computational breakdown due to grid distortion. Therefore, the development of arbitrary Lagrangian-Eulerian (ALE) methods \cite{HIRT1974227} becomes attractive. There are two types of ALE methods: direct and indirect. Direct ALE methods involve moving both the object and the computational grid simultaneously. Indirect ALE methods separate the motion of the object and the computational grid into three stages: the Lagrangian stage, the rezoning stage \cite{BRACKBILL1982342}, and the remapping stage \cite{KUCHARIK2014268}. During the Lagrangian stage, the flow field and grid are updated simultaneously. During the rezoning stage, a new computational grid is constructed to match the object's new position and shape, often involving techniques such as grid generation and deformation to ensure grid quality and accuracy.
In the remapping stage, the solution is transferred into the rezoned mesh. The current research focuses on the direct ALE method with a fixed grid topology, which is designed to handle moving boundaries and shock waves. Therefore, the remapping step can be omitted.

The gas-kinetic scheme (GKS) is a kinetic theory-based numerical method to solve the Euler and Navier-Stokes equations \cite{xuGKS2001}.
Under the initial condition of a generalized Riemann problem, a time-evolving gas distribution function is constructed in GKS to calculate the numerical fluxes and evaluate the time-dependent flow variables at a cell interface.
As a result, both cell-averaged flow variables and their gradients can be updated.
Therefore, the HWENO-type method and the two-step multi-resolution WENO reconstruction can be developed in the scheme for the high-order spatial data reconstruction \cite{zhuNewFiniteVolume2018,jiMultiresolution2021}.
At the same time, due to the time-accurate flux function, the multi-stage multi-derivative (MSMD) method can be used to update the solution with high-order temporary accuracy. Specifically, the two-stage fourth-order (S2O4) time-stepping method is used in the compact GKS (CGKS) \cite{liTwoStageFourthOrder2016,li2019two}.
The CGKS has been constructed on both structured and unstructured meshes in 2D and 3D cases  \cite{ji4order2018,jiThreedimensionalCompactHighorder2020,
zhaoCompactHigherorderGaskinetic2019,zhaoAcousticShockWave2020,zhaoCompactHighorderGaskinetic2022a}.
In order to further improve the robustness of the scheme in high-speed flow simulations,
further improvements have been incorporated into the scheme.
First, the evolution of possible discontinuous flow variables at different sides of a cell interface is constructed
for updating reliable cell averaged gradient of flow variables \cite{zhaoDirectModelingComputational2021}.
Second, the nonlinear limiting process is implemented on the high-order time derivative of the flux function under the MSMD framework.
Equipped with the above remedies, the CGKS on a 3D unstructured mesh is highly robust in hypersonic flow computation and
a large time step, such as CFL number $\approx 0.8$, can be used in the fourth-order compact scheme.
Alternatively, a gradient compression factor is designed to improve the robustness and efficiency of CGKS in case of low-quality mesh \cite{jiGC2021}, and further improvements are studied in \cite{zhang2023slidingmesh}.

In recent years, high-order GKS has tended to incorporate ALE formulation. Ren \cite{REN2016-ALE-DG} developed a one-stage DG-ALE gas-kinetic method for oscillating airfoil calculation. The variation of mesh velocity is considered in the time-dependent flux calculation along a cell interface, and the geometric conservation law is satisfied accurately. Pan developed a high-order ALE gas-kinetic method on structure mesh for two and three dimensions\cite{PAN2020ALE2d,PAN2021ALE3d}, where a non-compact WENO scheme and the two-stage fourth-order discretization are used.
A bilinear interpolation is used to parameterize both grid coordinates and grid moving velocity with the preservation of the geometric conservation law. However, this interpolation isn't a unified formulation for other types of mesh, such as triangular prism and pyramid. In this paper, our target is to develop a compact gas-kinetic scheme on unstructured mesh in ALE formulation. Instead of bilinear interpolation, we use a subdivision of arbitrary meshes into tetrahedrons and a linear interpolation in triangular surfaces to preserve the geometry conservation law. A compact third-order WENO reconstruction with the requirement of Neumann neighboring cells only is adopted in spatial reconstruction. The radius basic function method \cite{RBF2007784} and Lagrangian nodal solver \cite{Lagrangianvel} are used to determine grid velocity.

The structure of the paper is the following. Section 2 introduces the finite volume method and GKS in ALE formulation.
Section 3 discusses the grid velocity, the two-stage fourth-order time integrating method, and the geometric conservation law.
Section 4 focuses on the compact reconstruction. In Section 5, several test cases are used to validate the proposed method.
Finally, the conclusion is presented in the last section.

\section{Gas-Kinetic Scheme in ALE Formulation}
\subsection{Governing equation on moving mesh}
By using the finite volume method, the computational domain $\Omega$ is discretized into a series of non-overlap elements $\Omega_i$,
  \begin{equation*}
    \Omega =\bigcup \Omega_i,\ \Omega_i \bigcap \Omega_j=\phi (i \neq j).
  \end{equation*}
  The boundary of element $\Omega_i$ can be expressed as
  \begin{equation*}
    \partial \Omega_i=\bigcup_{p=1}^{N_f}\Gamma_{ip},
  \end{equation*}
where $N_f$ is the number of element surface and $\Gamma_{ip}$ is surface of the element.
Reynolds transport theorem gives the conservation laws in an arbitrary Lagrangian-Eulerian (ALE) formulation on a moving cell
\begin{equation*}
  \frac{\td}{\td t}\int_{\Omega_i (t)}\boldsymbol{W} \td \Omega + \int_{\partial \Omega_i(t)} (\boldsymbol{F}(\boldsymbol{W})-\boldsymbol{W}\boldsymbol{U})\cdot \boldsymbol{n}\td S,
\end{equation*}
where $\boldsymbol{U}$ is the mesh velocity, and $\boldsymbol{F}(\boldsymbol{W})$ is the flux pass though the element surfaces. For Navier-Stokes equation, $\boldsymbol{W}=[\rho,\rho V_1,\rho V_2,\rho V_3,\rho E]^T$ is the conservative values, and $\rho, \boldsymbol{V}=(V_1, V_2, V_3), E$ is the density, absolute velocity and energy of fluid.
\begin{itemize}
  \item If $\boldsymbol{U}=\boldsymbol{V}$, the governing equation becomes Lagrangian form;
  \item If $\boldsymbol{U}=\boldsymbol{0}$, the governing equation becomes Eulerian form.
\end{itemize}

\begin{figure}[hbt!]
  \centering
  \includegraphics[width=8cm]{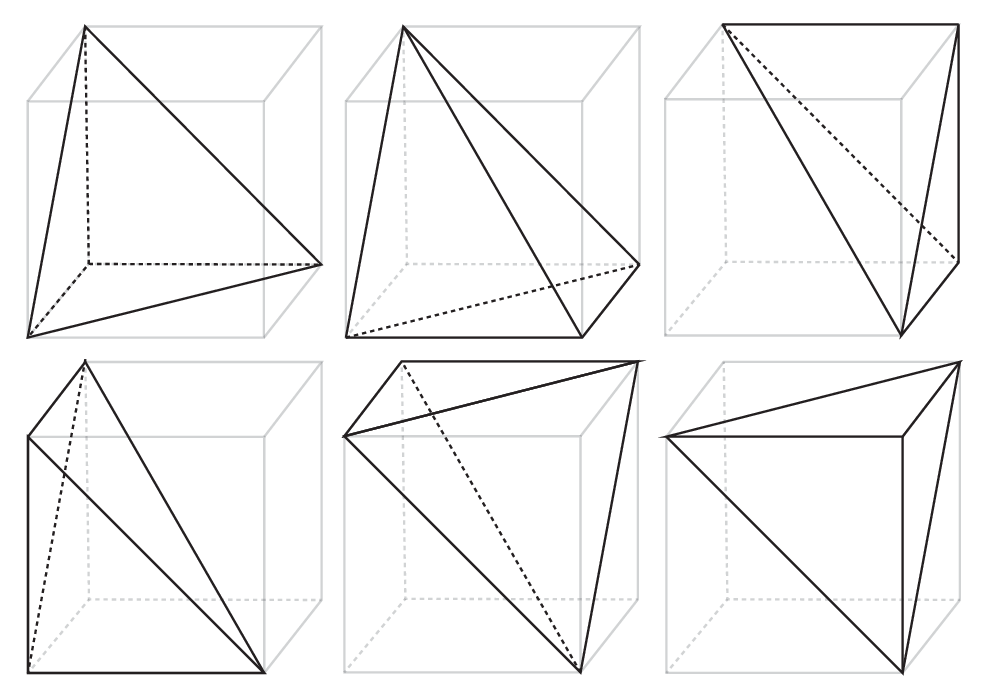}
  \caption{An example of mesh partition}\label{tetrahedrons}
\end{figure}

\begin{figure}[hbt!]
  \centering
  \includegraphics[width=6cm]{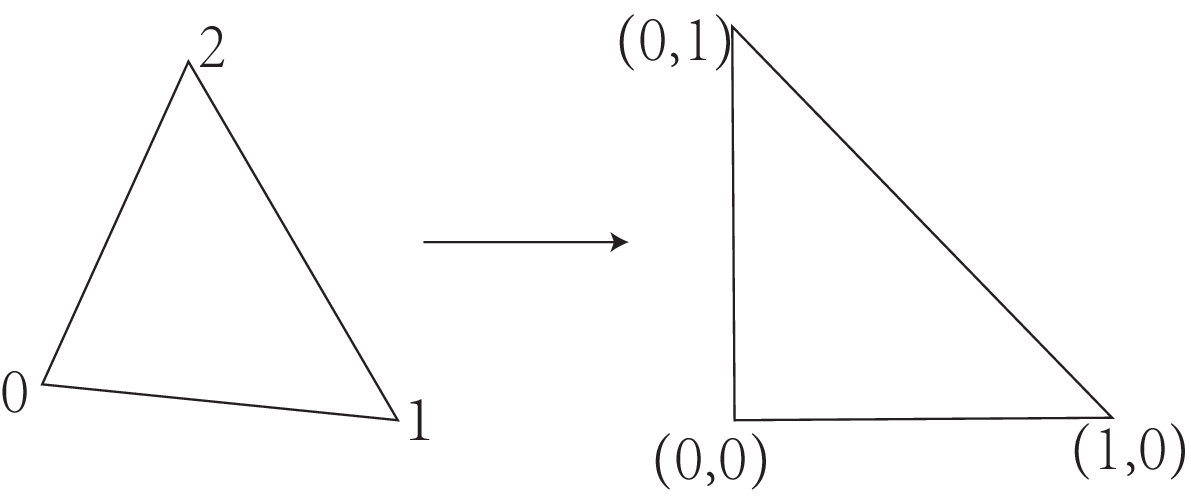}
  \caption{Translation a triangle surface to local coordinates}\label{triangles_trans}
\end{figure}

A unified partitioning approach is adopted to handle different types of grids and maintain consistency. Specifically, different types of grids (such as hexahedrons, prisms, and pyramids) are all partitioned into several tetrahedrons. The partitioning approach ensures computational consistency and facilitates the conversion between different types of grids. An example of partitioning a hexahedron into six tetrahedrons is illustrated in Figure \ref{tetrahedrons}.
To maintain consistency with the partitioning of elements, we divide the quadrilateral surface into two triangles. Under this condition, the conservation law in ALE formulation can be written as
\begin{equation*}
  \frac{\td}{\td t}\int_{\Omega (t)}\boldsymbol{W} \td \Omega +\sum_{p_t=1}^{N_t}\int_{\Sigma_{p_t}} (\boldsymbol{F}(\boldsymbol{W})-\boldsymbol{W}\boldsymbol{U})\cdot \boldsymbol{n}\td S=0,
\end{equation*}
where $N_t$ is the number of triangles after partitioning and $\Sigma_{p_t}$ is the individual triangle surfaces.
The translation from Eulerian coordinates to local two-dimensional coordinates $(\xi,\eta)$ as shown in Figure \ref{triangles_trans} is used to evaluate the surface integration.
Thus, a linear interpolation is introduced for each triangle
\begin{equation*}
  \boldsymbol{x}(\xi,\eta)=\boldsymbol{x}_0+\xi(\boldsymbol{x}_1-\boldsymbol{x}_0)+\eta(\boldsymbol{x}_2-\boldsymbol{x}_0),
\end{equation*}
where $\boldsymbol{x}_0,\boldsymbol{x}_1,\boldsymbol{x}_2$ is the coordinates of the triangle vertices. With this transformation, the normal vector of triangle surfaces can be written as
\begin{equation*}
  \boldsymbol{n}= \boldsymbol{x}_{,\xi}\times\boldsymbol{x}_{,\eta}=\frac{\partial \boldsymbol{x}}{\partial \xi} \times \frac{\partial \boldsymbol{x}}{\partial \eta}.
\end{equation*}
The surface integration can be written as
\begin{equation*}
  \int_{\Sigma_{p_t}}(\boldsymbol{F}(\boldsymbol{W})-\boldsymbol{W}\boldsymbol{U})\cdot \boldsymbol{n}\td S=\int_0^1\int_0^{1-\xi}(\boldsymbol{F}(\boldsymbol{W})-\boldsymbol{W}\boldsymbol{U})\cdot (\boldsymbol{x}_{,\xi}\times\boldsymbol{x}_{,\eta}) \td \eta \td \xi.
\end{equation*}
Assuming the grid velocity is time-independent, the coordinates of points on the surface can be written as $\boldsymbol{x}=\boldsymbol{x}^0+\boldsymbol{U}t$, where $\boldsymbol{x}^0$ is the initial position of the points. Then, we have
\begin{equation*}
  \begin{aligned}
  \int_0^1\int_0^{1-\xi}\boldsymbol{W}\boldsymbol{U}\cdot (\boldsymbol{x}_{,\xi}\times\boldsymbol{x}_{,\eta}) \td \eta \td \xi&=
  \int_0^1\int_0^{1-\xi}\boldsymbol{W}\boldsymbol{U} \cdot (\boldsymbol{x}^0_{,\xi}\times\boldsymbol{x}^0_{,\eta}) \td \eta \td \xi \\
  &+t\int_0^1\int_0^{1-\xi}\boldsymbol{W}\boldsymbol{U}\cdot (\boldsymbol{U}^0_{,\xi}\times\boldsymbol{x}^0_{,\eta}) \td \eta \td \xi\\
  &+t\int_0^1\int_0^{1-\xi}\boldsymbol{W}\boldsymbol{U}\cdot (\boldsymbol{x}^0_{,\xi}\times\boldsymbol{U}_{,\eta}) \td \eta \td \xi\\
  &+t^2\int_0^1\int_0^{1-\xi}\boldsymbol{W}\boldsymbol{U} \cdot (\boldsymbol{U}_{,\xi}\times\boldsymbol{U}_{,\eta}) \td \eta \td \xi
  \end{aligned}.
\end{equation*}
The total flux can be approximated as
\begin{equation*}
  \begin{aligned}
  \int_{\Sigma_{p_t}}(\boldsymbol{F}(\boldsymbol{W})-\boldsymbol{W}\boldsymbol{U})\cdot \boldsymbol{n}\td S
  &\approx\int_0^1\int_0^{1-\xi}(\boldsymbol{F}(\boldsymbol{W})-\boldsymbol{W}\boldsymbol{U}) \cdot (\boldsymbol{x}^0_{,\xi}\times\boldsymbol{x}^0_{,\eta}) \td \eta \td \xi \\
  &+t\int_0^1\int_0^{1-\xi}\boldsymbol{W}\boldsymbol{U}\cdot (\boldsymbol{U}^0_{,\xi}\times\boldsymbol{x}^0_{,\eta}) \td \eta \td \xi\\
  &+t\int_0^1\int_0^{1-\xi}\boldsymbol{W}\boldsymbol{U}\cdot (\boldsymbol{x}^0_{,\xi}\times\boldsymbol{U}_{,\eta}) \td \eta \td \xi\\
  &+t^2\int_0^1\int_0^{1-\xi}\boldsymbol{W}\boldsymbol{U} \cdot (\boldsymbol{U}_{,\xi}\times\boldsymbol{U}_{,\eta}) \td \eta \td \xi\\
  &=\mathcal{L}_{p_t}(\boldsymbol{W},\Omega) +\mathbb{V}_{p_t}(t)
  \end{aligned},
\end{equation*}
where $\mathcal{L}_{p_t}(\boldsymbol{W},\Omega) $ is the surface flux, and $\mathbb{V}_{p_t}(t)$ represents the flux correction due to the motion of grid, except the translation. Thus, the semi-discretized equation can be written as
\begin{equation}\label{fvm}
  \frac{\td}{\td t}\int_{\Omega_i (t)}\boldsymbol{W} \td \Omega + \mathcal{L}(\boldsymbol{W},\Omega) +\mathbb{V}(t)=0.
\end{equation}
The surface integration can be calculated by Gaussian integration. Three Gaussian points with local coordinates $(\frac{1}{6},\frac{1}{6}),(\frac{2}{3},\frac{1}{6}), (\frac{1}{6},\frac{2}{3})$ and weightes $\omega_1=\omega_2=\omega_3=\frac{1}{3}$ are used in each triangle surface. In this condition, the surface flux and correction can be written as
\begin{equation*}
  \begin{aligned}
   \mathcal{L}(\boldsymbol{W},\Omega)&=  \sum_{p_t=1}^{Nt}\mathcal{L}_{p_t}(\boldsymbol{W},\Omega) = \sum_{p_t=1}^{Nt}\sum_{k=1}^{3}\omega_k\left[ \boldsymbol{F}(\boldsymbol{W}_{p_t,k})-\boldsymbol{W}_{p_t,k}\boldsymbol{U}_{p_t,k}\right]\cdot(\boldsymbol{x}^0_{,\xi}\times\boldsymbol{x}^0_{,\eta})_{p_t},\\
  \mathbb{V}(t)&=  \sum_{p_t=1}^{Nt}\mathbb{V}_{p_t}(t)= \sum_{p_t=1}^{Nt}\sum_{k=1}^{3}\omega_k\boldsymbol{W}_{p_t,k}\boldsymbol{U}_{p_t,k}\cdot\left[t (\boldsymbol{U}^0_{,\xi}\times\boldsymbol{x}^0_{,\eta}) +t(\boldsymbol{x}^0_{,\xi}\times\boldsymbol{U}^0_{,\eta})+t^2(\boldsymbol{U}^0_{,\xi}\times\boldsymbol{U}^0_{,\eta})\right]_{p_t}.
  \end{aligned}
\end{equation*}

\subsection{BGK equation in ALE Formulation}
The gas-kinetic BGK equation can be written as
\begin{equation*}
  \frac{\partial f}{ \partial t}+ \boldsymbol{v}\cdot \nabla_x f  = \frac{g-f}{\tau},
\end{equation*}
where $f=f(\boldsymbol{x},t,\boldsymbol{v},\xi)$ is the gas distribution function, $g$ is the corresponding equilibrium state, and $\tau$ is the collision time.
 $\boldsymbol{v}=(v_1,v_2,v_3)$ is the particle velocity.
The collision term in the above equation describes the evolution process from a non-equilibrium state to an equilibrium one with
the satisfaction of compatibility condition
\begin{equation*}
  \int \frac{g-f}{\tau} \Psi \td\boldsymbol{w}\td\Xi = 0,
\end{equation*}
where $\Psi=(1,v_1,v_2,v_3,\frac{1}{2}(v_1^2+v_2^2+v_3^2+\xi^2))^T$ and $\td\Xi=\td\xi_1\cdots\td\xi_K$ ($K$ is the number of internal degree of freedom, i.e. $K=(5-3\gamma)/(\gamma-1)$ for three-dimensional flow and $\gamma$ is the specific heat ratio). According to the Chapman-Enskog Expansion for BGK equation, the macroscopic governing equations can be derived\cite{vki1998}. In the continuum region, the BGK equation can be rearranged, and the gas distribution function can be expanded as
\begin{equation*}
  f=g-\tau \mathcal{D}_{\boldsymbol{u}}g+\tau \mathcal{D}_{\boldsymbol{u}}(\tau\mathcal{D}_{\boldsymbol{u}})g-\tau \mathcal{D}_{\boldsymbol{u}}[ \tau \mathcal{D}_{\boldsymbol{u}}(\tau\mathcal{D}_{\boldsymbol{u}})]g+\cdots,
\end{equation*}
where $\mathcal{D}_{\boldsymbol{u}}=\frac{\partial }{\partial t}+ \boldsymbol{u}\cdot \nabla$. With the zeroth-order truncation $f=g$, the Euler equations can be obtained. For the first-order truncation $f=g-\tau \mathcal{D}_{\boldsymbol{u}}g$, the Navier-Stokes equations with the dynamic  viscosity coefficient $\mu=\tau p$ ($p$ is pressure) and Prandtl number $\text{Pr}=1$ can be obtained.

In this paper, a high-order gas-kinetic scheme will be constructed in the arbitrary Lagrangian-Eulerian (ALE) framework for three-dimensional flow. The particle speed related to grid $\boldsymbol{w}=(w_1,w_2,w_3)$ is
\begin{equation*}
  \boldsymbol{w} = \boldsymbol{v}-\boldsymbol{U},
\end{equation*}
where $\boldsymbol{U}=(U_1,U_2,U_3)$ is the grid velocity given aribitray. The BGK equation on moving reference can be written as
\begin{equation} \label{BGK}
  \frac{\partial f}{ \partial t}+ \boldsymbol{w}\cdot \nabla_x f  = \frac{g-f}{\tau}.
\end{equation}

The flux in the term $\mathcal{L}$ of ALE-formulation finite volume method (\ref{fvm}) can be evaluated by
  \begin{equation}
    \mathbb{F} (\boldsymbol{W}) =(\boldsymbol{F} (\boldsymbol{W})-\boldsymbol{W}\boldsymbol{U}) \cdot (\boldsymbol{x}^0_{,\xi}\times\boldsymbol{x}^0_{,\eta})=\int f(\boldsymbol{x},t,\boldsymbol{w},\xi)\boldsymbol{w}\cdot \boldsymbol{n} \boldsymbol{ \Psi} \td\boldsymbol{v}\td\Xi.
  \end{equation}
In the actual computation, the numerical flux can be obtained by taking moments of gas distribution function with related velocity
  \begin{equation}\label{flux}
    \mathbb{F}(\boldsymbol{W})^\prime=\int f(\boldsymbol{x},t,\boldsymbol{w},\xi)\boldsymbol{w}\cdot \boldsymbol{n} \boldsymbol{ \Psi}^\prime \td\boldsymbol{v}\td\Xi,
  \end{equation}
where ${\boldsymbol{ \Psi}} =(1,{w}_1,{w}_2,{w}_3,\frac{1}{2}({w}_1^2+{w}_2^2+{w}_3^2+\xi^2))^T$. The component of $\mathbb{F}$ can be given by the combination of $\mathbb{F}^\prime$ as follows
\begin{equation*}
  \begin{cases}
    \mathbb{F}_\rho=\mathbb{F}_\rho^\prime \\
    \mathbb{F}_{\rho V_1}=\mathbb{F}_{\rho W_1}^\prime + U_1 \mathbb{F}_\rho^\prime \\
    \mathbb{F}_{\rho V_2}=\mathbb{F}_{\rho W_2}^\prime + U_2 \mathbb{F}_\rho^\prime \\
    \mathbb{F}_{\rho V_3}=\mathbb{F}_{\rho W_3}^\prime + U_3 \mathbb{F}_\rho^\prime \\
    \mathbb{F}_{\rho E}=\mathbb{F}_{\rho E}^\prime+U_1\mathbb{F}_{\rho W_1}^\prime +U_2\mathbb{F}_{\rho W_2}^\prime +U_3\mathbb{F}_{\rho W_3}^\prime + \frac{1}{2}(U_1^2+U_2^2+U_3^2) \mathbb{F}_\rho^\prime \\
  \end{cases}.
\end{equation*}
By defining the rotating matrix $\boldsymbol{T}=\text{diag}(1,\boldsymbol{T}^\prime,1)$ and
\begin{equation*}
  \boldsymbol{T}^\prime =\left( \begin{matrix}
    n_1&n_2&n_3\\
    -n_2&n_1+\frac{n_3^2}{1+n_1}&-\frac{n_2n_3}{1+n_1}\\
    -n3&-\frac{n_2n_3}{1+n_1}&1-\frac{n_3^2}{1+n_1}
  \end{matrix}\right),n_1\neq -1,
\end{equation*}
and when $n_1=-1$, $\boldsymbol{T}^\prime$ becomes $\text{diag}(-1,-1,1)$, the flux (Eq. \ref{flux}) can be evaluated by
  \begin{equation}\label{flux}
    \mathbb{F}(\boldsymbol{W})^\prime=\boldsymbol{T}^{-1}\int f(\tilde{\boldsymbol{x}}_{p_t,k},t,\tilde{\boldsymbol{w}},\xi) \tilde{w}_1 \tilde{\boldsymbol{ \Psi}}^\prime \td\tilde{\boldsymbol{w}}\td\Xi,
  \end{equation}
where the origin point of local coordinate is $\tilde{\boldsymbol{x}}=(0,0,0)$  with $x$-direction in $\boldsymbol{n}$ and $\widetilde{\boldsymbol{ \Psi}}^\prime =(1,\tilde{w}_1,\tilde{w}_2,\tilde{w}_3,\frac{1}{2}(\tilde{w}_1^2+\tilde{w}_2^2+\tilde{w}_3^2+\xi^2))^T$. The microscopic velocities in local coordinate are given by $\tilde{\boldsymbol{w}} = \mathbf{T}^\prime \boldsymbol{w}$ and $\tilde{w}_1= n_1w_1 +n_2w_2+n_3w_3$.

  \subsection{Gas evolution model}
  In order to construct the numerical fluxes at $\boldsymbol{x} = (0,0,0)^T$, the integral solution of the BGK equation Eq.(\ref{BGK}) is
  \begin{equation}\label{solbgk}
    f(\boldsymbol{x},t, \boldsymbol{w},\xi)=\frac{1}{\tau}\int_0^t g(\boldsymbol{x}^{\prime},t^{\prime}\boldsymbol{w},\xi)e^{-(t-t^{\prime})/\tau}\td t^{\prime}+ e^{t/\tau}f_0(\boldsymbol{x}_{0},\boldsymbol{w}),
  \end{equation}
with the trajectory
  \begin{equation*}
    \boldsymbol{x}=\boldsymbol{x}^{\prime}+\boldsymbol{w}(t-t^{\prime}).
  \end{equation*}
In Eq.(\ref{solbgk}), $f_0$ is the initial gas distribution function, and $g$ is the corresponding equilibrium state. $\boldsymbol{x}_0$ are the initial position by tracing back particles $\boldsymbol{x}$ at time $t$ back to $t=0$.

Before constructing the initial distribution function $f_0$ and equilibrium state $g$, we first denote
  \begin{equation*}
    \boldsymbol{a}=(a_1,a_2,a_3) = \nabla_x g/g, A = g_t/g.
  \end{equation*}
In the following derivation, quadratic terms of time will be ignored directly.
With the consideration of possible discontinuity at an interface, the initial distribution is constructed as
  \begin{equation}\label{ini_gas}
    f_0(\boldsymbol{x}_{0},\boldsymbol{w}_0)=f_0^l(\boldsymbol{x}_{0},\boldsymbol{w}) (1 -\mathbb{H}(x_1)) +f_0^r(\boldsymbol{x}_{0},\boldsymbol{w})\mathbb{H}(x_1),
  \end{equation}
  where $\mathbb{H}$ is the Heaviside function. $f_0^{l}$ and $f_0^{r}$ are the initial gas distribution functions on the left and right sides of the interface, which are determined by corresponding initial macroscopic variables and their spatial derivatives.
  With the second-order accuracy, $f_0^k(\boldsymbol{x},\boldsymbol{w})$ is constructed by Taylor expansion around $(\boldsymbol{x},\boldsymbol{w})$
  \begin{equation}\label{taylor_ex}
    f_0^k(\boldsymbol{x}_{0},\boldsymbol{w})=f_G^k (\boldsymbol{x},\boldsymbol{w}) - \boldsymbol{w}t\cdot \nabla_xf_G^k
  \end{equation}
 for $k=l,r$. Based on the Chapman-Enskog expansion, $f_G^k$ is given by
  \begin{equation}\label{ce_ex}
    f_G^k= g^k[1-\tau(A^k +  \boldsymbol{a}^k \cdot \boldsymbol{w})],
  \end{equation}
where $g^k$ is the equilibrium distribution function defined by the macroscopic variables $\boldsymbol{W}^k$ at both sides of a cell interface, $\boldsymbol{a}^k$ are defined by the spatial derivatives of $g^k$
\begin{equation*}
	\begin{aligned}
a_i^k&= (\frac{\partial g^k}{\partial \rho^k }\frac{\partial \rho^k}{\partial x_i}+ \frac{\partial g^k}{\partial W_1^k }\frac{\partial W_1^k}{\partial x_i} + \frac{\partial g^k}{\partial W_2^k }\frac{\partial W_2^k}{\partial x_i} + \frac{\partial g^k}{\partial W_3^k }\frac{\partial W_3^k}{\partial x_i}+ \frac{\partial g^k}{\partial \lambda^k}\frac{\partial \lambda^k}{\partial x_i})/g^k
\\&= a_{i1}^k +a_{i2}^k w_1 + a_{i3}^k w_2 + a_{i4}^k w_3 +a_{i5}^k \frac{1}{2}(w_1^2+w_2^2+w_3^2\xi^2),
	\end{aligned}
\end{equation*}
and
  \begin{equation*}
A^k=A_{1}^k +A_{2}^k w_1 + A_{3}^k w_2 + A_{4}^k w_3 +A_{5}^k \frac{1}{2}(w_1^2+w_2^2+w_3^2\xi^2)
\end{equation*}
are determined  by compatibility condition
  \begin{equation}\label{compatibility}
\int(f^k_G-g^k) \Psi\td\boldsymbol{w}\td\Xi=0.
\end{equation}
  Substituting Eq.(\ref{taylor_ex}) and (\ref{ce_ex}) into (\ref{ini_gas}), the initial gas distribution has following form
  \begin{equation}\label{inidis}
    \begin{aligned}
    f_{0}= \begin{cases}g^{l}\left[1-\left(\boldsymbol{a}^{l} \cdot \boldsymbol{w}\right) t-\tau\left(A^{l}+\boldsymbol{a}^{l} \cdot \boldsymbol{w} \right)\right], & x_1<0 , \\
      g^{r}\left[1-\left(\boldsymbol{a}^{r} \cdot \boldsymbol{w}\right) t-\tau\left(A^{r}+\boldsymbol{a}^{r} \cdot \boldsymbol{w}\right)\right], & x_1 \geq 0 ,\end{cases}
    \end{aligned}
    \end{equation}
Then, the equilibrium distribution is defined by the Taylor expansion
    \begin{equation}\label{equdis}
      \begin{aligned}
        g(\boldsymbol{x^{\prime}},t^{\prime},\boldsymbol{w})
        =&\overline{g}(\boldsymbol{x},0,\boldsymbol{w}) + \nabla_x \overline{g}\cdot (\boldsymbol{x^{\prime}} -\boldsymbol{x}) + \overline{g}_t t^{\prime} \\
        =&\overline{g}(\boldsymbol{x},0,\boldsymbol{w}) - \nabla_x \overline{g}\cdot \boldsymbol{w}(t-t^{\prime})   + \overline{g}_t t^{\prime} \\
        =&\overline{g}(\boldsymbol{x},0,\boldsymbol{w}) \left\{ 1 - \overline{\boldsymbol{a}}\cdot \boldsymbol{w}(t-t^{\prime})+A t^{\prime} \right\} , \\
       \end{aligned}
    \end{equation}
where $\overline{g}$ and $\overline{\boldsymbol{a}}$ are determined from the reconstruction of macroscopic flow variables presented in section \ref{equ_res}, and $A$ is obtained by compatibility condition (\ref{compatibility}).

In smooth flow region, the collision time for viscous flow is determined by $\tau=\mu/p$, where $\mu$ is the dynamic viscosity coefficient and $p$ is the pressure at the cell interface, and for inviscid flow, the collision time should be set as zero. In order to properly capture the un-resolved shock structure, additional numerical dissipation is needed. The physical collision time $\tau$ in the exponential function part can be replaced by a numerical collision time $\tau_n$. For inviscid flow, it is set as
$$\tau_n=C_1\Delta t + C_2\frac{|p_l-p_r|}{p_l+p_r}\Delta t ,$$
and for viscous flow, it is
$$\tau_n=\tau + C_2\frac{|p_l-p_r|}{p_l+p_r}\Delta t ,$$
where $p_l$ and $p_r$ denote the pressure on the left and right sides of the cell interface. In this paper, we have $C_1=0.01, C_2=5.0$.

By substituting  Eq. (\ref{inidis}) and Eq. (\ref{equdis}) into Eq. (\ref{solbgk}) with $\tau$ and $\tau_n$,
the second-order time-dependent gas distribution function at a cell interface becomes
  \begin{equation}\label{solution}
    \begin{aligned}
    f\left(\boldsymbol{x}, t, \boldsymbol{w}, \xi\right)
    &=(1-e^{-t / \tau_n})\bar{g}+e^{-t / \tau_n}\left[\mathbb{H}\left(w_1\right) g^{l}+\left(1-\mathbb{H}\left(w_1\right)\right) g^{r}\right] +t\bar{A} \bar{g}\\
    &-\tau\left(1-e^{-t / \tau_n}\right)\bar{g}\left(\overline{\boldsymbol{a}} \cdot \boldsymbol{w} +\bar{A} \right)\\
    &-\tau e^{-t / \tau_n} \mathbb{H}\left(w_1\right)g^{l}\left(\boldsymbol{a}^{l}\cdot  \boldsymbol{w} + A^{l}\right)\\
    &-\tau e^{-t / \tau_n} \left(1-\mathbb{H}\left(w_1\right)\right)g^{r} \left(\boldsymbol{a}^{r} \cdot \boldsymbol{w} + A^{r}\right)\\
    &+t e^{-t / \tau_n}\left[ \left(\overline{\boldsymbol{a}} \cdot \boldsymbol{w}\right)\bar{g}- \mathbb{H}\left(w_1\right)\left(\boldsymbol{a}^{l} \cdot \boldsymbol{w}\right) g^{l}-\left(1-\mathbb{H}\left(w_1\right)\right)\left(\boldsymbol{a}^{r} \cdot \boldsymbol{w}\right) g^{r}\right].\\
  \end{aligned}
    \end{equation}
The fluxes in Eq.(\ref{flux}) can be obtained by taking the moments of the above distribution function. The calculation of moments can be found in \cite{xuGKS2001}.

In order to account for the Prandtl number not equal to 1, additional correction heat flux should be added in energy flux:
\begin{equation*}
  (\frac{1}{\text{Pr}}-1)q,
\end{equation*}
where $q = F_{v4}-W_1F_{v1}-W_2F_{v2}-W_3F_{v3}$ can be obtained through the viscous flux $\boldsymbol{F}_{vis}=[0,F_{v1},F_{v2},F_{v3},F_{v4}]$. The viscous flux can be obtained by
\begin{equation*}
  \boldsymbol{F}_{vis}=-\tau \int w_1\bar{g}\left[ \left(\overline{\boldsymbol{a}} \cdot \boldsymbol{w}\right) +\bar{A} \right]\td{\boldsymbol{w}}\td\Xi.
\end{equation*}

  \subsection{Evolution of the cell-averaged spatial gradients}
 By taking moments of the above gas distribution function in Eq. (\ref{solution}), the time-accurate conservative flow variables 
 in the moving frame at the Gaussian points can also be obtained
\begin{equation}\label{conval}
  \boldsymbol{W}_{p_t,k}^\prime (t^{n+1})= \mathbf{T}^\prime\left(\int  \widetilde{\boldsymbol{ \Psi}} f(\tilde{\boldsymbol{x}}_{p_t,k},t^{n+1},\tilde{\boldsymbol{w}},\xi) \td\boldsymbol{w}\td\Xi \right),
\end{equation}
by which the corresponding conservative flow variables in the absolute inertia frame $\boldsymbol{W}_{p_t,k} (t^{n+1})$ can be obtained.
According to the Divergence theorem, the cell averaged gradients over cell $\Omega_i$ at time $t^{n+1}$ are
\begin{equation}\label{slop1}
  \overline{\nabla \boldsymbol{W}}^{n+1}_i = \frac{1}{|\Omega_i^{n+1}|}\sum_{p_t=1}^{N_t}\int_{\Sigma_{p_t}} \boldsymbol{W}^{n+1} \boldsymbol{n}_p dS,
\end{equation}
where the surface integration can be calculated by Gaussian quadrature
\begin{equation}\label{slop2}
  \int_{\Sigma_{p_t}} \boldsymbol{W}^{n+1}  \boldsymbol{n}_{p_t} dS\approx \sum_{k=1}^{3}|S_{p_t}| \omega_k \boldsymbol{W}_{p_t,k}(t^{n+1})\boldsymbol{n}_{p_t}.
\end{equation}
Besides evaluating the cell averaged gradients,  the solution updates of the scheme are presented next.

\section{Temporal discretization and Geometrical Conservation Law}
\subsection{Mesh Velocity}
This article considers two types of mesh motion when the mesh velocity is not given directly. The first method is the radial basic function (RBF) method, which is used to address moving boundary problems. The second method is the Lagrangian velocity method, which is used to handle moving shock problems.
\subsubsection{Radial Basic Function (RBF) Method}
For moving boundary problems, we can obtain the boundary moving vector in the next step on boundary mesh, and the moving vectors of interior grids can be obtained by radial basic function interpolation \cite{RBF2007784}. We chose the radial basic function as
\begin{equation*}
  \psi(\xi)=(1-\xi)^4(1+4\xi),
\end{equation*}
where $\xi=|\boldsymbol{x}|/R$ is the related distance, and $R$ is the compactly supported radius. $|\boldsymbol{x}|$ is calculated by $\sqrt{x_1^2+x_2^2+x_3^2}$. Then, we can define an interpolation function
\begin{equation*}
  f(\boldsymbol{x})=\sum_i\lambda_i\psi(|\boldsymbol{x}-\boldsymbol{x_i}|/R),
\end{equation*}
where the $\boldsymbol{x}_i$ is a series of given points, and 
\begin{equation*}
  f(\boldsymbol{x_j})=\sum_i\lambda_i\psi(|\boldsymbol{x_j}-\boldsymbol{x_i}|/R)=g_j,
\end{equation*}
so we can solve $\lambda_i$ in above equation. The whole process of RBF interpolation is shown as Algorithm \ref{interpolation_rbf}.
\begin{algorithm}[H]
  \caption{RBF interpolation}
  \label{interpolation_rbf}
  \begin{algorithmic}[1]
  \REQUIRE This algorithm requires a series of points $\boldsymbol{x}_i$ and their corresponding data $g_i$, as well as the number of points $N$ to be selected for interpolation, a compactly supported radius $R$, and a target interpolation point $\boldsymbol{x}_0$.
  \ENSURE This algorithm ensures that an interpolated result $g_0$ is returned.
  \STATE Select the closest $N$ points to point $\boldsymbol{x}_0$.
  \STATE Calculate an $N \times N$ matrix $M_{ij}$ using the formula $M_{ij}=\psi(|\boldsymbol{x}_i-\boldsymbol{x}_j|/R)$.
  \STATE Solve the equation $M_{ij} \lambda_j =g_i$ by obtaining the values of $\lambda_j$.
  \STATE Initialize $g_0$ to 0.
  \FOR{$i$ in range($N$)}
      \STATE Compute $\xi_i=|\boldsymbol{x}_i-\boldsymbol{x}_0|/R$.
      \IF{ $\xi_i<1.0$ }
          \STATE Add $\lambda_i\psi(\xi_i)$ to $g_0$.
      \ENDIF
  \ENDFOR
  \RETURN $g_0$
  \end{algorithmic}
\end{algorithm}
\subsubsection{Lagrangian Velocity Method}
The mesh velocity can be determined by the cell-centered Lagrangian nodal solver \cite{Lagrangianvel}. The basic idea is to solve two half one-dimensional Riemann problems at each cell interface by assuming that its velocity equals the point velocity. We denote $C(p)$ as the set of cells $c$ that share the common vertex $p$ and $F_p(c)$ as the set of faces of cell $c$ that share the common vertex $p$. The grid speed $\boldsymbol{U}_p$ can be given by
\begin{equation*}
  \boldsymbol{U}_p=\mathbb{M}_p^{-1}\sum_{c\in C(p)}\sum_{f\in F_p(c)}[S_fp_c\boldsymbol{N}_f^c+\mathbb{M}_{pcf}\boldsymbol{V}_c],
\end{equation*}
where $p_c$ and $\boldsymbol{V}_c$ are the pressure and velocity of cell $c$ and $S_f, \boldsymbol{N_f}$ is the surface area and normal direction. The matrix $\mathbb{M}_{pcf}$ is calculated by
\begin{equation*}
  \mathbb{M}_{pcf}=S_f\rho_c a_c(\boldsymbol{N}_f^c \otimes \boldsymbol{N}_f^c),
\end{equation*}
where $a_c$ and $\rho_c$ are the sound speed and density of cell $c$ and $\otimes$ is the operator of tensor production, and matrix $\mathbb{M}_p$ is the sum of $\mathbb{M}_{pcf}$ for faces which is a symmetric positive definite matrix. In this case, the coordinate of points in the next time step can be calculated by
\begin{equation*}
  \boldsymbol{x}_p^{n+1}=\boldsymbol{x}_p^n+\boldsymbol{U}_p\Delta t.
\end{equation*}

With the Lagrangian velocity, the meshes become distorted, and computation may break down. Therefore, the meshes need to be smoothed by
\begin{equation*}
  \widetilde{\boldsymbol{x}}_p=(1-\omega)\boldsymbol{x}_p+\omega \frac{1}{N_p}\sum_{ip\in N(p)}\boldsymbol{x}_{ip},
\end{equation*}
where $N(p)$ is the set of vertexes jointing with vertex $p$, $N_p$ is the size of $N(p)$, and $\omega$ is relaxation coefficient.
\subsection{Solution Update and Geometrical Conservation Law}
According to the semi-discretization equation Eq. (\ref{fvm}), the two-stage fourth-order (S2O4) temporal discretization is adopted here for the solution updates \cite{liTwoStageFourthOrder2016},
\begin{equation*}
  \begin{aligned}
    (\Omega_i\boldsymbol{W}_{i})^{*}=& (\Omega_i\boldsymbol{W}_{i})^{n}+\frac{1}{2} \Delta t \mathcal{L}\left(\boldsymbol{W}_{i}^{n},\Omega^n\right)+\frac{1}{8} \Delta t^{2} \frac{\partial}{\partial t} \mathcal{L}\left(\boldsymbol{W}_{i}^{n},\Omega^n\right) +\int_0^{1/2\Delta t} \mathbb{V}(t)\td t,\\
    (\Omega_i\boldsymbol{W}_{i})^{(n+1)}=& (\Omega_i\boldsymbol{W}_{i}) ^{n}+\Delta t \mathcal{L}\left(\boldsymbol{W}_{i}^{n},\Omega^n\right) +\frac{1}{6} \Delta t^{2}\left(\frac{\partial}{\partial t} \mathcal{L}\left(\boldsymbol{W}_{i}^{n},\Omega^n\right)+2 \frac{\partial}{\partial t} \mathcal{L}\left(\boldsymbol{W}_{i}^{*},\Omega^*\right)\right)+\int_0^{\Delta t} \mathbb{V}(t)\td t,\\
  \end{aligned}
  \end{equation*}
  which is a two-step time integration method. And constant grid velocities are considered during the time-integrating process. For simplicity, the conservative value at Gaussian points appears in correction term $\mathbb{V}(t)$ set as the moments of equilibrium state $\bar{g}$.

  With a uniform flow field, the time evolution of the cell moving volume can be calculated as
 \begin{equation*}
  \begin{aligned}
  (\Omega_i)^{*}&= (\Omega_i)^{n}+\frac{1}{2} \Delta t  \sum_{p_t=1}^{Nt}\sum_{k=1}^{3}\omega_k\boldsymbol{U}_{p_t,k}\cdot(\boldsymbol{x}^0_{,\xi}\times\boldsymbol{x}^0_{,\eta})_{p_t}\\
  &+\frac{1}{4}\Delta t^2 \sum_{p_t=1}^{Nt}\sum_{k=1}^{3}\omega_k\boldsymbol{U}_{p_t,k}\cdot\left[\frac{1}{2}(\boldsymbol{U}^0_{,\xi}\times\boldsymbol{x}^0_{,\eta}) +\frac{1}{2}(\boldsymbol{x}^0_{,\xi}\times\boldsymbol{U}^0_{,\eta})+\frac{1}{6} \Delta t(\boldsymbol{U}^0_{,\xi}\times\boldsymbol{U}^0_{,\eta})\right]_{p_t},\\
  (\Omega_i)^{n+1}&= (\Omega_i)^{n}+ \Delta t  \sum_{p_t=1}^{Nt}\sum_{k=1}^{3}\omega_k\boldsymbol{U}_{p_t,k}\cdot(\boldsymbol{x}^0_{,\xi}\times\boldsymbol{x}^0_{,\eta})_{p_t}\\
  &+\Delta t^2 \sum_{p_t=1}^{Nt}\sum_{k=1}^{3}\omega_k\boldsymbol{U}_{p_t,k}\cdot\left[\frac{1}{2}(\boldsymbol{U}^0_{,\xi}\times\boldsymbol{x}^0_{,\eta}) +\frac{1}{2}(\boldsymbol{x}^0_{,\xi}\times\boldsymbol{U}^0_{,\eta})+\frac{1}{3} \Delta t(\boldsymbol{U}^0_{,\xi}\times\boldsymbol{U}^0_{,\eta})\right]_{p_t},\\
  \end{aligned}
\end{equation*}
which automatically satisfies the Geometrical Conservation Law (GCL) when the volume of the element is taken as the sum of the tetrahedral volumes.

The gas distribution function at Gaussian points Eq. (\ref{solution}) can be written as a form
\begin{equation}
  f=f+tf_t,
\end{equation}
which denotes the second-order time accuracy of the gas distribution function.
A two-step evolution of the gas distribution function can be constructed, which is updated by
  \begin{equation*}
    \begin{aligned}
    f^{*} &=f^{n}+\frac{1}{2} \Delta t f_{t}^{n}, \\
    f^{n+1} &=f^{n}+\Delta t f_{t}^{*}.
    \end{aligned}
    \end{equation*}
Then, the time-dependent conservative values at each Gauss point can be obtained by Eq. (\ref{conval}). Then, by Eq. (\ref{slop1}) and Eq. (\ref{slop2}), the cell-averaged slopes can be updated.

\subsection{The Algorithm of ALE Method}
Algorithm \ref{alg1} presents the complete computational algorithm, where the bold text indicates the special treatments compared to the Eulerian formulation. The mesh velocity is determined once per time step, but the geometry requires the update in every substep of the two-stage fourth-order temporal discretization. This necessitates the updating of the volume and reconstruction matrix of cells, as well as the normal vector, area, and Gaussian points coordinates of interfaces in every substep.
\begin{algorithm}
	\renewcommand{\algorithmicrequire}{\textbf{Input:}}
	\renewcommand{\algorithmicensure}{\textbf{Output:}}
  \caption{CGKS in ALE formulation}
  \label{alg1}
  \begin{algorithmic}[1]
    \WHILE{the computation uncompleted}
    \STATE calculate the time step $\Delta t$ according to CFL number
      \STATE {\bf{ determine mesh velocity}}
      \FOR{i=1,2(for S2O4)}
      \STATE define boundary condition for ghost cell
      \STATE reconstruct the cell distribution
      \STATE evolution for interfaces
      \STATE {\bf{update the geometry }}
      \STATE update cell average conservative values and the first-order spatial derivatives of conservative values
      \ENDFOR
    \ENDWHILE
  \end{algorithmic}
  \end{algorithm}

\section{HWENO Reconstruction}
The 3rd-order compact reconstruction \cite{jiHWENOReconstructionBased2020} is adopted here with cell-averaged values and cell-averaged first-order spatial derivative. In order to capture shock, WENO weights \cite{zhu2020} and gradient compression factor (CF) \cite{jiGC2021} are used. The reconstruction was further improved \cite{zhang2023slidingmesh}, where only one large stencil and one sub-stencil are involved in the new WENO procedure.
\subsection{Third-order compact reconstruction for large stencil}
Firstly, a linear reconstruction is presented. A quadratic polynomial $p^2$ with a third-order accuracy in space is constructed as follows
\begin{equation*}
  \begin{aligned}
    p^2&=a_0+\frac{1}{h}[a_1(x-x_0)+a_2(y-y_0)+a_3(z-z_0)]\\
    &+\frac{1}{2h^2}[a_4(x-x_0)^2+a_5(y-y_0)^2+a_6(z-z_0)^2]\\
    &+\frac{1}{h^2}[a_7(x-x_0)(y-y_0)+a_8(y-y_0)(z-z_0)+a_9(x-x_0)(z-z_0)],
  \end{aligned}
\end{equation*}
where $h=\frac{|\Omega_0|}{\max_j S_j}$($|\Omega_0|$ is the volume of cell and $S_j$ is the area of cell's surface ) is the cell size, and $(x_0,y_0,z_0)$ is the coordinate of cell center.

The $p^2$ on $\Omega_0$  is constructed on the compact stencil $S$ including $\Omega_0$ and its all von Neumann neighbors $\Omega_m$ ($m=1,\cdots,N_f$, where $N_f=6$ for hexahedron cell or $N_f=5$  for triangular prism). The cell averages $\overline{Q}$ over $\Omega_0$ and $\Omega_m$ and cell averages of space partial derivatives $\overline{Q}_x,\overline{Q}_y $ and $\overline{Q}_z$ over $\Omega_m$ are used to obtain $p^2$.

The polynomial $p^2$ is required to exactly satisfy cell averages over both $\Omega_0$ and $\Omega_m$ ($m=1,\cdots,N_f$)
\begin{equation*}
  \iiint_{\Omega_0} p^2 \text{d}V = \overline{Q}_0|\Omega_0|,\iiint_{\Omega_m} p^2 \text{d}V = \overline{Q}_m|\Omega_m| ,
\end{equation*}
with the following condition satisfied in a least-square sense
\begin{equation*}
  \begin{aligned}
    \iiint_{\Omega_m}\frac{\partial}{\partial x} p^2 \text{d}V = \left(\overline{Q}_x\right)_m|\Omega_m|\\
    \iiint_{\Omega_m}\frac{\partial}{\partial y} p^2 \text{d}V = \left(\overline{Q}_y\right)_m|\Omega_m|\\
    \iiint_{\Omega_m}\frac{\partial}{\partial z} p^2 \text{d}V = \left(\overline{Q}_z\right)_m|\Omega_m|.\\
  \end{aligned}
\end{equation*}
The constrained least-square method is used to solve the above system.
\subsection{Green-Gauss reconstruction for the sub stencil}
The classical Green-Gauss reconstruction with only cell-averaged values is adopted to provide the linear polynomial $p^1$ for the sub-stencil.
\begin{equation*}
  p^1=\overline{Q}+\frac{1}{|\Omega_0|}\boldsymbol{x}\cdot \sum_{m=1}^{N_f}\frac{\overline{Q}_m+\overline{Q}_0}{2}S_m\boldsymbol{n} _m,
\end{equation*}
where $S_m$ is the area of the cell's surface and $\boldsymbol{n}_m$ is the surface's normal vector.
\subsection{Gradient compression Factor}
The CF was first proposed in \cite{jiGC2021}, and several improvements have been made in \cite{zhang2023slidingmesh}. Denote $\alpha_i \in[0,1]$ as gradient compression factor at targeted cell $\Omega_i$
\begin{equation*}
  \alpha_i = \prod_{p_t=1}^{n_t}\prod_{k=0}^{M_p}\alpha_{p_t,k},
\end{equation*}
where $\alpha_{p_t,k}$ is the CF obtained at the $k$th Gaussian point of the interface $p_t$ around cell $\Omega_i$, which can be calculated by
\begin{equation*}
  \begin{aligned}
    &\alpha_{p_t,k}=\frac{1}{1+A^2}, \\
    &A=\frac{|p^l-p^r|}{p^l} +\frac{|p^l-p^r|}{p^r}+(\text{Ma}^{l}_n-\text{Ma}^{r}_n)^2+(\text{Ma}^{l}_t-\text{Ma}^{r}_t)^2,
  \end{aligned}
\end{equation*}
where $p$ is pressure, $\text{Ma}_n$ and $\text{Ma}_t$ are the Mach numbers defined by normal and tangential velocity, and superscript $l,r$ denote the left and right values of the Gaussian points.

Then, the updated slope is modified by
\begin{equation*}
  \widetilde{\overline{\nabla \boldsymbol{W}}}_i^{n+1} = \alpha_i\overline{\nabla \boldsymbol{W}}_i^{n+1},
\end{equation*}
and the Green-Gauss reconstruction is modified as
\begin{equation*}
  p^1=\overline{Q}+\alpha \frac{1}{|\Omega_0|}\boldsymbol{x}\cdot \sum_{m=1}^{N_f}\frac{\overline{Q}_m+\overline{Q}_0}{2}S_m\boldsymbol{n} _m.
\end{equation*}
\subsection{Non-linear WENO weights}
In order to deal with discontinuity, the idea of multi-resolution WENO reconstruction is adopted \cite{jiGC2021,zhu2020}.
Here only two polynomials are chosen
\begin{equation*}
  P_2=\frac{1}{\gamma_2}p^2-\frac{\gamma_1}{\gamma_2}p^1  ,P_1 =p^1.
\end{equation*}
Here, we choose $\gamma_1=\gamma_2=0.5$. So, the quadratic polynomial $p^2$ can be written as
\begin{equation}\label{weno-linear}
  p^2 = \gamma_1P_1 + \gamma_2 P_2.
\end{equation}
Then, we can define the smoothness indicators
\begin{equation*}
  \beta_{j}=\sum_{|\alpha|=1}^{r_{j}}\Omega^{\frac{2}{3}|\alpha|-1} \iiint_{\Omega}\left(D^{\alpha} p^{j}(\mathbf{x})\right)^{2} \mathrm{~d} V,
  \end{equation*}
where $\alpha$ is a multi-index and $D$ is the derivative operator, and $r_1=1$ and $r_2=2$. Special care is given for $\beta_1$ for better robustness
  \begin{equation*}
    \beta_1=\min(\beta_{1,\text{Green-Gauss}},\beta_{1,\text{least-square}}),
  \end{equation*}
where $\beta_{1,\text{Green-Gauss}}$ is the smoothness indicator defined by Green-Gauss reconstruction, and $\beta_{1,\text{least-square}}$ is the smoothness indicator defined by second-order least-square reconstruction. Then, the smoothness indicators $\beta_i$ are non-dimensionalized by
  \begin{equation*}
    \tilde{\beta}_i=\frac{\beta_i}{Q_0^2+\beta_1+10^{-40}}.
  \end{equation*}
The nondimensionalized global smoothness indicator $\tilde{\sigma}$ can be defined as
  \begin{equation*}
    \tilde{\sigma}=\left|\tilde{\beta}_{1}-\tilde{\beta}_{0}\right|.
    \end{equation*}
Therefore, the corresponding nonlinear weights are given by
    \begin{equation*}
      \tilde{\omega}_{m}=\gamma_{m}\left(1+\left(\frac{\tilde{\sigma}}{\epsilon+\tilde{\beta}_{m}}\right)^{2}\right),\epsilon=10^{-5},
      \end{equation*}
      \begin{equation*}
        \bar{\omega}_{m}=\frac{\tilde{\omega}_{m}}{\sum \tilde{\omega}_{m}}, m=1,2.
      \end{equation*}
Replacing $\gamma_m$ in equation (\ref{weno-linear}) by $\bar{\omega}_{m}$ , the final non-linear reconstruction can be obtained
      \begin{equation*}
        R(\boldsymbol{x})=\bar{\omega}_2P_2+\bar{\omega}_1 P_1.
      \end{equation*}
The desired non-equilibrium states at Gaussian points become
\begin{equation*}
  Q_{p, k}^{l, r}=R^{l, r}\left(\boldsymbol{x}_{p, k}\right), \left(Q_{x_{i}}^{l, r}\right)_{p, k}=\frac{\partial R^{l, r}}{\partial x_{i}}\left(\boldsymbol{x}_{p, k}\right).
  \end{equation*}
\subsection{Reconstruction of equilibrium states}\label{equ_res}
After reconstructing the non-equilibrium state, a kinetic weighted average method can be used to get equilibrium states
and tangential derivatives \cite{jiHWENOReconstructionBased2020},
\begin{equation*}
  \int \bar{g}\Psi\td\boldsymbol{v}\td\Xi =\boldsymbol{W}_0 = \int_{\tilde{w}_1>0} g^{l} \Psi\td\boldsymbol{v}\td\Xi +\int_{\tilde{w}_1<0} g^{r}\Psi\td\boldsymbol{v}\td\Xi,
\end{equation*}
\begin{equation*}
  \int \bar{a}_i \bar{g}\Psi\td\boldsymbol{v}\td\Xi = \frac{\partial \boldsymbol{W}_{0} }{\partial \tilde{x}_i}= \int_{\tilde{w}_1>0}a_i^l g^{l} \Psi\td\boldsymbol{v}\td\Xi +\int_{\tilde{w}_1<0} a_i^r g^{r}  \Psi\td\boldsymbol{v}\td\Xi  , i=2,3.
\end{equation*}
For the normal derivatives, the above solution is further modified according to the idea in linear diffusive generalized Riemann problem (dGRP) \cite{dgrp}
\begin{equation}\label{dgrp}
  \int \bar{a}_1\bar{g}\Psi\td\boldsymbol{v}\td\Xi = \frac{\partial \boldsymbol{W}_{0} }{\partial \tilde{x}_1}=  \int_{\tilde{w}_1>0} a_1^l g^{l}  \Psi\td\boldsymbol{v}\td\Xi +\int_{\tilde{w}_1<0} a_1^r g^{r}  \Psi\td\boldsymbol{v}\td\Xi + \frac{\boldsymbol{W}^r-\boldsymbol{W}^l}{(\boldsymbol{x}_{rc}-\boldsymbol{x}_{lc}) \cdot \boldsymbol{n}},
\end{equation}
where $\boldsymbol{x}_{rc}$ and $\boldsymbol{x}_{lc}$ are the coordinates of left and right cell centroid,$\boldsymbol{W}^l$ and $\boldsymbol{W}^r$ are the reconstructed left and right conservative value at Gaussian points and $\boldsymbol{n}$ is the normal vector of interface. By adding a penalty term in Eq. (\ref{dgrp}), the whole scheme is essentially free from the odd-even decoupling phenomenon \cite{BLAZEK201573}.
\section{Numerical Ecperiments}
Both two-dimensional and three-dimensional cases are tested in this section. For two-dimensional problems, the three-dimensional solver is used with one layer and mirror boundary condition in the $z$ direction. The time step is given by $\Delta t= \min \Delta t_i$, where $\Delta t_i$ is defined in each cell
\begin{equation*}
  \Delta t_i =C_{CFL}\min(\frac{h}{|V|_i+c_i},\frac{h^2}{3\nu_i}),
\end{equation*}
where $C_{\text{CFL}}$ is the CFL number, $|V|_i, c_i$ and $\nu_i=(\mu/\rho)_i$ is the magnitude of related velocities, sound speed and kinematic viscosity coefficient of cell $i$. Here, we set the CFL number as 0.5 if not specified.
\subsection{Accuracy Test}
The advection of density perturbation for three-dimensional flow is presented to test the order of accuracy. The computation domain is $[0,1]\times[0,1]\times[0,1]$. Three uniform meshes with $10^3, 20^3, 40^3$ tetrahedral cells are used in this case. The flow at time $t$ is
\begin{equation*}
\begin{aligned}
    \rho(x,y,z) = 1+0.2\sin(2\pi x_r)\sin(2\pi y_r)\sin(2\pi z_r) ,\\
    p(x,y,z) =1,V_1=V_2=V_3=1,
\end{aligned}
\end{equation*}
where the $(x_r,y_r,z_r)$ depends on time
\begin{equation*}
  (x_r,y_r,z_r)=(x,y,z)-(1,1,1)t.
\end{equation*}
In order to validate the order of accuracy with moving meshes, the time-dependent moving meshes are considered in this form
\begin{equation*}
    \begin{cases}
    x&=x_0+0.05\sin(\pi t)\sin(2\pi x_0)\sin(2\pi y_0)\sin(2\pi z_0) \\
    y&=y_0+0.05\sin(\pi t)\sin(2\pi x_0)\sin(2\pi y_0)\sin(2\pi z_0) \\
    z&=z_0+0.05\sin(\pi t)\sin(2\pi x_0)\sin(2\pi y_0)\sin(2\pi z_0). \\
\end{cases}
\end{equation*}
The errors and numerical orders are shown in Table \ref{er_r}, which shows that third-order accuracy is achieved under moving mesh.
  \begin{table}[htb!]
    \centering
    \caption{The errors and accuracy in moving mesh}\label{er_r}
    \begin{tabular}{c|c|c|c|c|c|c}
      \hline
  mesh    &  $Error_{L^1}$& $O_{L^1}$  &  $Error_{L^2}$   &$O_{L^2}$       & $Error_{L^\infty}$&$O_{L^\infty}$  \\ \hline
$10^3$   &1.40E-2&		    &1.83E-02&		    &5.17E-02&	\\
$20^3$   &2.25E-3&	2.64	&2.96E-03&	2.63	&8.02E-03&	2.69\\
$40^3$  &3.04E-4&	2.88	&3.94E-04&	2.91	&1.08E-03&	2.89\\
\hline
    \end{tabular}
  \end{table}
As a reference, the accuracy test result on stationary mesh is shown in Table \ref{ersta}, which indicates that the errors of moving and fixed mesh are compatible with each other.
    \begin{table}[htb!]
    \centering
    \caption{The errors and accuracy in stationary mesh}\label{ersta}
    \begin{tabular}{c|c|c|c|c|c|c}
      \hline
  mesh    &  $Error_{L^1}$& $O_{L^1}$  &  $Error_{L^2}$   &$O_{L^2}$       & $Error_{L^\infty}$&$O_{L^\infty}$  \\ \hline
 $10^3$&1.38E-02&	&	1.81E-02&&		5.16E-02\\	
 $20^3$&2.07E-03&	2.74	&2.79E-03&	2.70	&7.92E-03&	2.70\\
 $40^3$&2.68E-04&	2.95	&3.61E-04&	2.95	&1.05E-03&	2.92\\
\hline
    \end{tabular}
  \end{table}

Next, geometric conservation law, which is about the maintenance of uniform flow passing through the moving mesh, is tested. The same mesh and same mesh moving method are set in the test. The initial condition is set as
\begin{equation*}
  \rho(x,y,z) = 1,
  p(x,y,z) =1,V_1=V_2=V_3=1.
\end{equation*}
 The density errors at $t=1$ are shown in Table \ref{errgcl}. The results indicate that the errors are reduced to machine zero, implying the satisfaction of the geometric conservation law.
  \begin{table}[htb!]
    \centering
    \caption{The errors of geometric conservation law in moving mesh}\label{errgcl}
    \begin{tabular}{c|c|c|c}
      \hline
  mesh    &  $Error_{L^1}$&  $Error_{L^2}$        & $Error_{L^\infty}$  \\ \hline
 $10^3$ &1.20148e-15  &1.49810e-15  & 6.21725e-15\\
 $20^3$ &  1.92325e-15 & 2.49943e-15   &1.43219e-14\\
 $40^3$ &4.19724e-15  &6.13169e-15  & 9.50351e-14\\
\hline
    \end{tabular}
  \end{table}

\subsection{Flow inside a cylinder}
\begin{figure}[hbt!]
  \centering
  \subfigure[The origin mesh]{  \includegraphics[width=5cm]{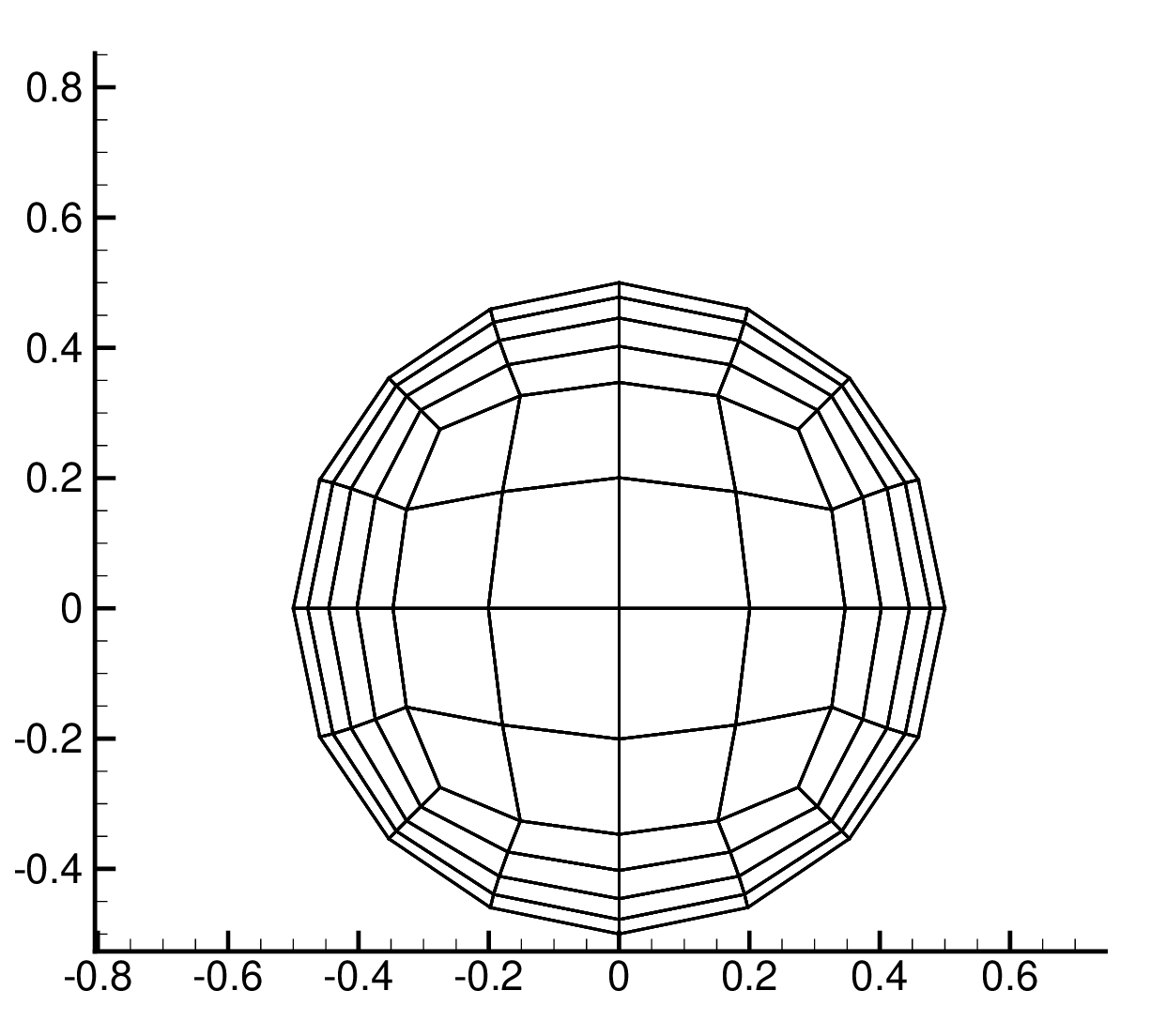}}
  \subfigure[The deformed mesh: $\alpha=0$]{  \includegraphics[width=5cm]{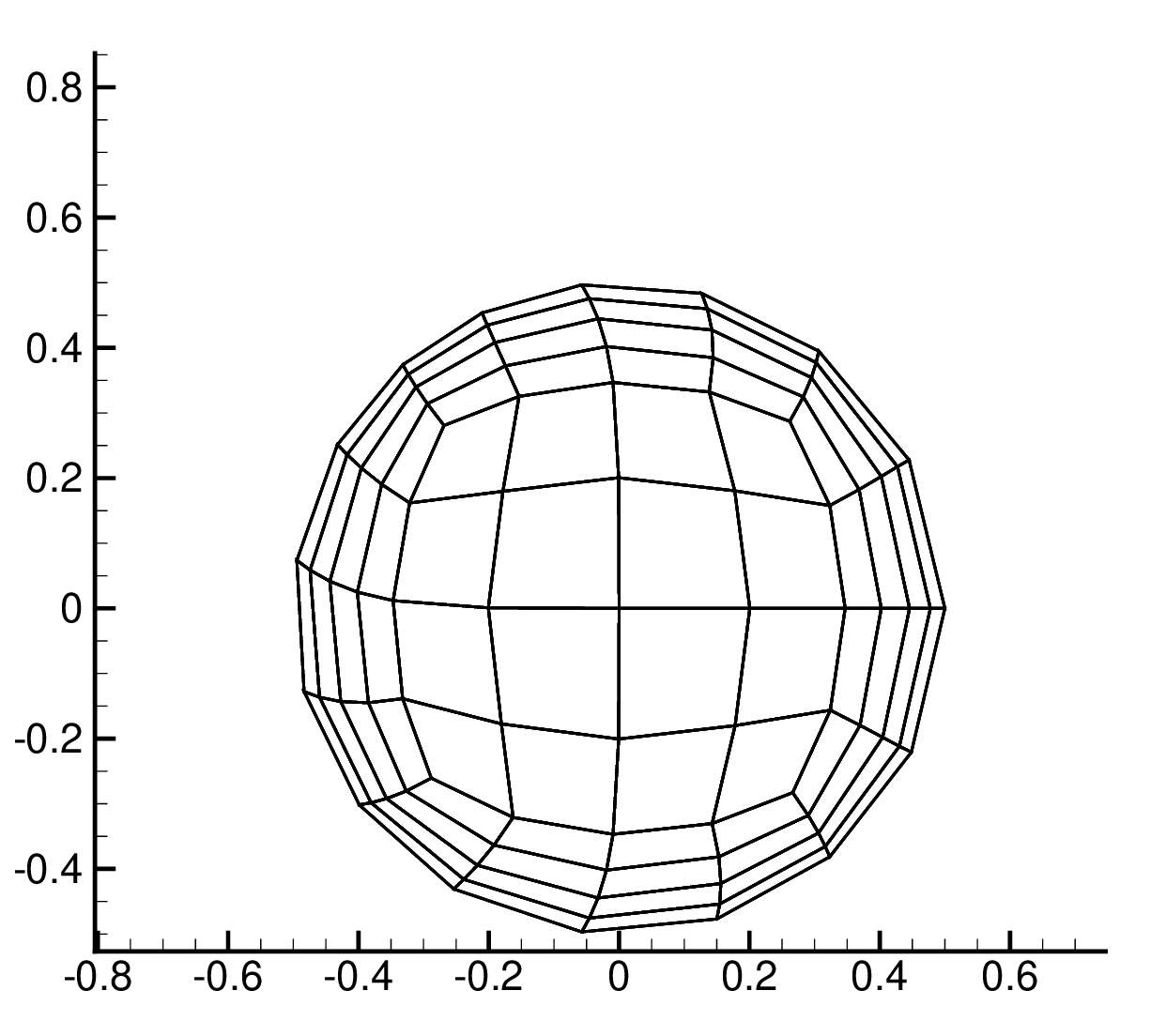}}
  \subfigure[The deformed mesh: $\alpha=1$]{  \includegraphics[width=5cm]{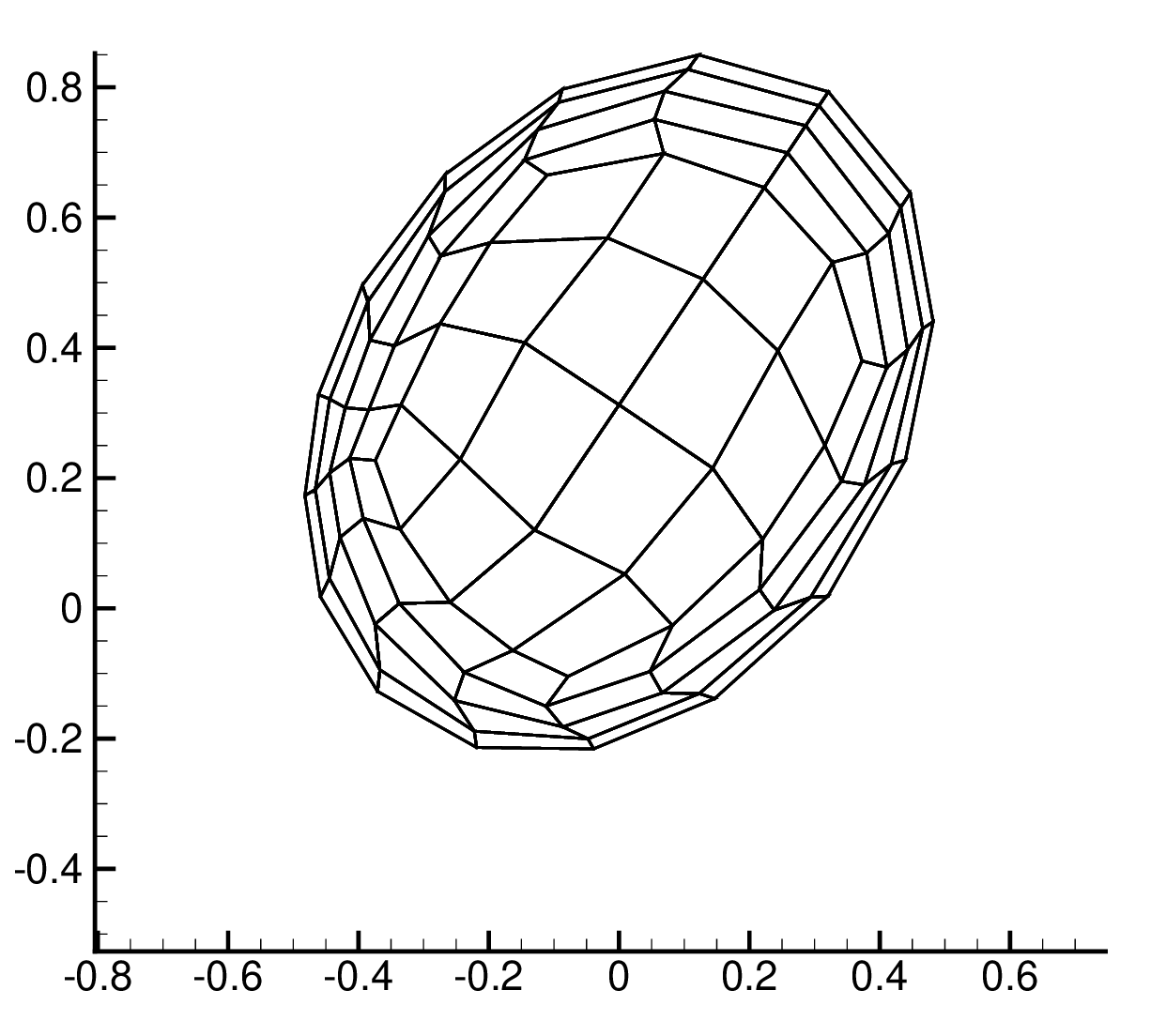}}
  \caption[]{An example mesh with cell number $8\times8\times5$}\label{cylinder_mesh}
\end{figure}

This case is provided by High-Fidelity CFD Verification Workshop \cite{high_Fidelity}.
It is about the flow inside a cylinder's interior volume. Five meshes (with mesh number $4\times4\times5, 8\times8\times5,16\times16\times5,32\times32\times5,64\times64\times5$) are tested in this case. The center of the cylinder is initially located on the original point, and the initial radius of the cylinder is $r_0=0.5$. The motion of meshes is defined by
\begin{equation*}
\left[\begin{array}{l}
x \\
y \\
1
\end{array}\right]=\left[\begin{array}{ccc}
1 & 0 & 0 \\
0 & 1 & \alpha(t) \\
0 & 0 & 1
\end{array}\right]\left[\begin{array}{ccc}
\cos \left(A_\theta \alpha(t)\right) & -\sin \left(A_\theta \alpha(t)\right) & 0 \\
\sin \left(A_\theta \alpha(t)\right) & \cos \left(A_\theta \alpha(t)\right) & 0 \\
0 & 0 & 1
\end{array}\right]\left[\begin{array}{ccc}
\psi(t) & 0 & 0 \\
0 & \frac{1}{\psi(t)} & 0 \\
0 & 0 & 1
\end{array}\right]\left[\begin{array}{c}
r_0 \cos \left(\theta_g(t)\right) \\
r_0 \sin \left(\theta_g(t)\right) \\
1
\end{array}\right],
\end{equation*}
as the composite of four motions (translation, rotation, const-volume deformation, and non-unit geometry mapping Jacobian). The translation motion is a motion in $y$ direction as $\alpha(t)$. And the rotation motion is about the center of the cylinder with a rotation amplitude $A_\theta=\pi$. The transformation of the cylinder deforming into an ellipse such that the interior area remains constant during deformation is facilitated by the function
\begin{equation*}
\psi(t)=1+\left(A_a-1\right) \alpha(t),
\end{equation*}
which varies from 1 to $A_a=1.5$ over $t=[0,2]$. Next, a parameterized function $\eta$ is defined as
\begin{equation*}
\eta(\lambda, \omega, \epsilon)=\sin (\omega \lambda+\epsilon (1-\cos (\omega \lambda))),
\end{equation*}
which is designed to break spatial and temporal symmetries due to the fact that the integral of $\eta$ over a period $\omega$ does not equal 0 for appropriate values of $\epsilon$. The parameter $\lambda$ represents the independent variable (e.g. $t$ or $\theta$), whereas $\omega$ and $\epsilon$ represent the function frequency and shape characteristics. A subsequent function $f_g$ is defined here to prescribe a deformation that ensures non-unit geometry mapping Jacobians ($g\neq 1$) while utilizing the function $\eta$  to break spatial and temporal symmetries. This is given by
\begin{equation*}
f_g\left(t, r_0, \theta_0\right)=\left(16 r_0^4+\frac{t^6}{t^6+0.01} \eta(t, 10,0.7)\left(\cos \left(32 \pi r_0^4\right)-1\right)\right) \eta\left(\theta_0, 1,0.7\right)
\end{equation*}
The prescribed deformation takes the form of a perturbation in $\theta$ as
\begin{equation*}
\theta_g\left(t, r_0, \theta_0\right)=\theta_0+A_g f_g\left(t, r_0, \theta_0\right),
\end{equation*}
which the volume deformation amplitude $A_g$ is set as 0.15. The origin mesh, deforming mesh when $\alpha=0$ and $\alpha=1$ are shown in Figure \ref{cylinder_mesh}, which indicates the large deforming and non-symmetry during the simulation.

The cylinder interior is prescribed with a no-slip, adiabatic wall boundary condition. The Reynolds number defined by cylinder diameter $d=1$ and reference velocity $V=1.0$ is $\text{Re}=1000$, a Prandtl number of 0.72, and constant viscosity is used. The initial condition is given by a constant zero velocity field, which $\rho=1$ and $\rho E=50$.
\subsubsection{Short Time Study}
\begin{figure}[hbt!]
  \centering
  \includegraphics[width=5cm]{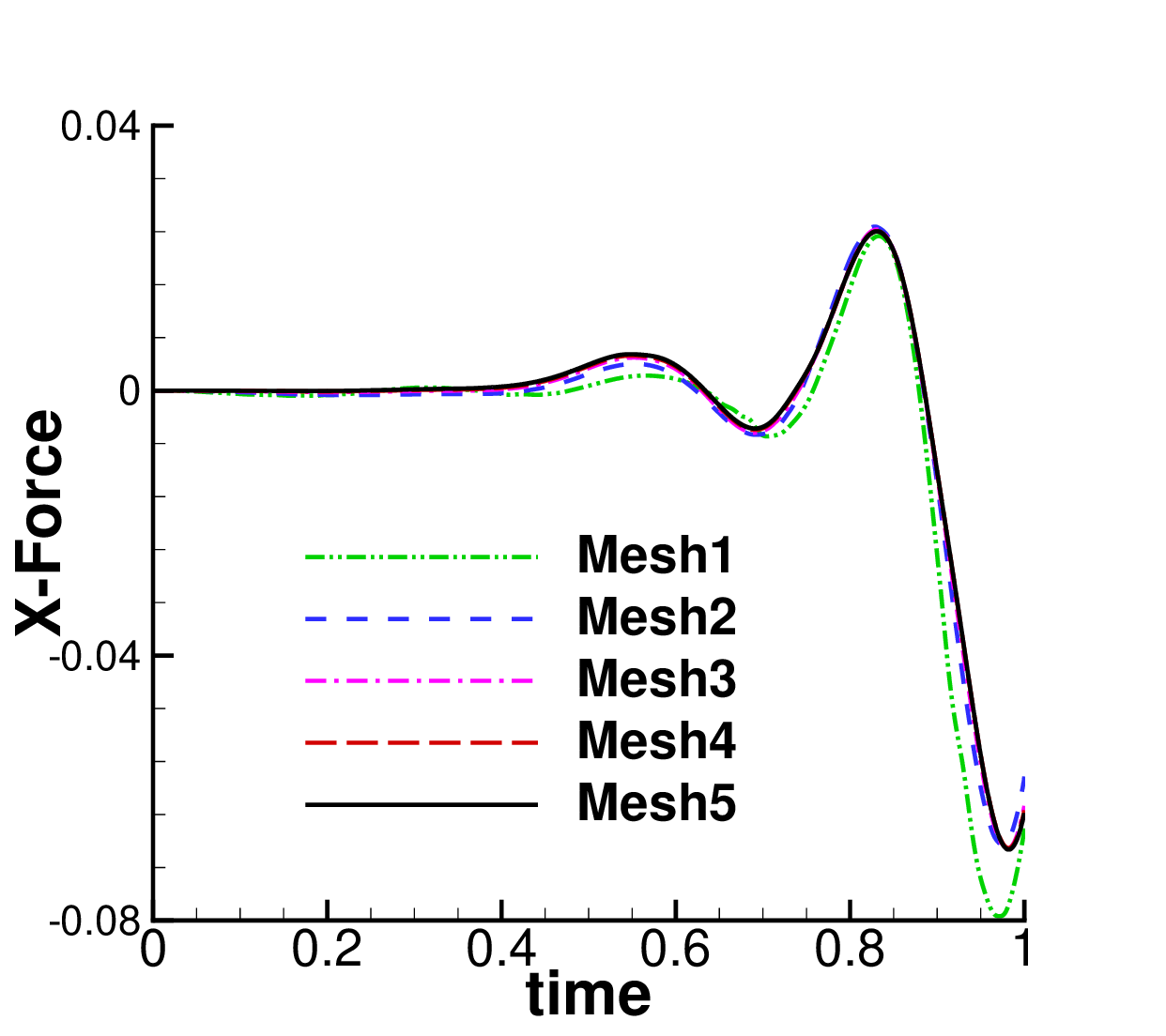}
    \includegraphics[width=5cm]{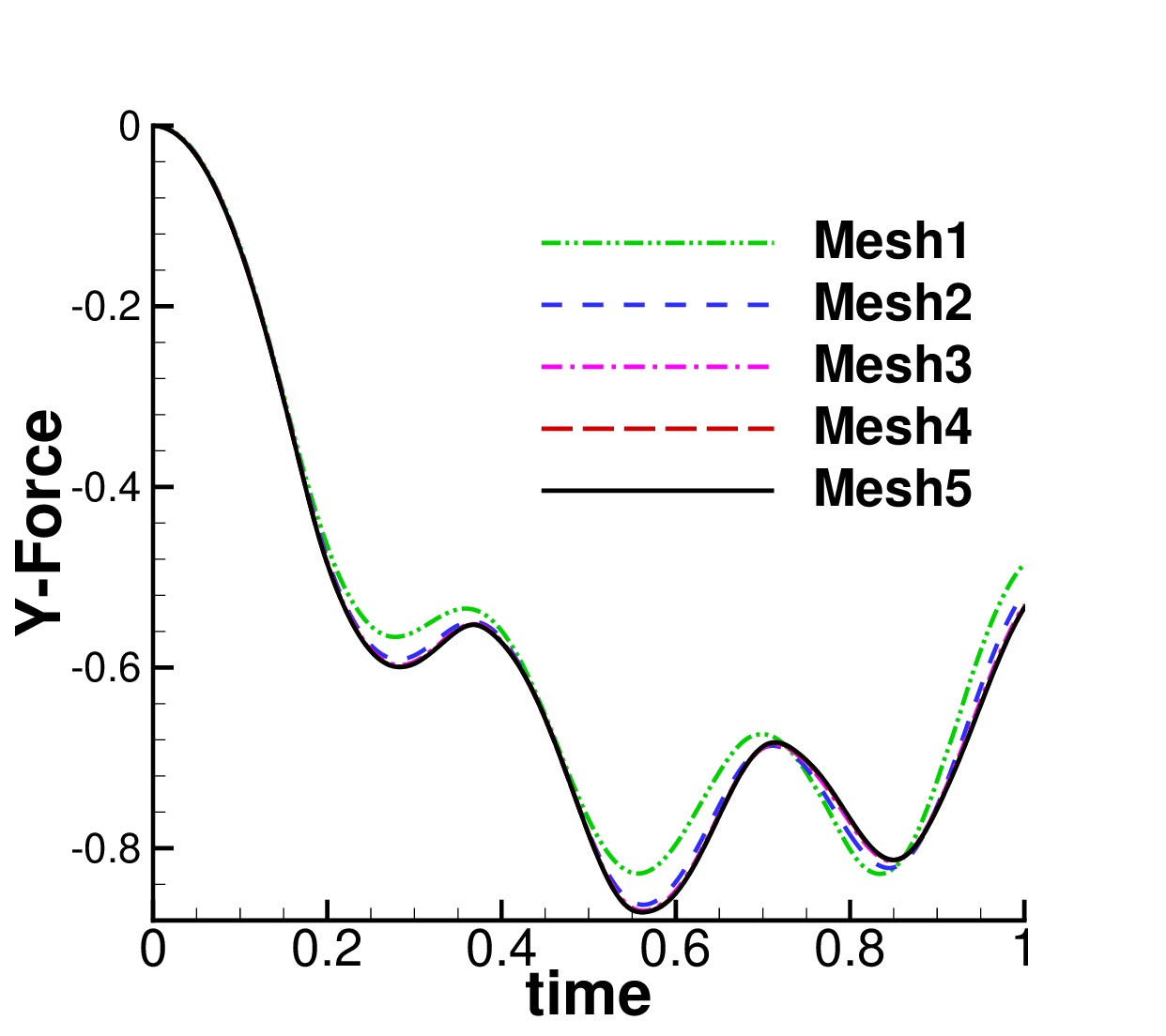}
      \includegraphics[width=5cm]{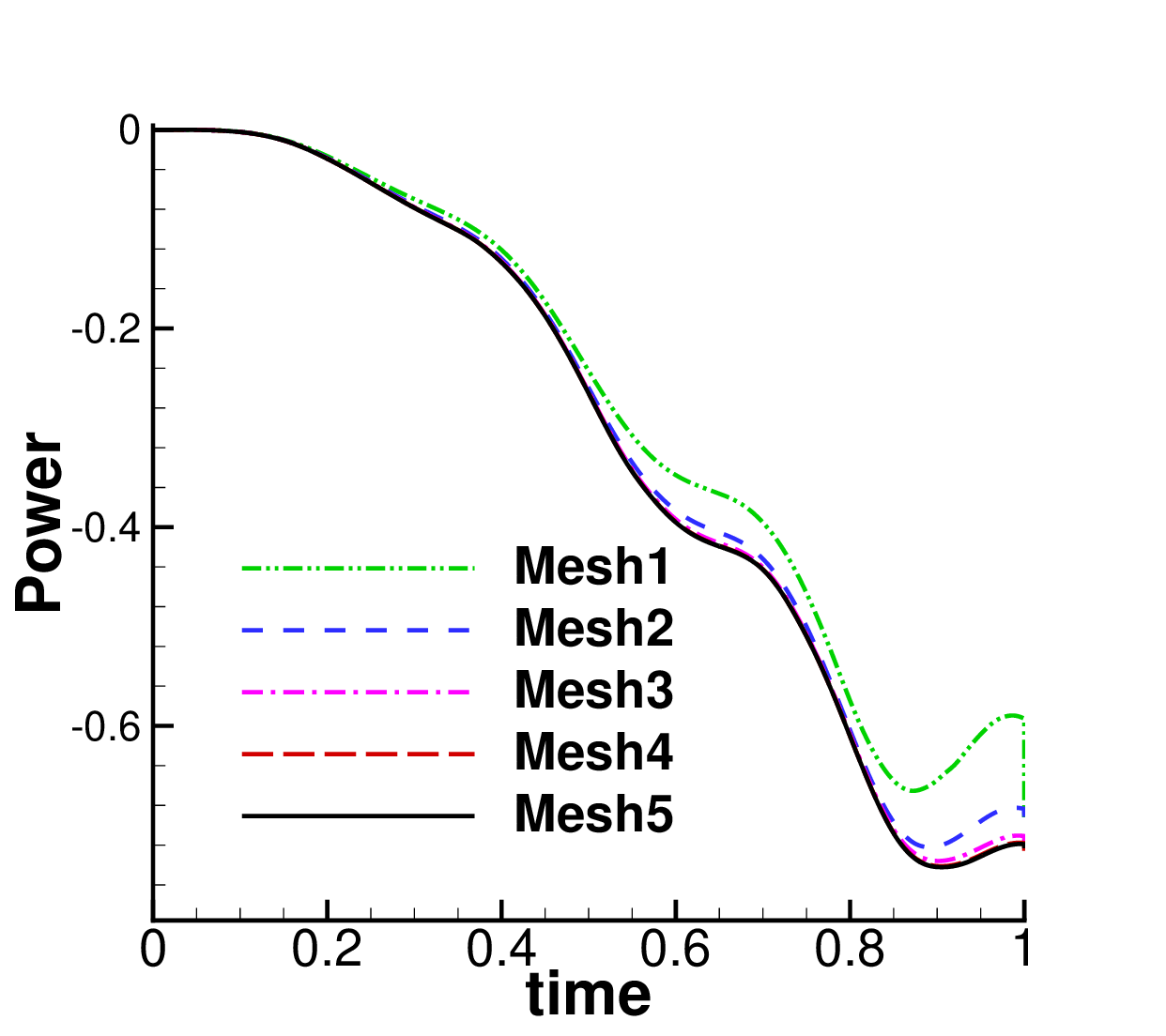}
  \caption[]{The time history force and power of short-time motion}\label{short_time_cylinder}
\end{figure}

The time-activation function $\alpha(t)$ for this case is defined by
\begin{equation*}
    \alpha (t)=t^3(8-3t)/16,
\end{equation*}
which varies from 0 to 1 on the interval $t=[0,2]$. This case runs from $t=0$ until $t=1$. The time history of force in $x$ and $y$ direction on the surface of the cylinder and power defined by
\begin{equation*}
  P=\int_{surface} \boldsymbol{U}\cdot \boldsymbol{f}_{surf} \td s,
\end{equation*}
where $\boldsymbol{f}_{surf}$ is the surface stress vector, are shown in Figure \ref{short_time_cylinder}.
The results indicate that the solutions obtained from Mesh 3, Mesh 4, and Mesh 5 are in good agreement with each other. This suggests that we have achieved mesh-independent solutions.
\subsubsection{Long-time Study of Conservation}
\begin{figure}[hbt!]
  \centering
    \includegraphics[width=5cm]{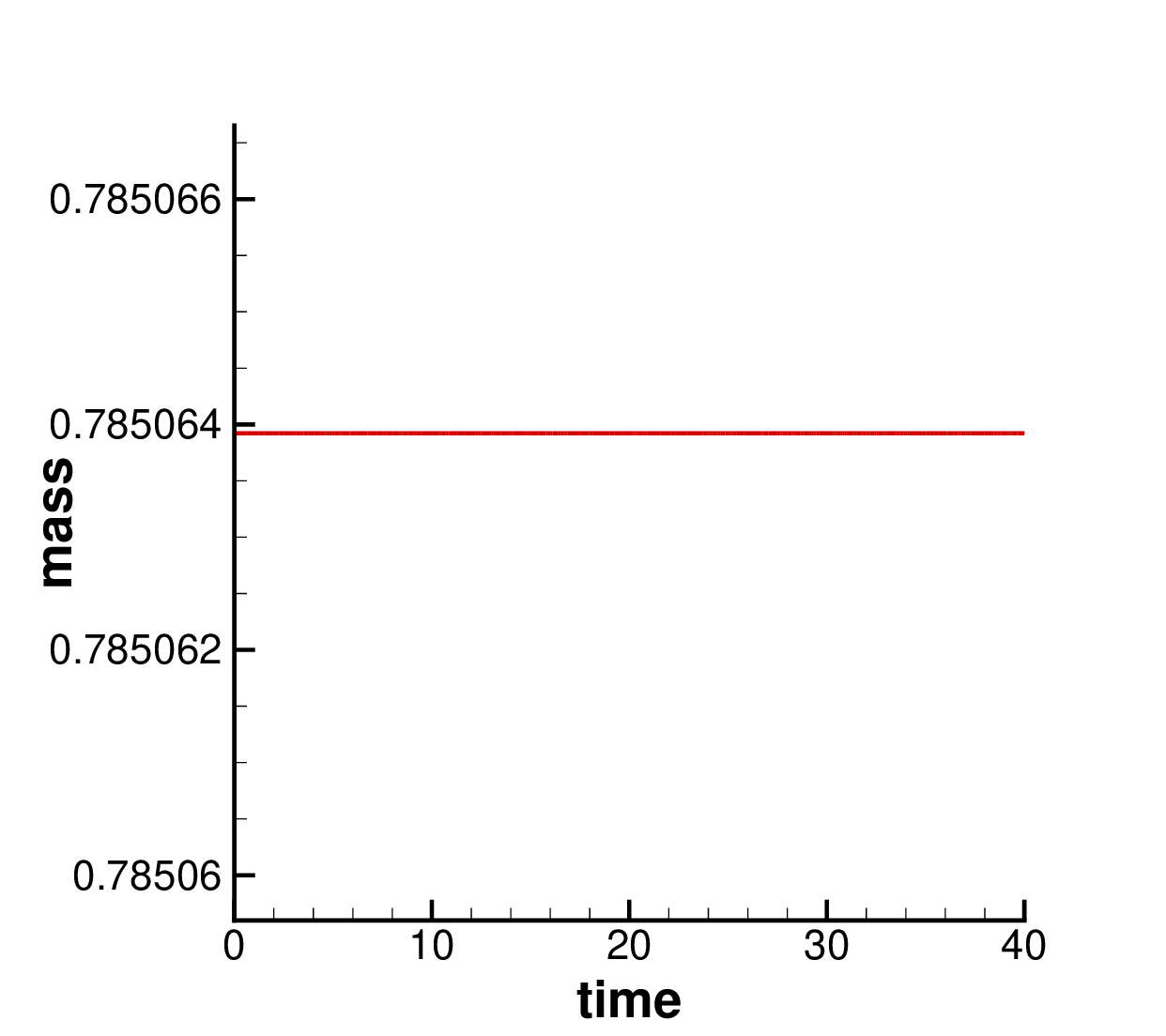}
    \includegraphics[width=5cm]{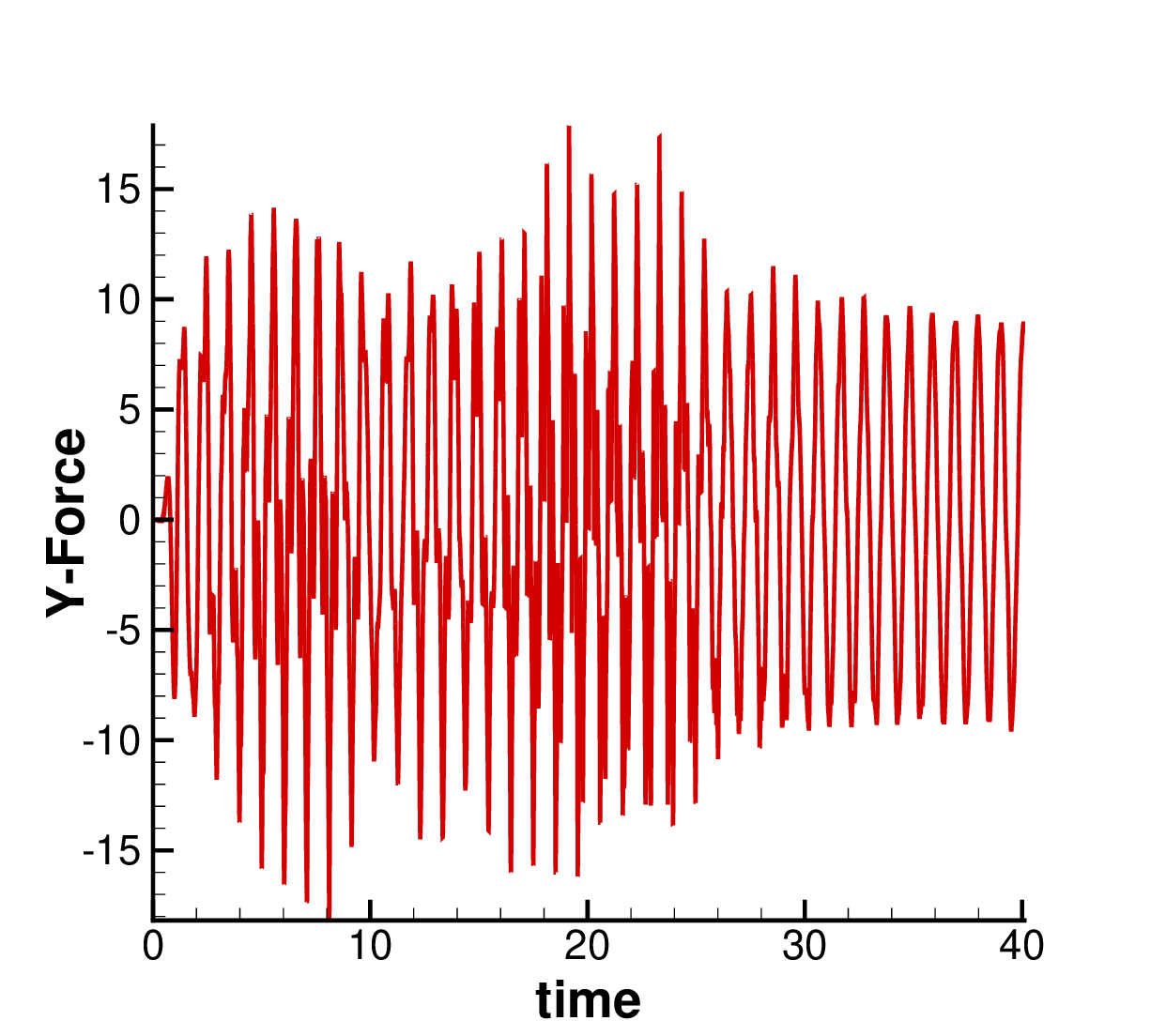}
    \includegraphics[width=5cm]{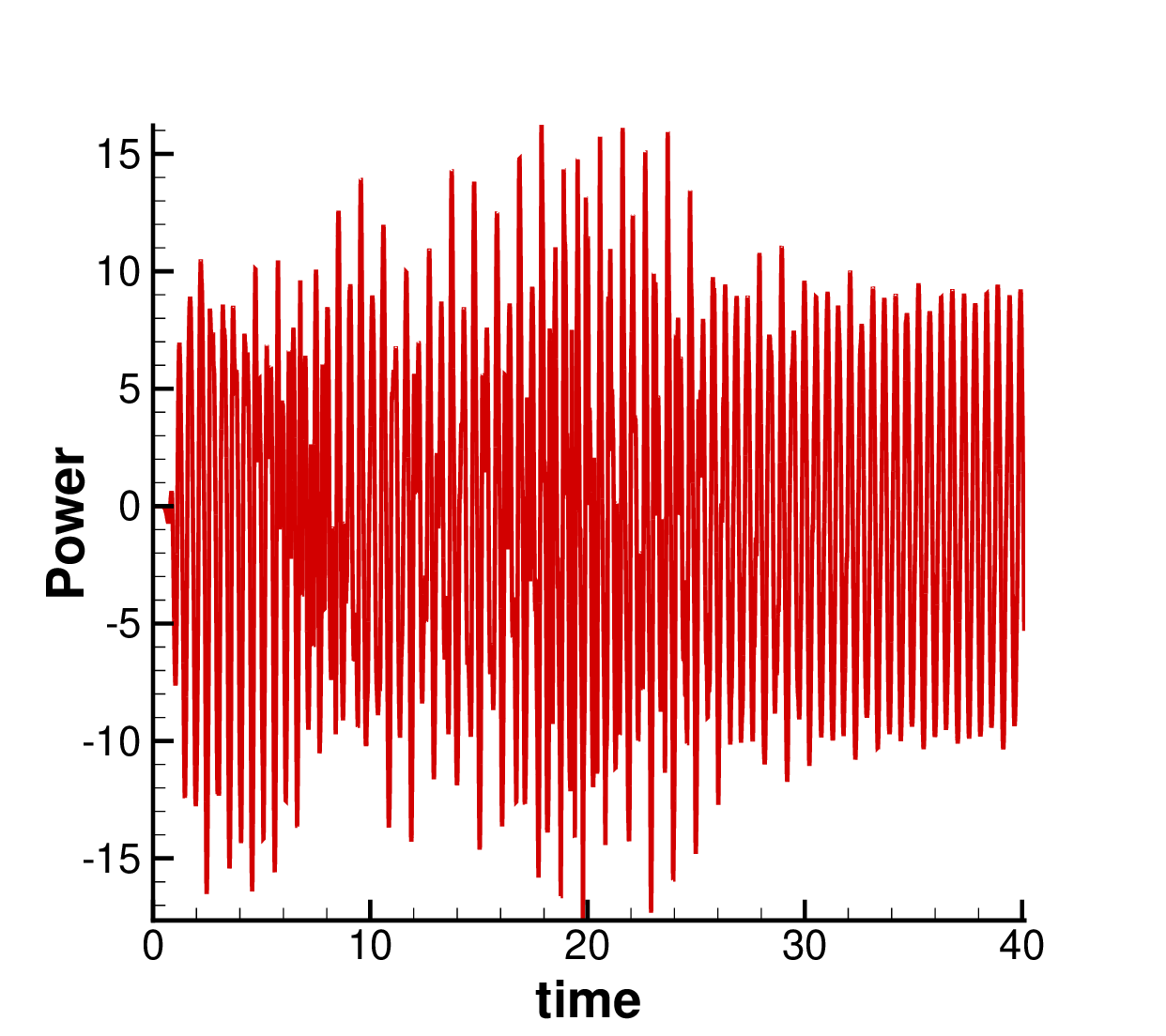}
  \caption[]{The time history force and power of long-time motion}\label{long_time_cylinder}
\end{figure}

The motion of the long-time study has the same form as that of the composite motion for the short-time study but with a different $\alpha(t)$ function. The $\alpha(t)$ is selected as
\begin{equation*}
    \alpha (t)=0.3 \sin (6t)\frac{t^6}{t^6+1}.
\end{equation*}
The purpose of the second term in the equation is to start the sinusoidal motion slowly at $t=0$. We simulate the process from $t=0$ to $t=40$. The mesh with $32\times32\times5$ cells is used in this study. The time history of mass, force in $y$ direction on the surface of the cylinder, and power are shown in Figure \ref{long_time_cylinder}.
The total mass of the entire domain remains constant at 0.785065 throughout the process, which indicates the mass conservation properties of our method. Additionally, the power and Y-Force exhibit periodic behavior after $t=30$.
\subsection{Heaving-pitching airfoil}
This case considers a NACA0012 airfoil undergoing a smooth flapping-type motion. The airfoil moves smoothly upward for a distance of one chord length over a duration of $T=2$  time units by heaving and pitching about a point located at the airfoil's 1/3 chord location. The functions governing the motion are provided below:
\begin{equation*}
    \begin{aligned}
        \Delta h(t)&=t^3(8-3t)/16, \\
        \Delta \theta(t)&=\frac{80\pi}{180}(-t^6+6t^5-12t^4+8t^3).
    \end{aligned}
\end{equation*}
Two types of motion are considered in this case. The first type of motion is defined solely by the heaving function $\Delta h(t)$, while the second type of motion is defined by both the heaving function $\Delta h(t)$ and the pitching function $\Delta \theta (t)$.

\begin{figure}[hbt!]
  \centering
    \subfigure[Mesh 1]{ \includegraphics[width=5cm]{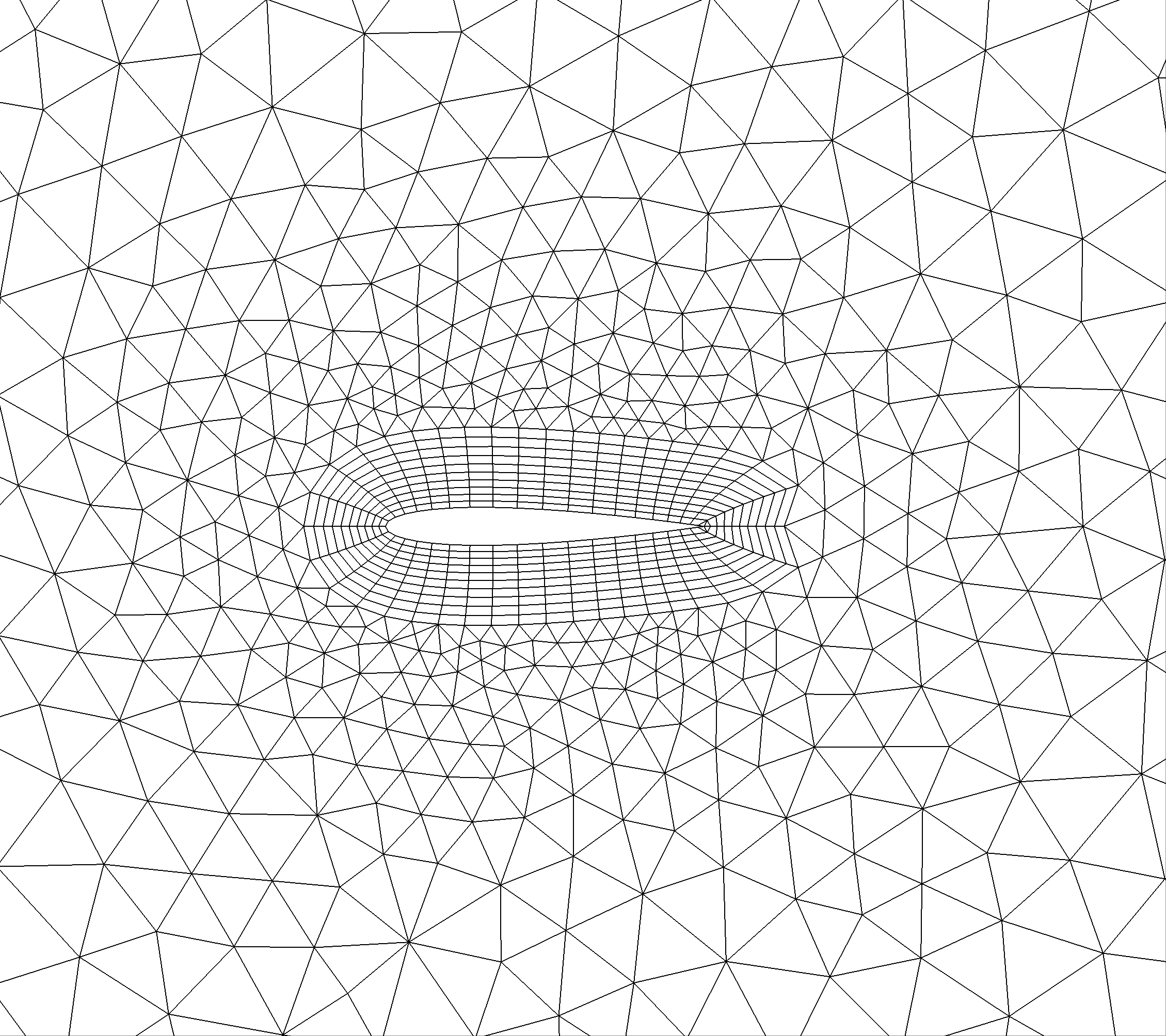}}
    \subfigure[Mesh 2]{ \includegraphics[width=5cm]{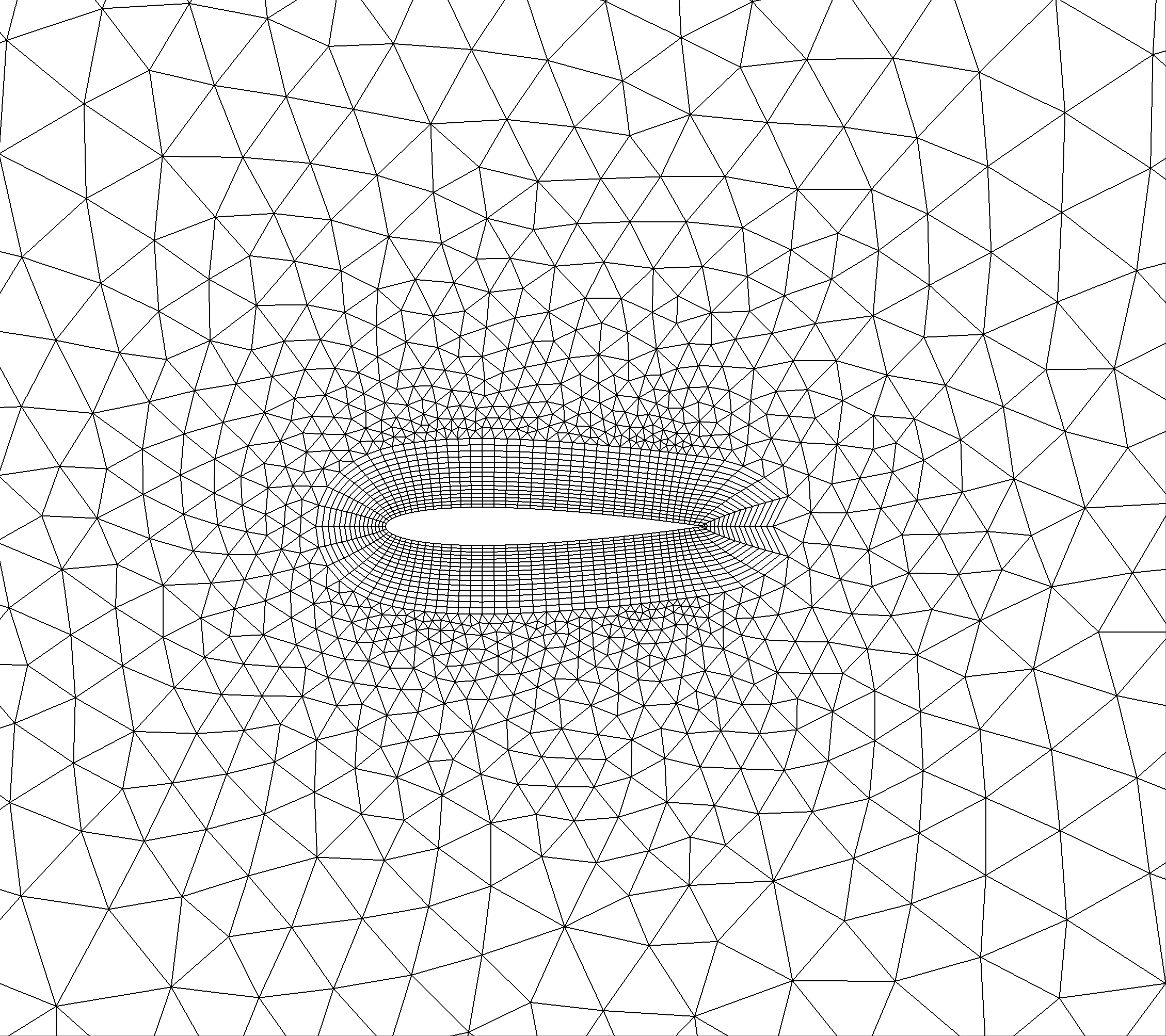}}
    \subfigure[Mesh 3]{ \includegraphics[width=5cm]{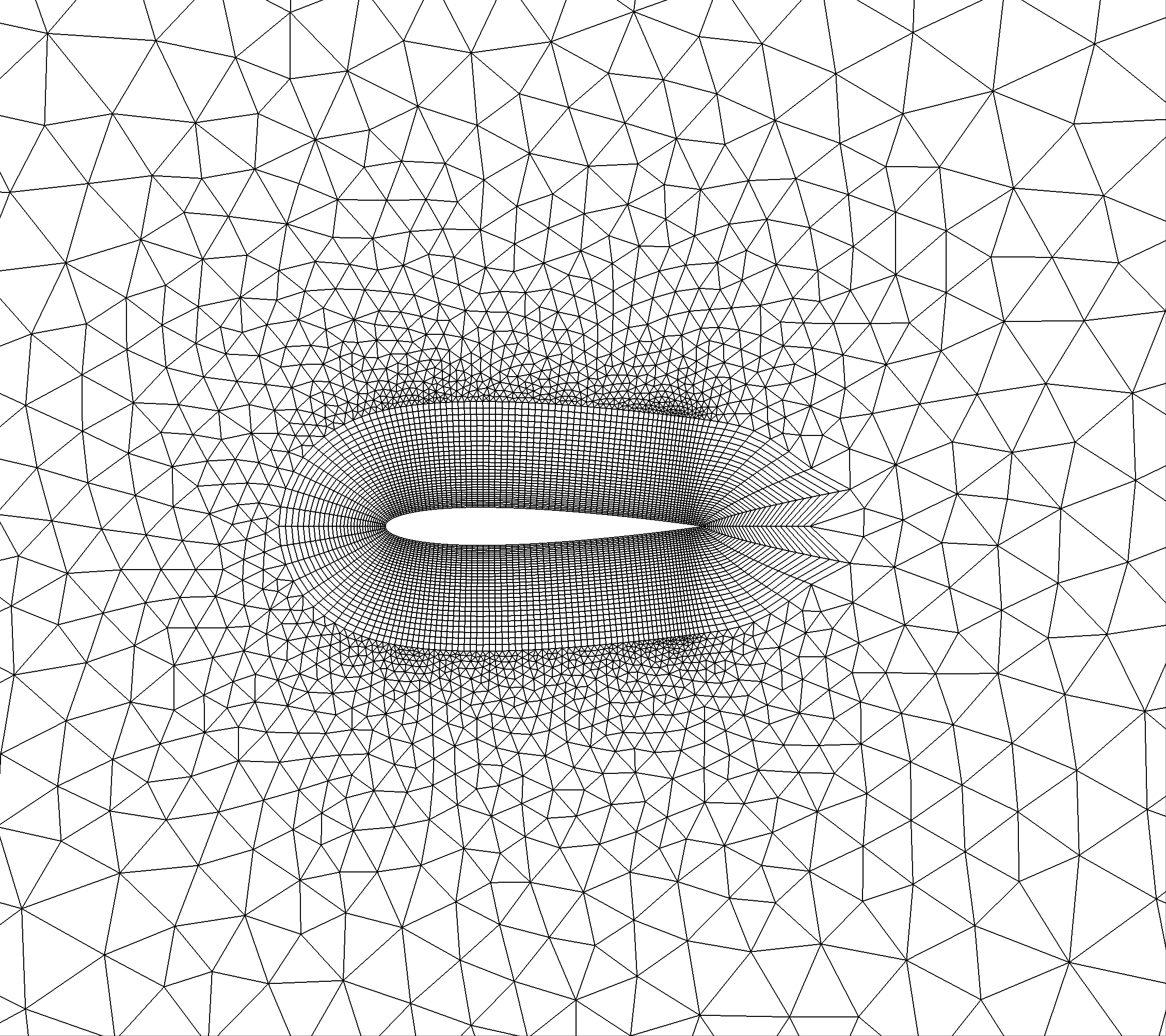}}
  \caption{Mesh of NACA 0012}\label{naca0012mesh}
\end{figure}

The income flow is set as $\rho_\infty=1.0, T_\infty=25, U_\infty=1$, which has a Mach number of 0.2. The Reynolds number based on the chord of the airfoil and incoming velocity is $\text{Re}=1000$, a Prandtl number of 0.72, and a constant viscosity is used. As shown in figure \ref{naca0012mesh}, three meshes are used in this case.
The adiabatic non-slip wall is set on the airfoil surface, and the far-field boundary condition is set as the outer boundary. The initial condition is the steady-state solution for the initial position $\Delta h=\Delta \theta=0$. The radius basic function method is used for mesh motion.

\begin{figure}[hbt!]
  \centering
    \includegraphics[width=5cm]{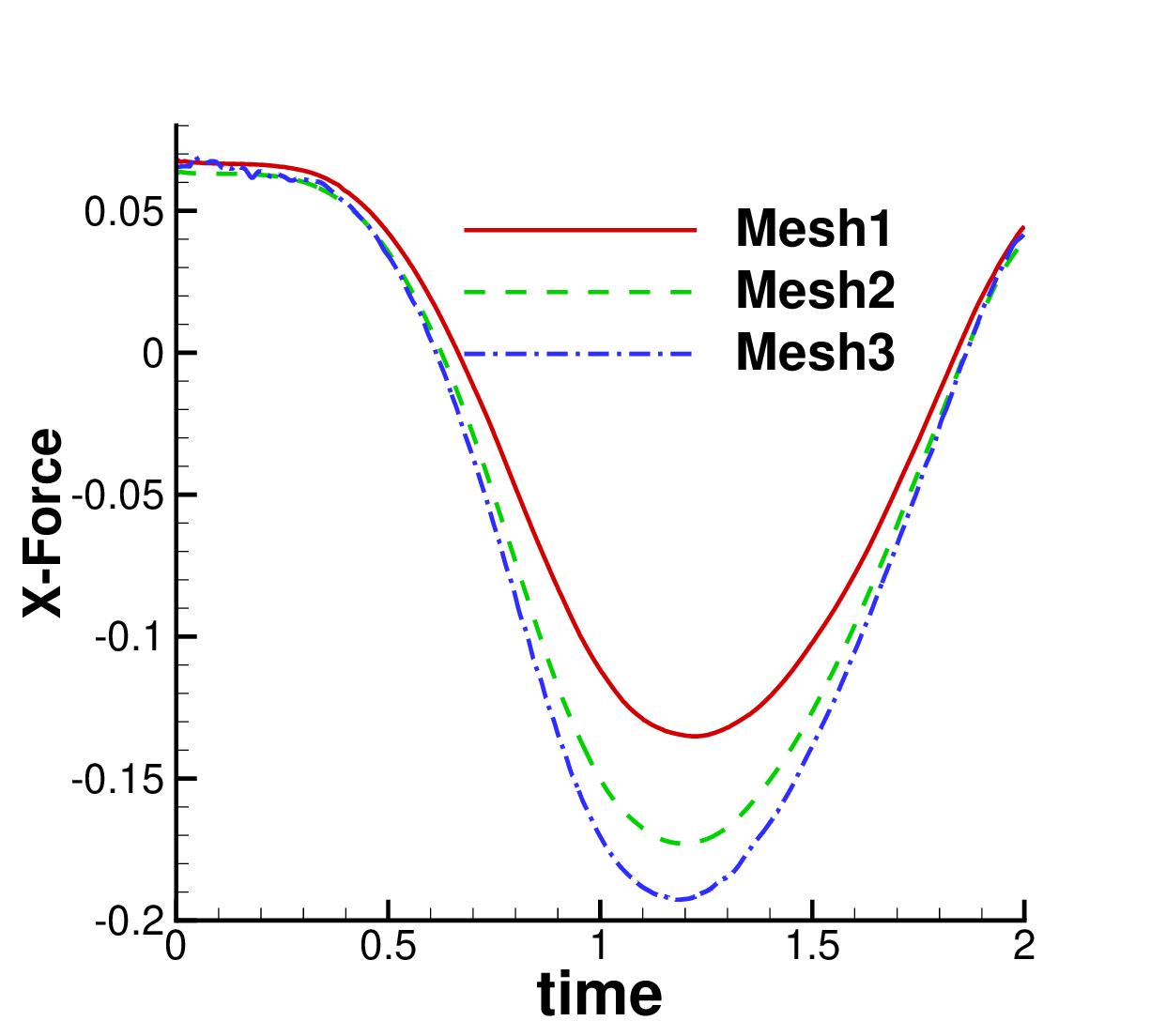}
    \includegraphics[width=5cm]{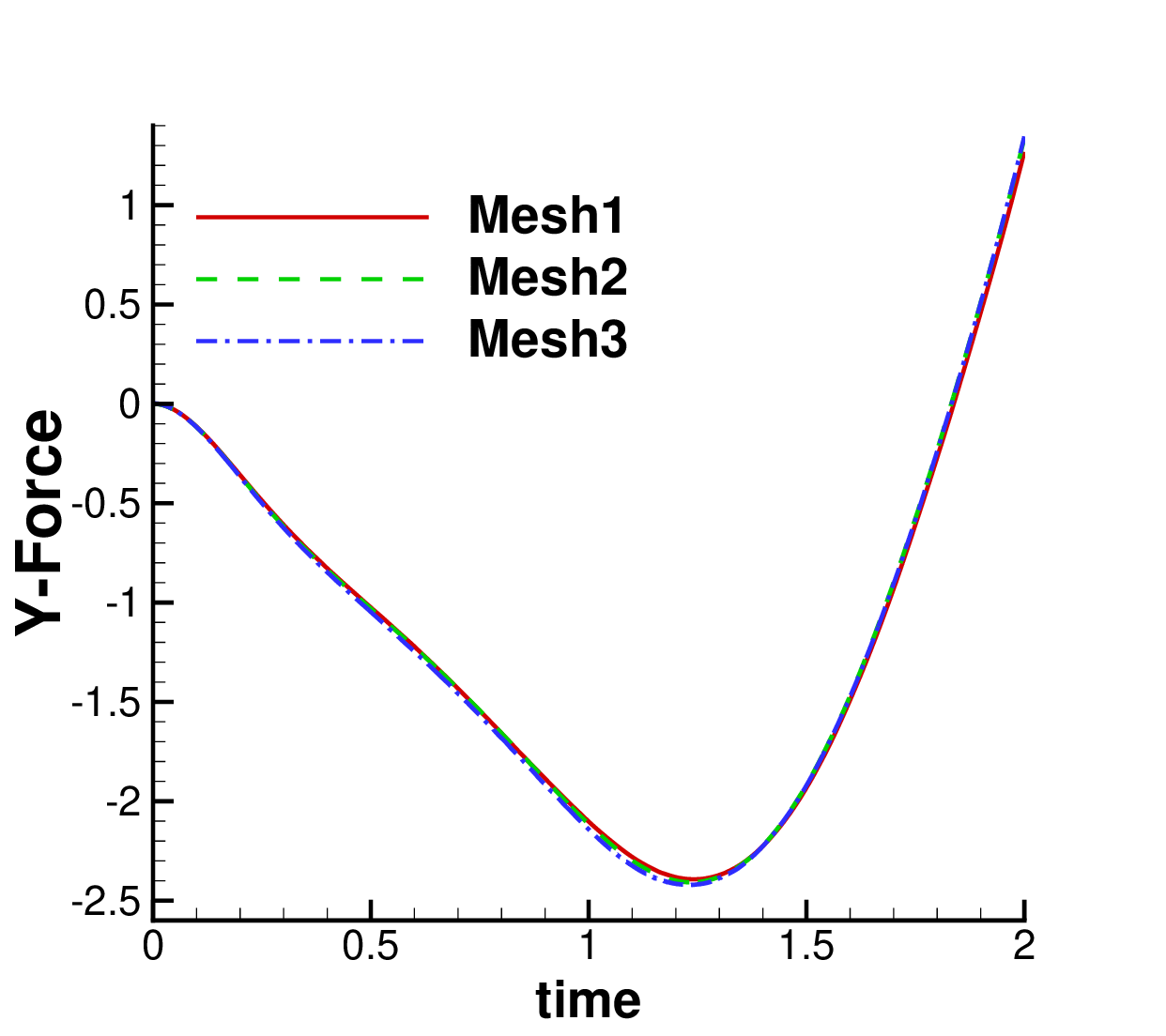}
    \includegraphics[width=5cm]{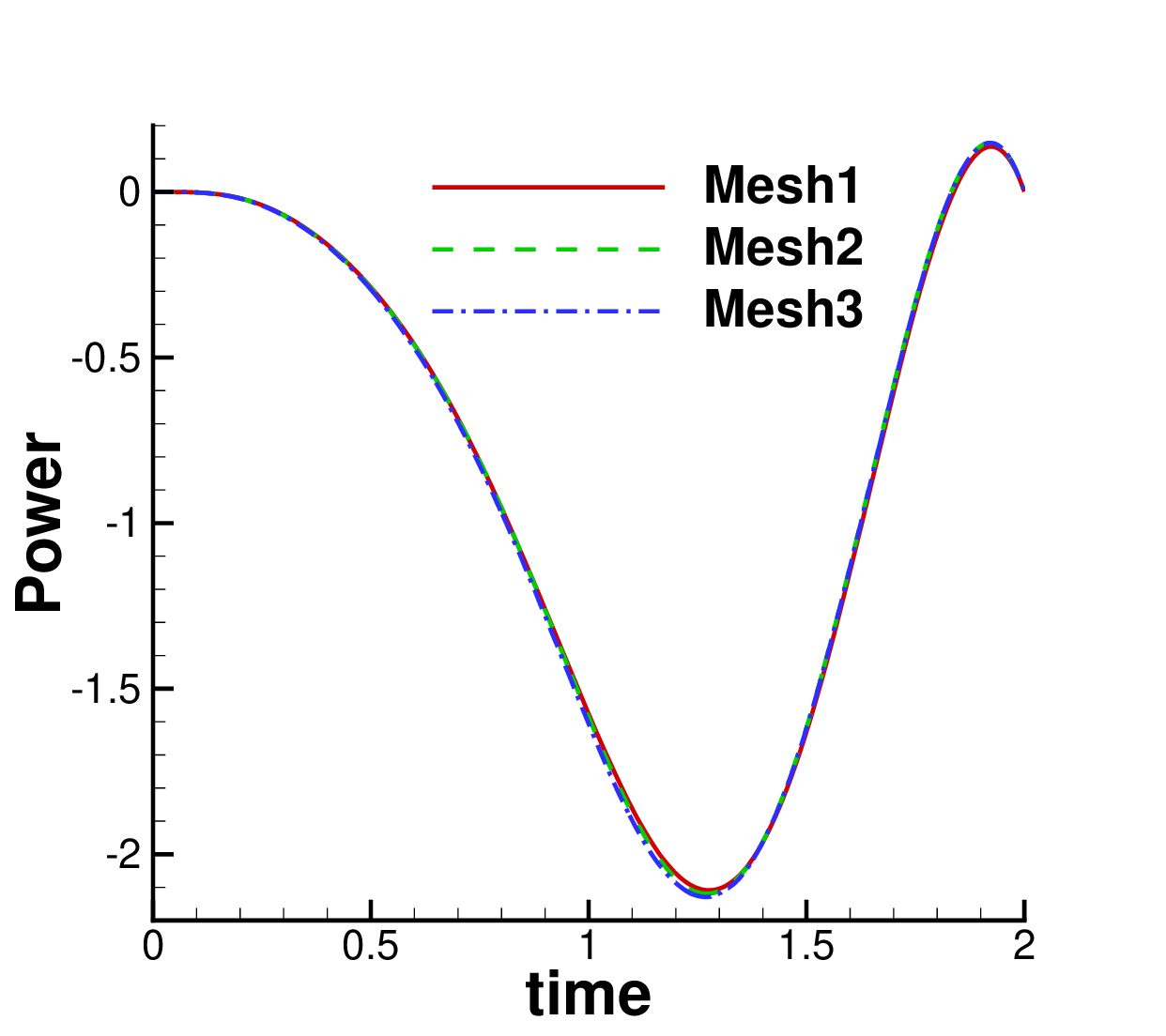}
  \caption[]{NACA0012: force and power history of Motion 1 (translation)}\label{naca0012_motion1_history}
\end{figure}

\begin{figure}[hbt!]
  \centering
    \subfigure[Vorticity]{ \includegraphics[width=6cm]{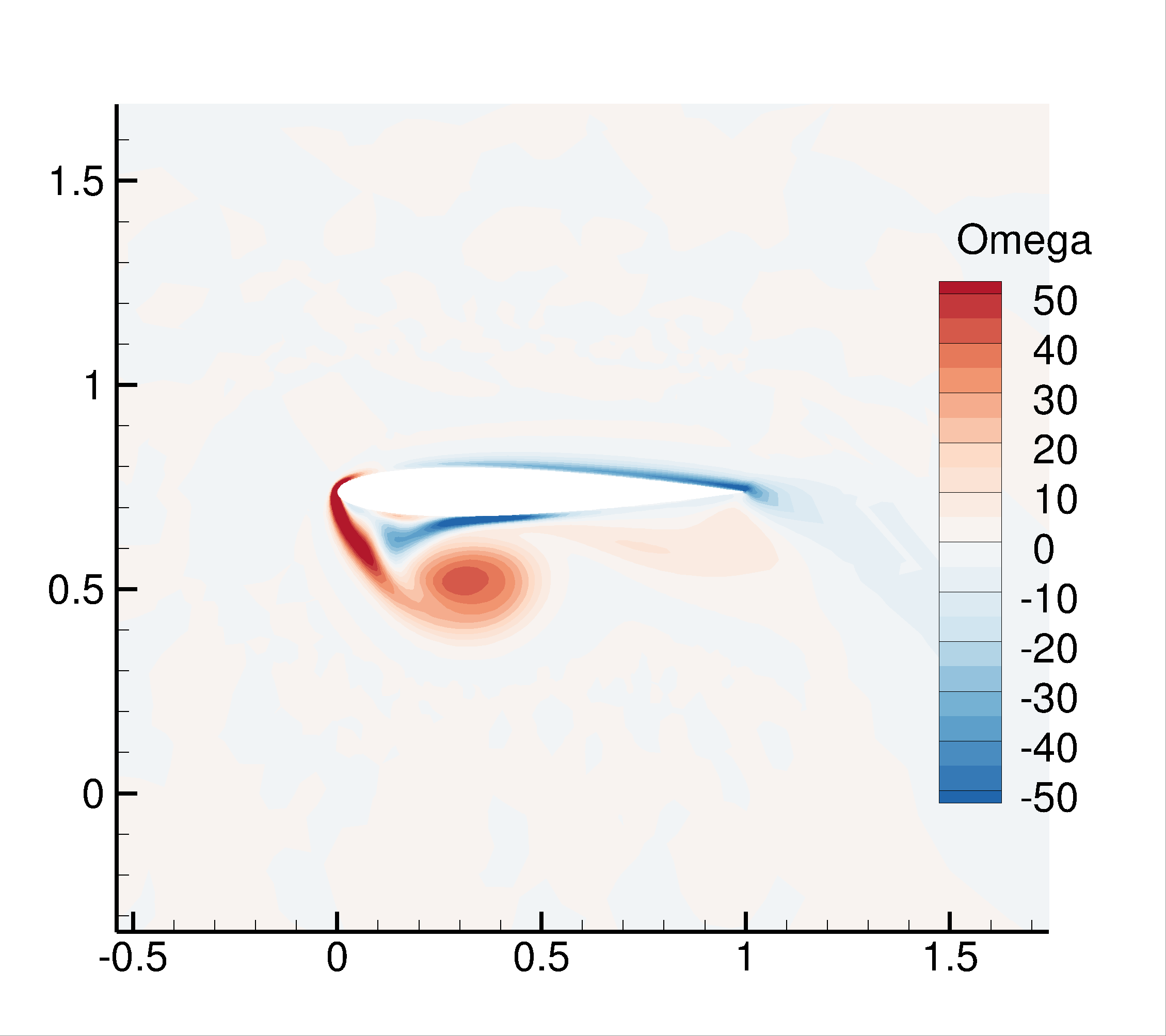}}
    \subfigure[Mesh]{ \includegraphics[width=6cm]{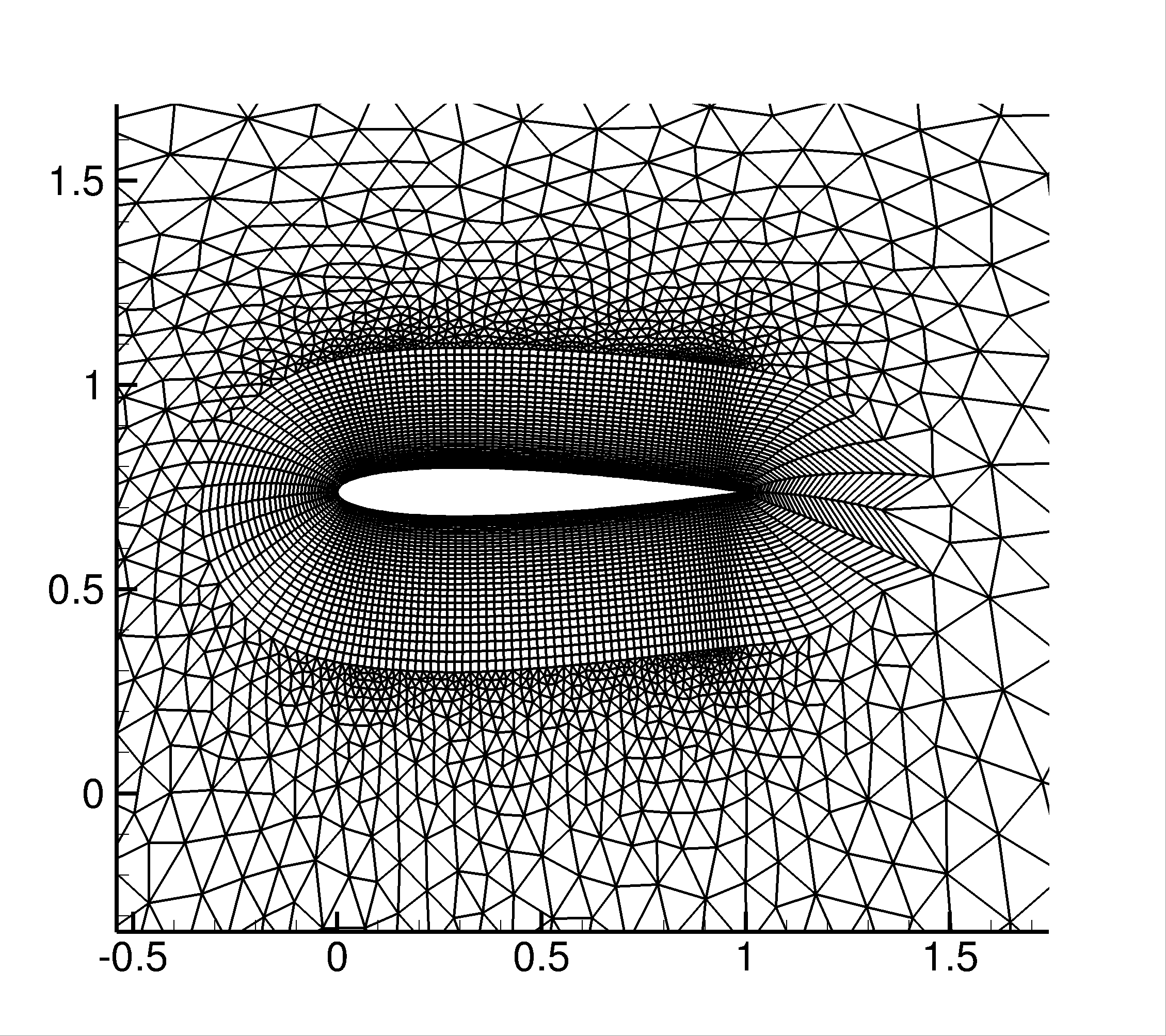}}
  \caption[]{NACA0012: the solution of Motion 1 (translation) at time $t=1.5$}\label{naca0012_motion1_t15}
\end{figure}

Figure \ref{naca0012_motion1_history} displays the time history of the force in the $x$ and $y$ directions, as well as the power. The results indicate that the three meshes produce nearly identical solutions for the force in the $y$ direction and power. Although the solutions for the force in the $x$ direction appear to differ slightly among the meshes, the difference is negligible since the force in this direction is very small.
Figure \ref{naca0012_motion1_t15} illustrates the vorticity contour and mesh distribution at $t=1.5$. At this time, the airfoil has moved approximately 0.738 times its chord length in the $y$ direction, while the mesh surrounding the airfoil remains of good quality. The vorticity contour indicates the generation of a vortex at the front edge of the airfoil.

\begin{figure}[hbt!]
  \centering
    \includegraphics[width=5cm]{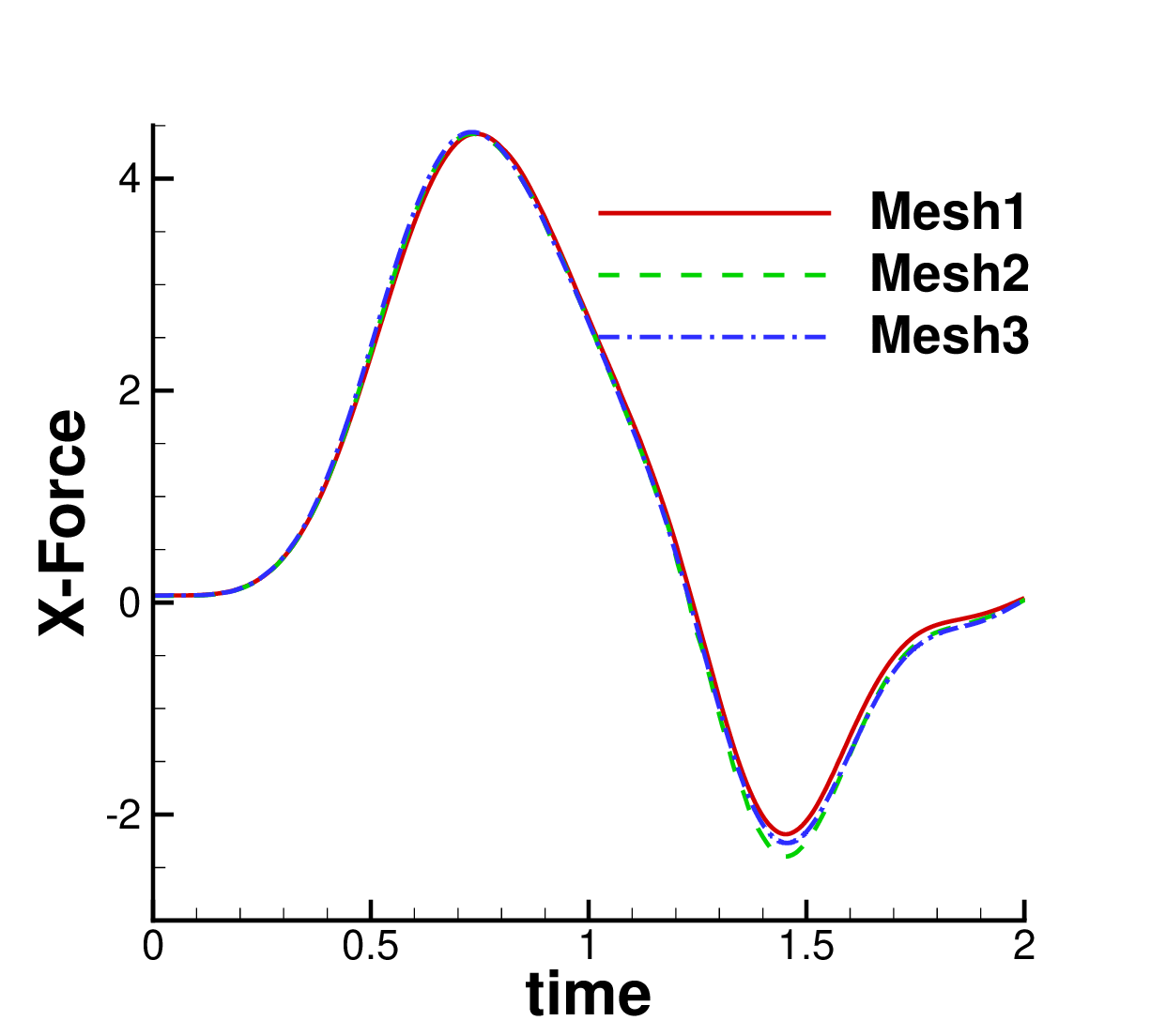}
    \includegraphics[width=5cm]{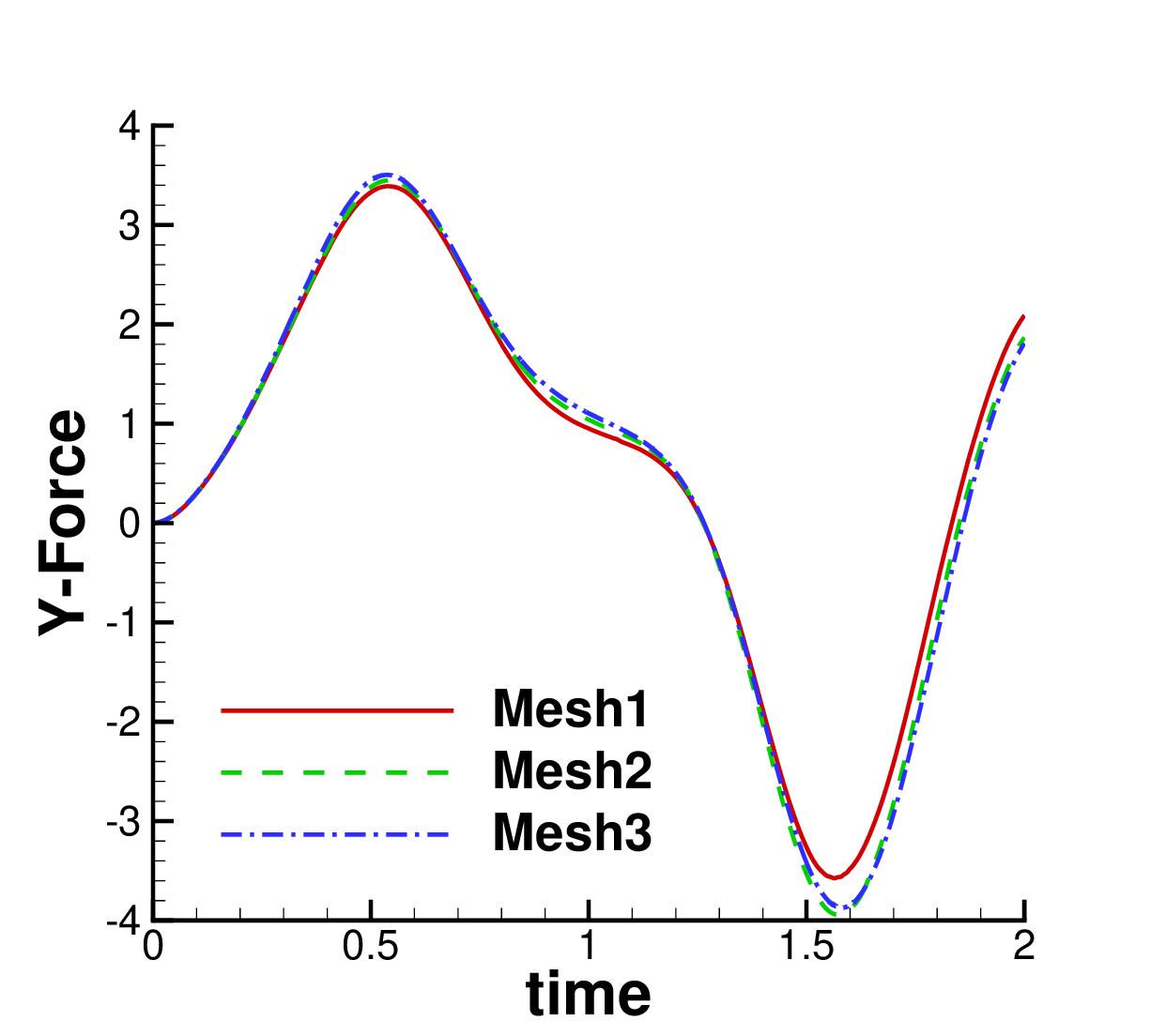}
    \includegraphics[width=5cm]{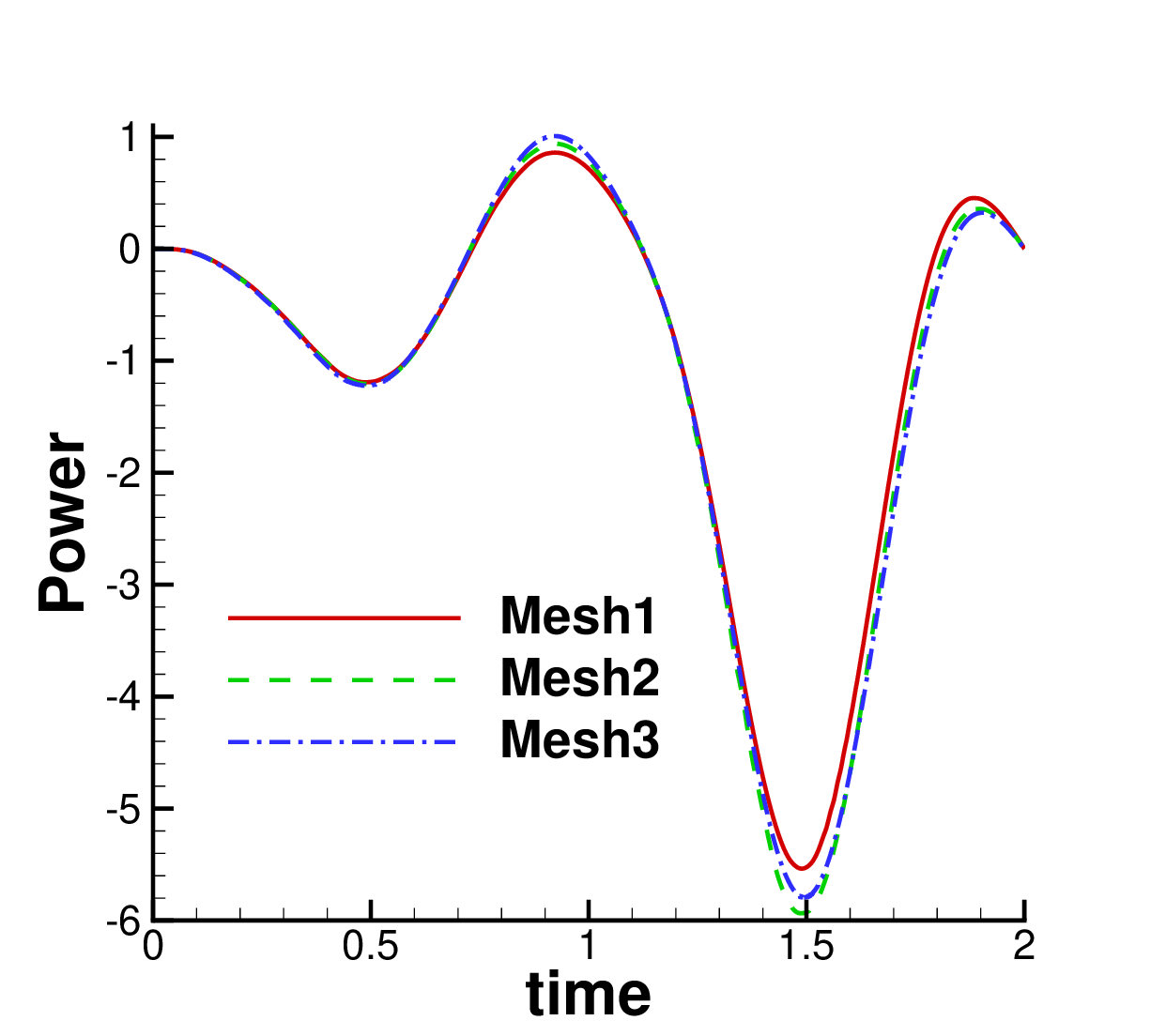}
  \caption[]{NACA0012: force and power history of Motion 2 (translation + rotation)}
  \label{naca0012_motion2_history}
\end{figure}

\begin{figure}[hbt!]
  \centering
    \subfigure[Vorticity]{ \includegraphics[width=6cm]{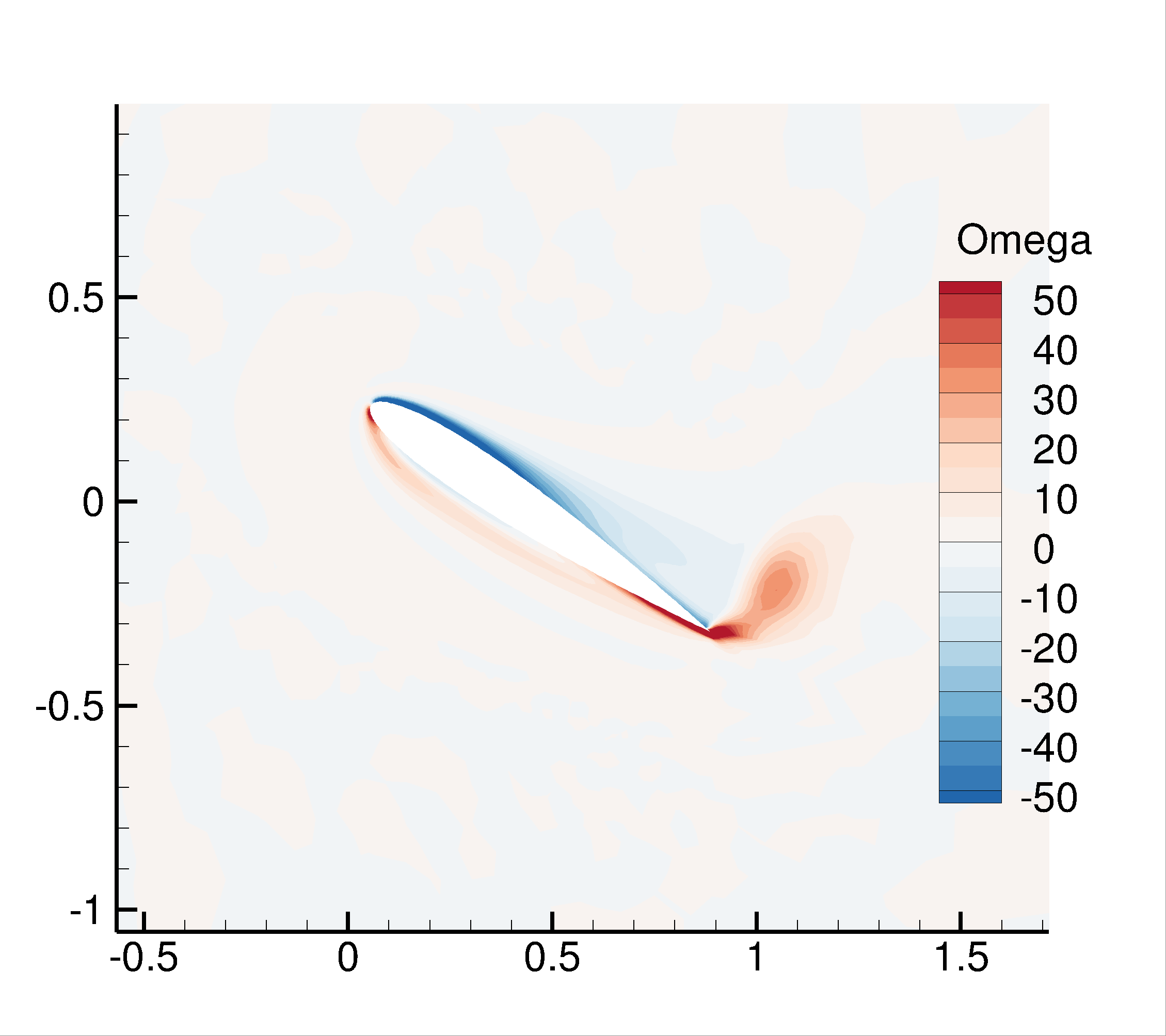}}
    \subfigure[Mesh]{ \includegraphics[width=6cm]{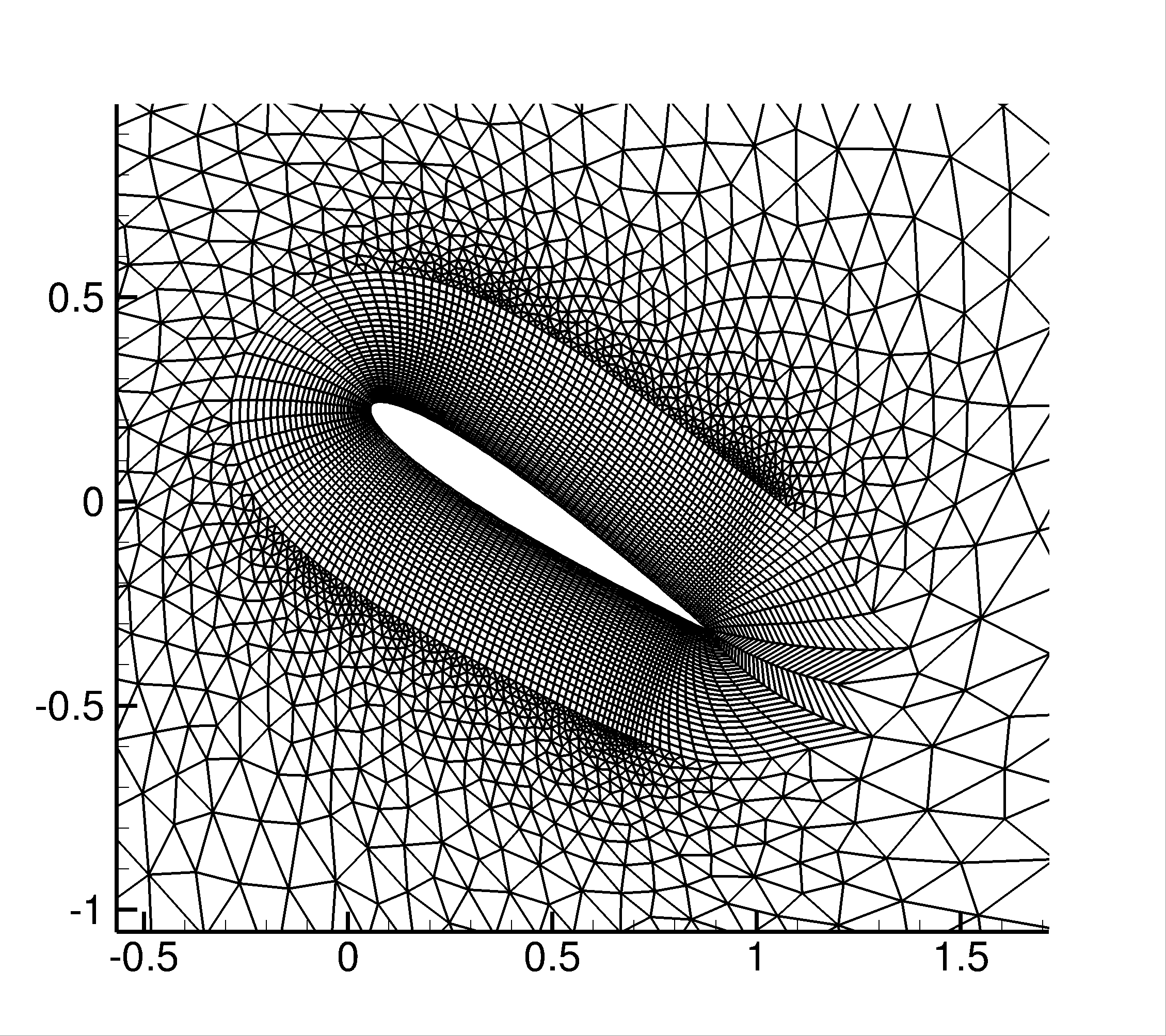}}
  \caption[]{NACA0012: the solution of Motion 2 (translation + rotation) at time $t=0.5$}
  \label{naca0012_motion2_t05}
\end{figure}

Figure \ref{naca0012_motion2_history} displays the force and power history of Motion 2. The results indicate that the solutions obtained from Mesh 2 and Mesh 3 are in good agreement, suggesting that the solutions are mesh-independent.
Figure \ref{naca0012_motion2_t05} depicts the solution at time $t=0.5$ when the airfoil has rotated 33.75 degrees. At this time, the mesh remains of high quality with only slight distortion at the trailing edge. The vorticity contour shows the shedding of a vortex at the trailing edge.

\subsection{2D Noh Problem}
\begin{figure}[hbt!]
  \centering
    \subfigure[Initial mesh]{ \label{noh_ini_mesh} \includegraphics[width=5cm]{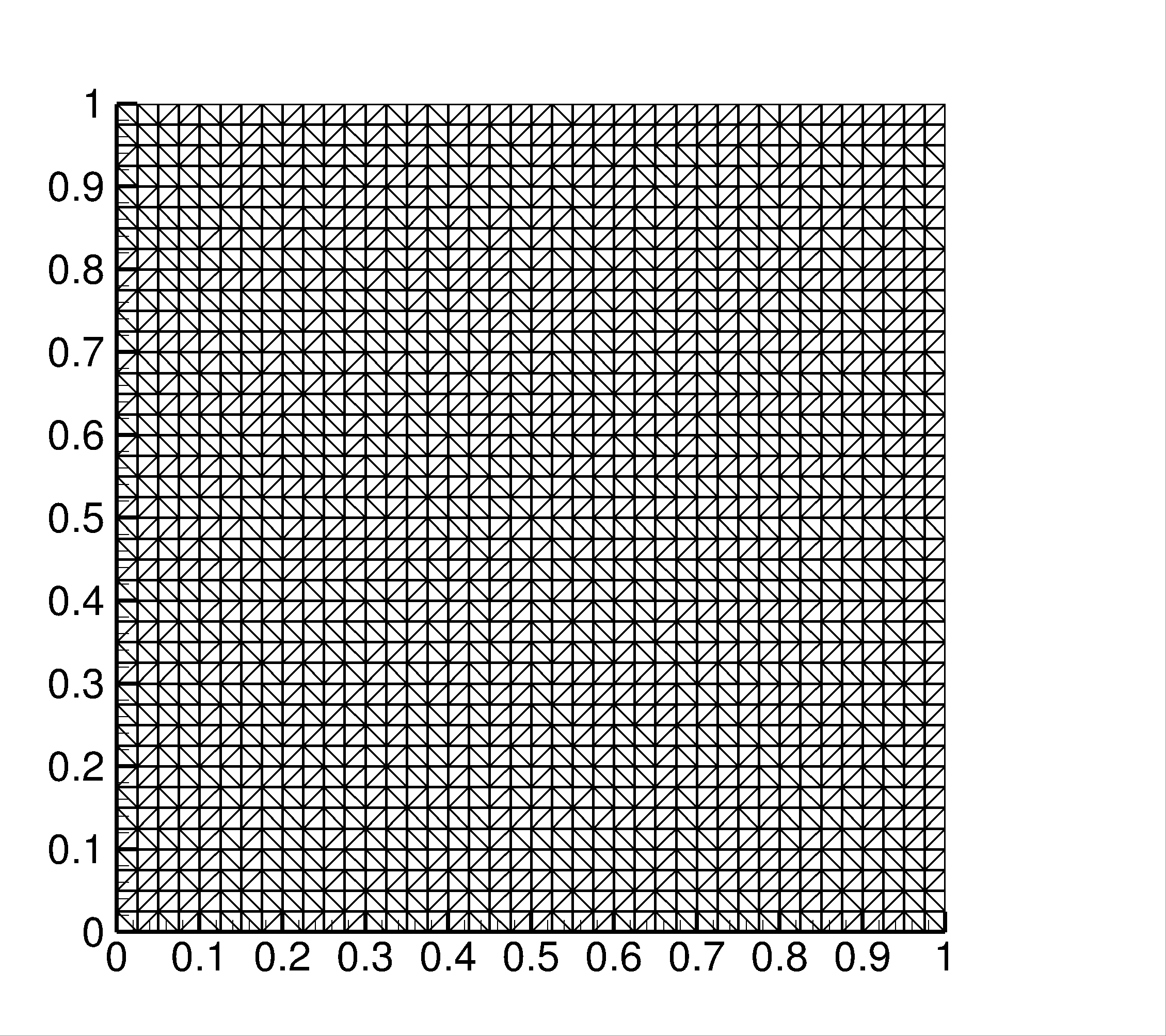}}
    \subfigure[Final mesh]{ \label{noh_final_mesh} \includegraphics[width=5cm]{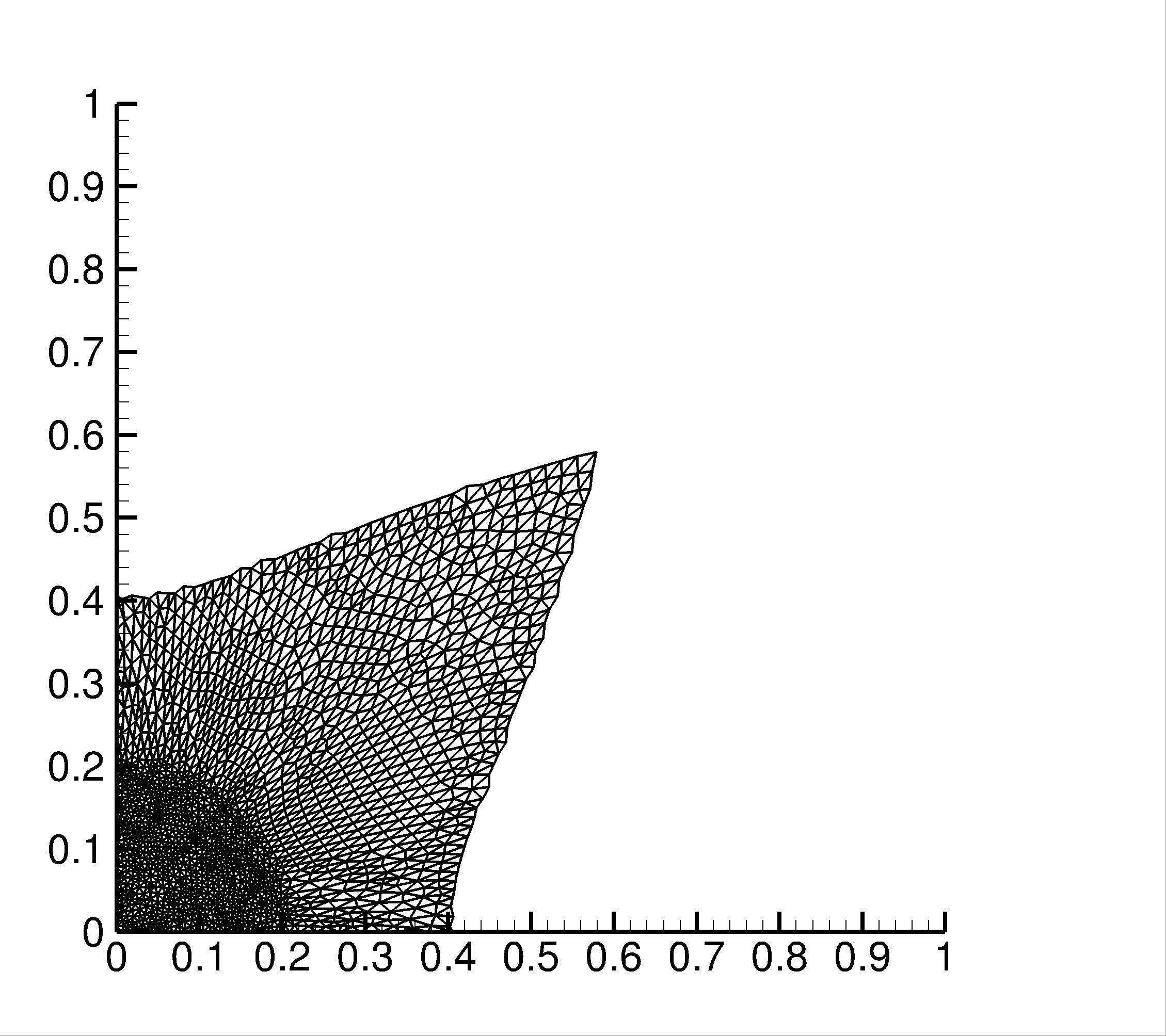}}
  \caption[]{The mesh used in Noh Problem}
\end{figure}

\begin{figure}[hbt!]
  \centering
    \subfigure[The density contour]{\label{noh_dens_contour} \includegraphics[width=5cm]{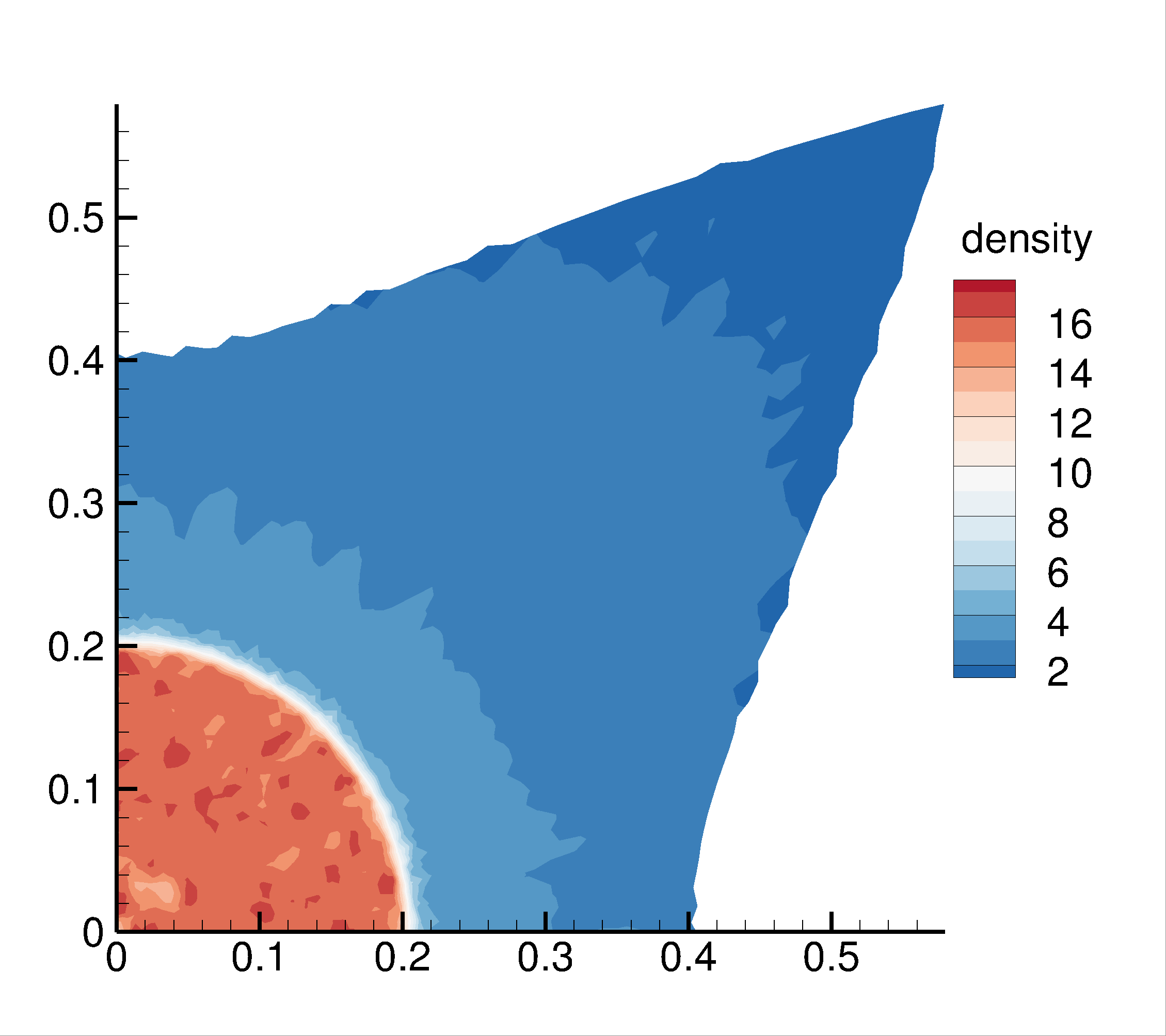}}
    \subfigure[The density distribution with respect to $r$]{\label{noh_dens_dist} \includegraphics[width=5cm]{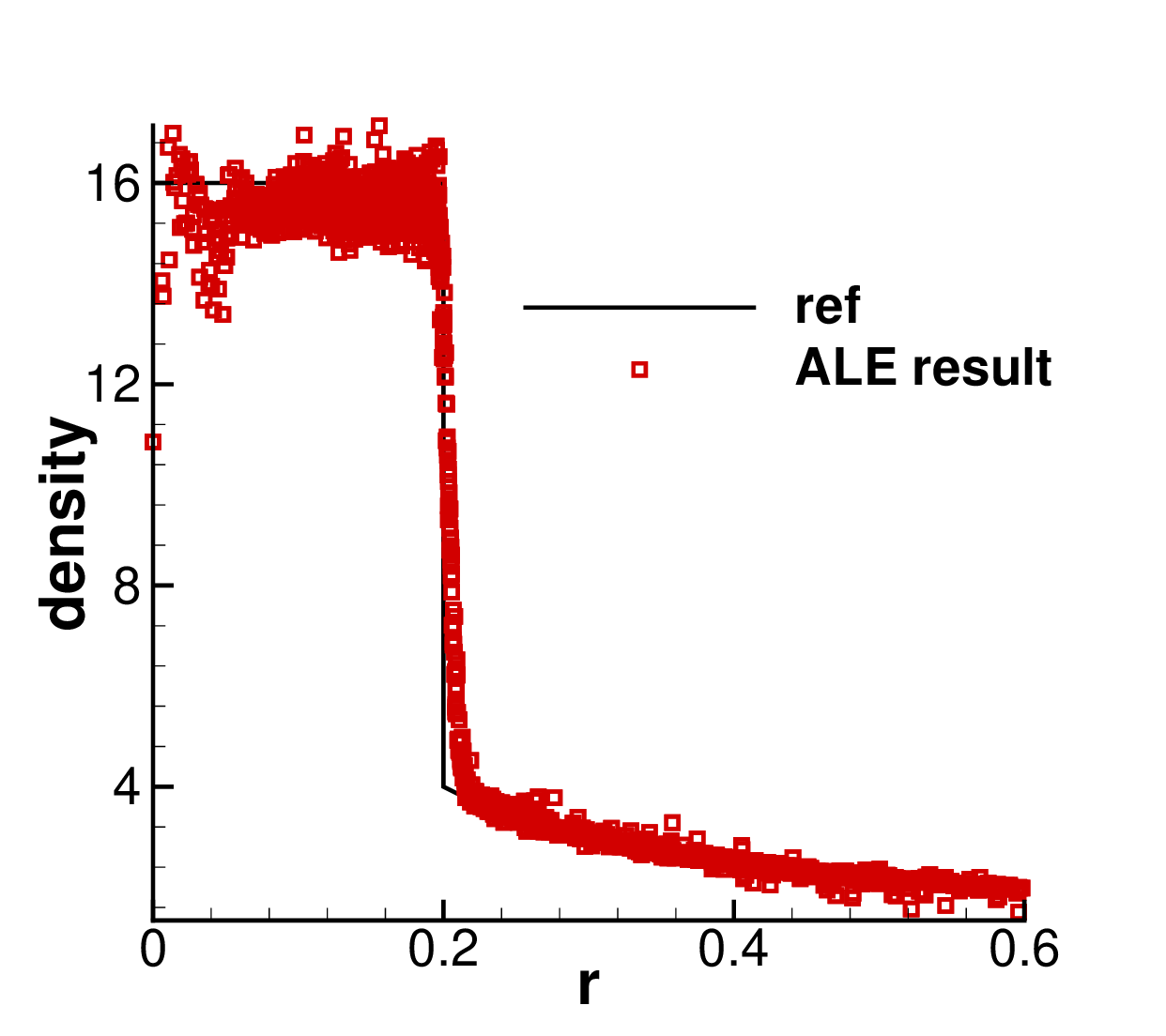}}
  \caption[]{The result of Noh problem at $t=0.6$}
\end{figure}

This case is a typical test case for verifying ALE code. A two-dimensional domain with dimensions $[0,1]\times[0,1]$ is considered. An unstructured mesh with mesh size $h=1/40$ is used for the computation, as shown in Figure \ref{noh_ini_mesh}. The initial conditions are set to uniform density $\rho=1$ and internal energy $e=\frac{p}{\gamma-1}=1\times10^{-4}$, with $\gamma=5/3$. The initial velocity is given by $\boldsymbol{V}=(-x/r,-y/r)$, where $r$ is the distance from the origin. An axially symmetric shock is generated at the origin and propagates at a constant speed. At time $t=0.6$, the shock is located at $r=0.2$. An exact solution of density \cite{NOH1987} is
\begin{equation*}
  \rho=\begin{cases}
    16,&r<0.2 \\ 1+t/r,&r>0.2
  \end{cases}.
\end{equation*}
An asymmetric boundary condition is enforced at the lines $x=0$ and $y=0$, while a non-reflective boundary condition is used at all other boundaries. Mesh movement is achieved using Lagrangian velocity, and a smoothing process is applied every 20 time-steps with a relaxation coefficient $\omega=0.4$. The CFL number is set to 0.3 for this case. The final mesh is shown in Figure \ref{noh_final_mesh}, and it is observed that the mesh is concentrated in the post-shock region.
The density contour at time $t=0.6$ is shown in Figure \ref{noh_dens_contour}, and the distribution with respect to $r$ of the entire domain is presented in Figure \ref{noh_dens_dist}, which agrees well with the exact solution.

\subsection{Sedov Problem}
\begin{figure}[hbt!]
  \centering
    \subfigure[Mesh $20^3$]{ \includegraphics[width=5cm]{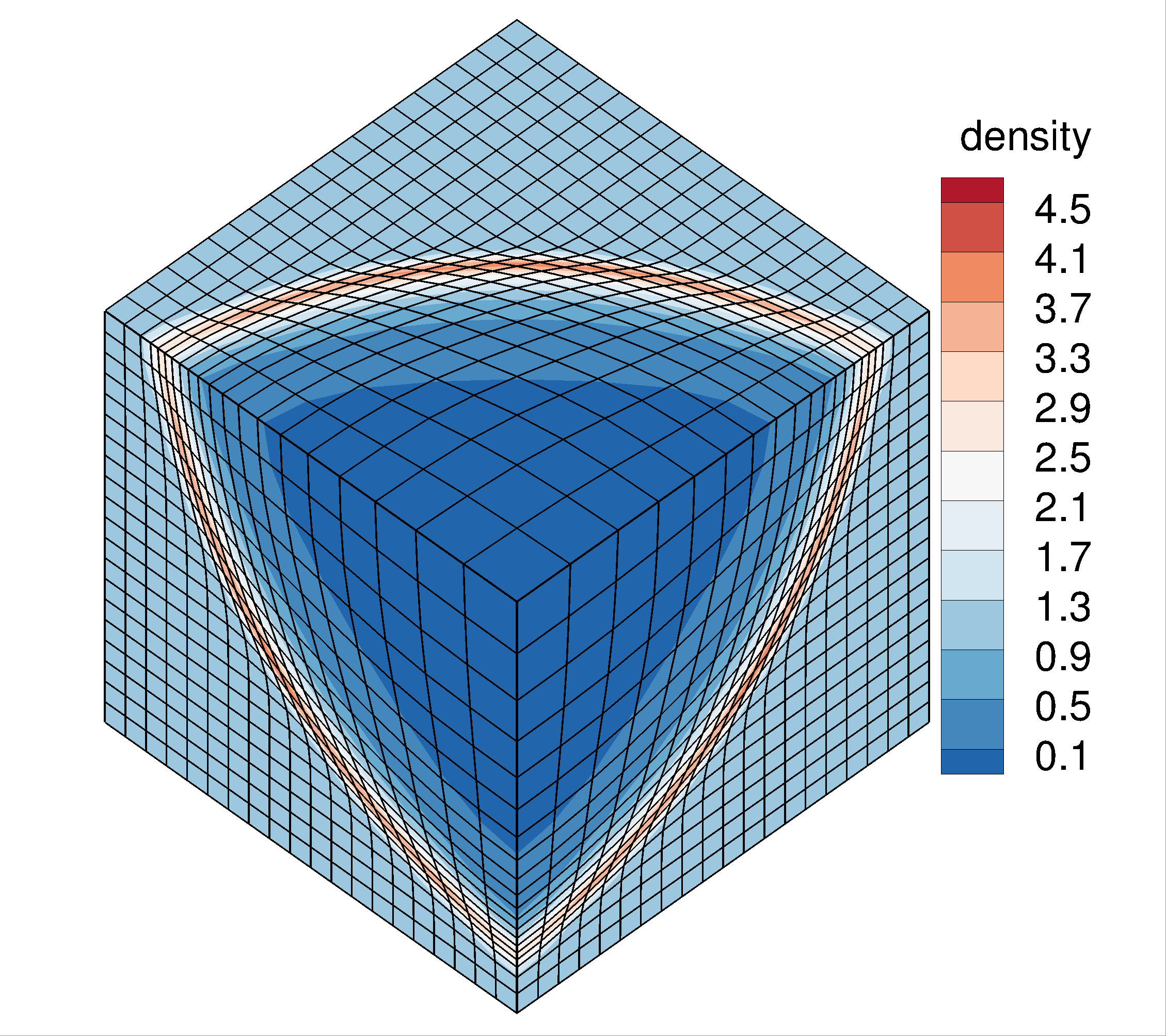}}
    \subfigure[Mesh $40^3$]{ \includegraphics[width=5cm]{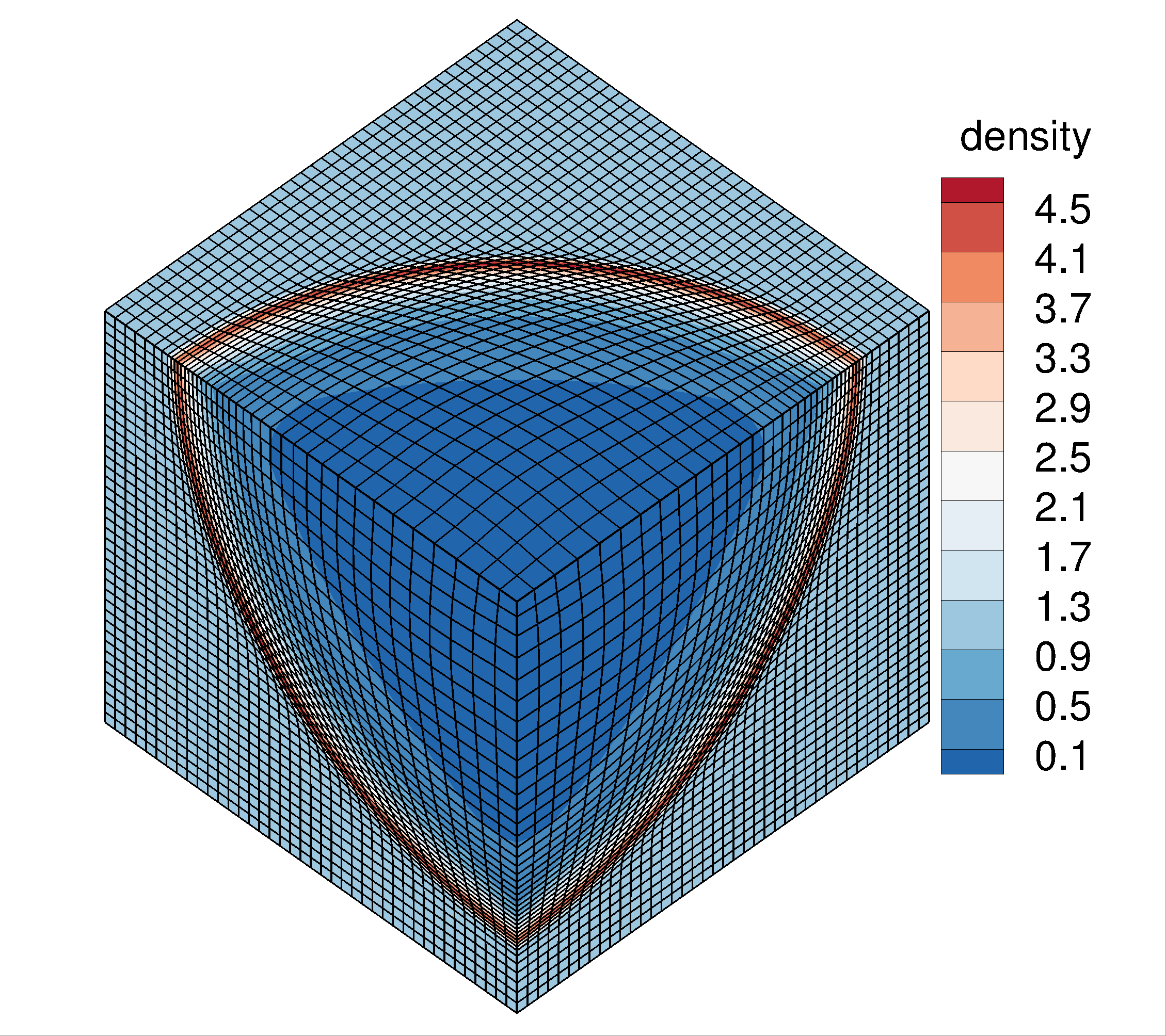}}
  \caption{Sedov problem: the mesh distribution and density contour at $t=1$}\label{sedov_contour}
\end{figure}

\begin{figure}[hbt!]
  \centering
    \subfigure[Density]{ \includegraphics[width=5cm]{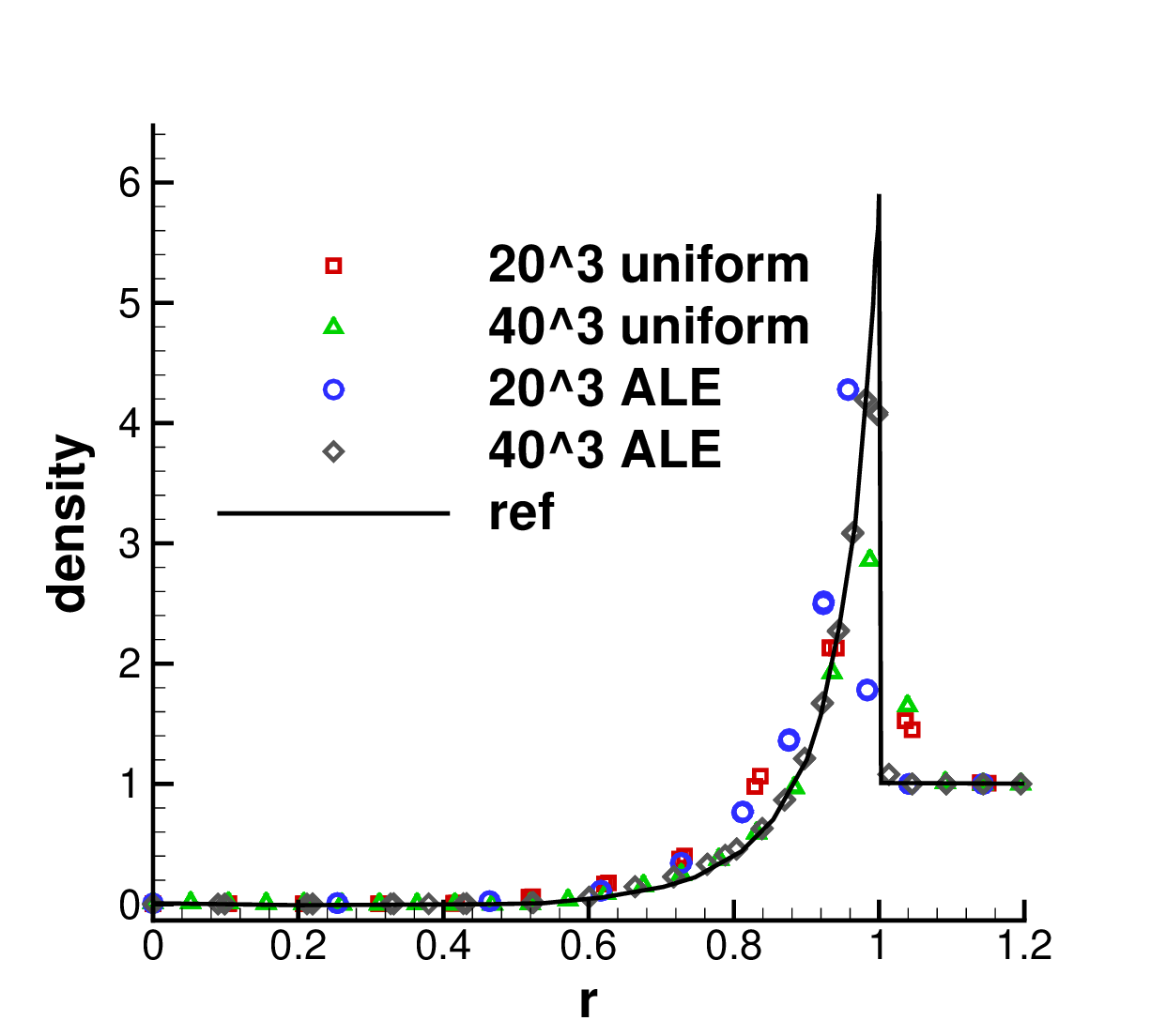}}
    \subfigure[Pressure]{ \includegraphics[width=5cm]{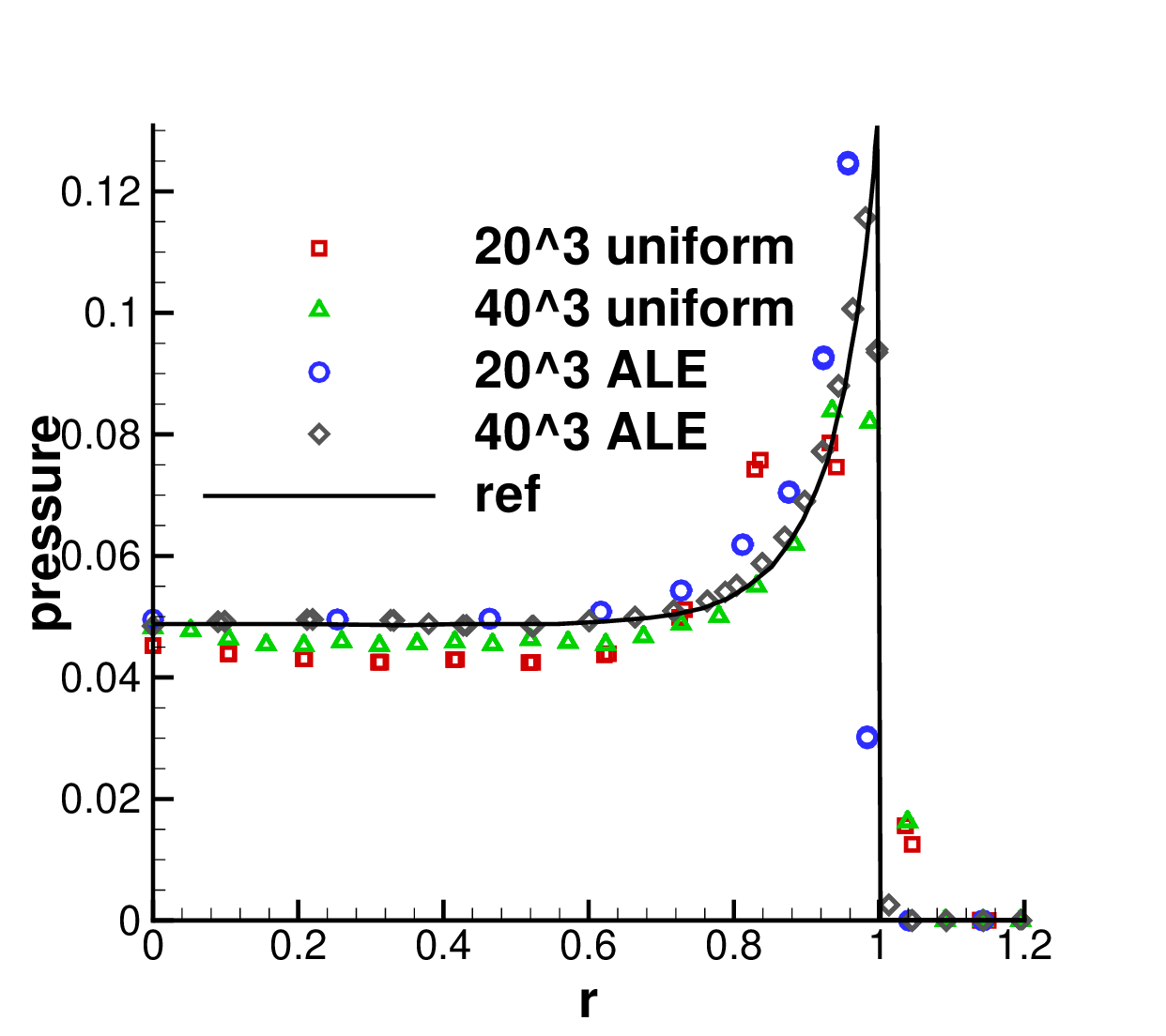}}
  \caption[]{Sedov problem: the density and pressure distribution along diagonal line}\label{sedov_line}
\end{figure}

The Sedov blast wave is a standard hydrodynamics test problem that models a blast wave emanating from a singular energy point. It is a benchmark problem for both the Lagrangian velocity method and the ALE method. The computational domain is $[0,1.2]^3$, and two uniform grids with $20^3$ and $40^3$ cells are used in this test case. The fluid is modeled by the ideal gas equation of state with $\gamma=1.4$. The initial condition is a uniform static field with density $\rho=1$ and pressure $p=1e-6$, except for the cell containing the origin. In this cell, the initial pressure is defined by $p=(\gamma-1)\epsilon_0/V$, where $\epsilon_0= 0.106384$ is the total amount of released energy and $V$ is the cell volume. Asymmetric boundary condition is enforced at the plane $x=0$,  $y=0$and $z=0$, while non-reflecting boundary conditions are used at the plane $x=1.2$,  $y=1.2$and $z=1.2$. Mesh movement is accomplished using Lagrangian velocity, and a smoothing process is applied every ten time-steps with a relaxation coefficient of $\omega=0.8$. A small CFL number of 0.01 is used to avoid the instability caused by the initial singularity at the origin. After ten steps, the simulation transitions to a normal CFL number of 0.3. Figure \ref{sedov_contour} displays the mesh distribution and density contour, which reveal that the mesh is refined adaptively around the shock.
 Figure \ref{sedov_line} displays the density and pressure distribution along the diagonal from $(0,0,0)$ to $(1.2,1.2,1.2)$, as well as the solutions of a uniform stationary mesh.
The results show that the solution of the ALE method agrees much more closely with the exact solution than the static method. The solutions at the extreme points of the ALE method are closer to the exact solution than those of the uniform mesh. Additionally, the fine mesh ALE solution provides a more accurate solution for the shock position than the coarse mesh one.
\subsection{Saltzmann Problem}
\begin{figure}[hbt!]
  \centering
    \includegraphics[width=12cm]{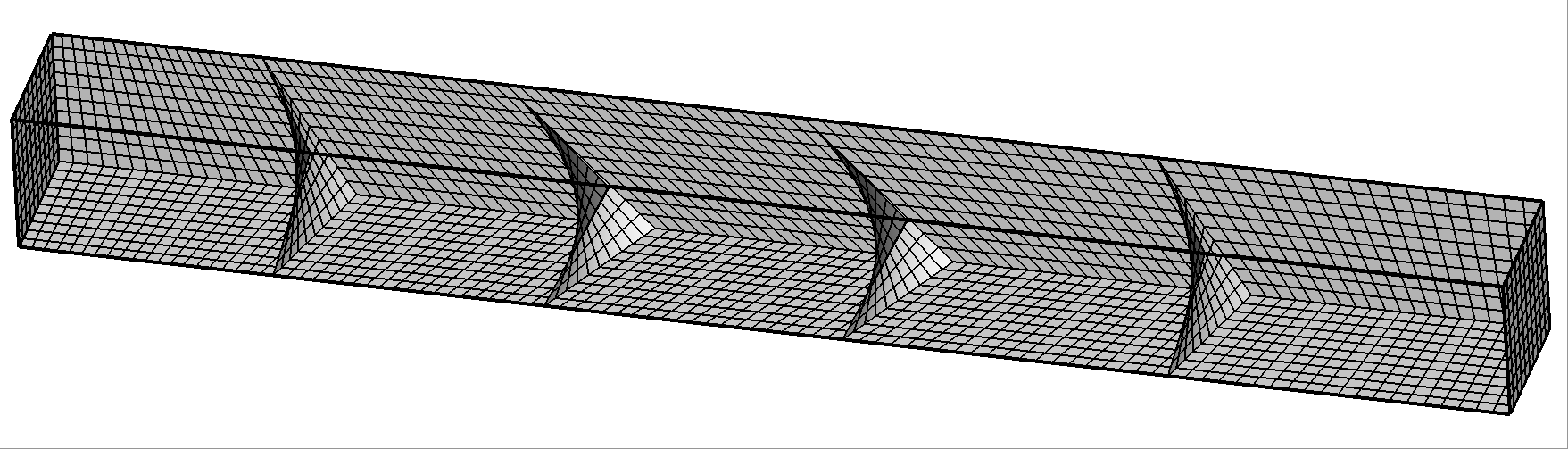}
  \caption[]{The initial mesh of Saltzmann problem} \label{Saltzmann_mesh}
\end{figure}

\begin{figure}[hbt!]
  \centering
  \subfigure[Density]{ \includegraphics[width=5cm]{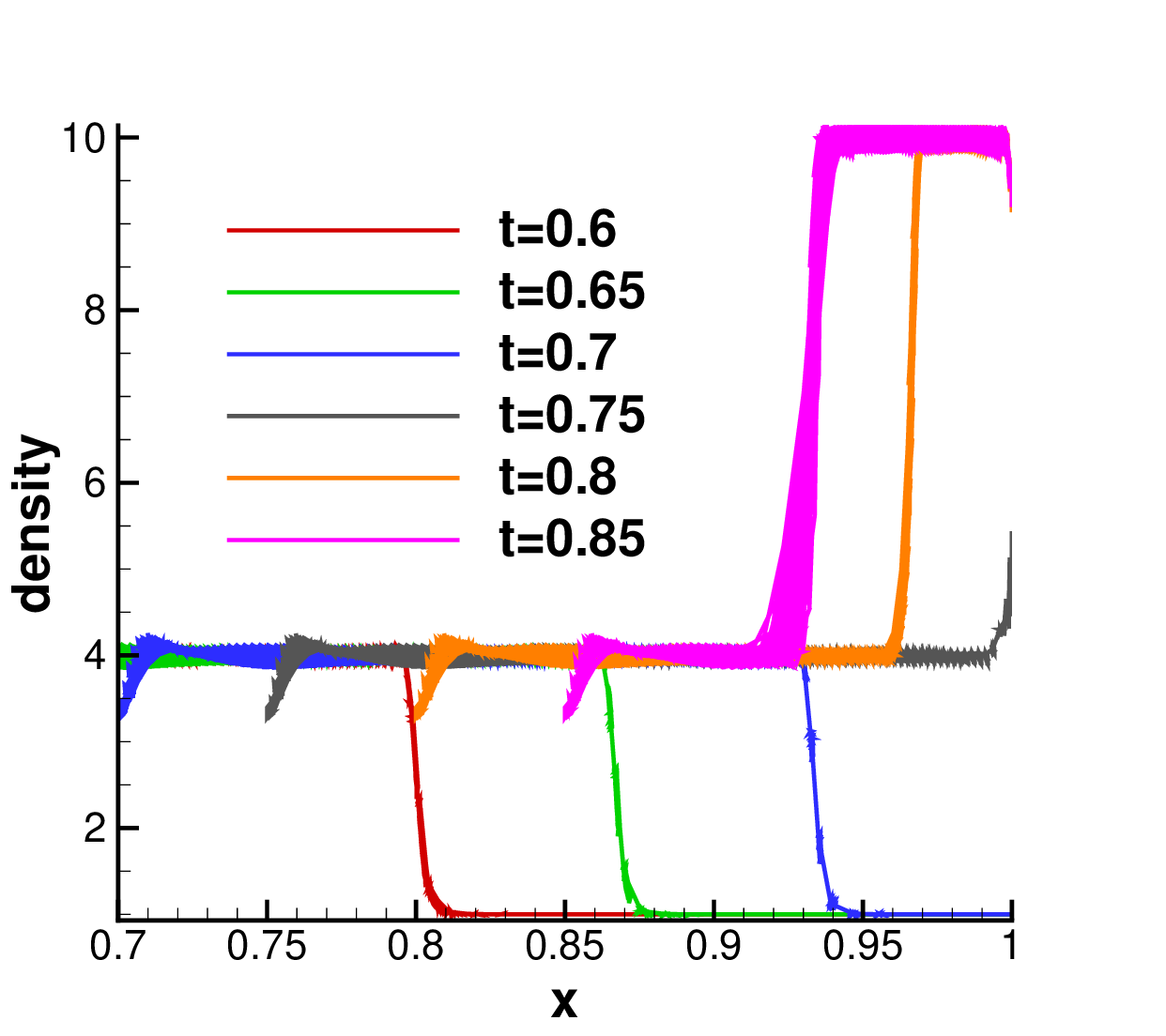}}
    \subfigure[Pressure]{ \includegraphics[width=5cm]{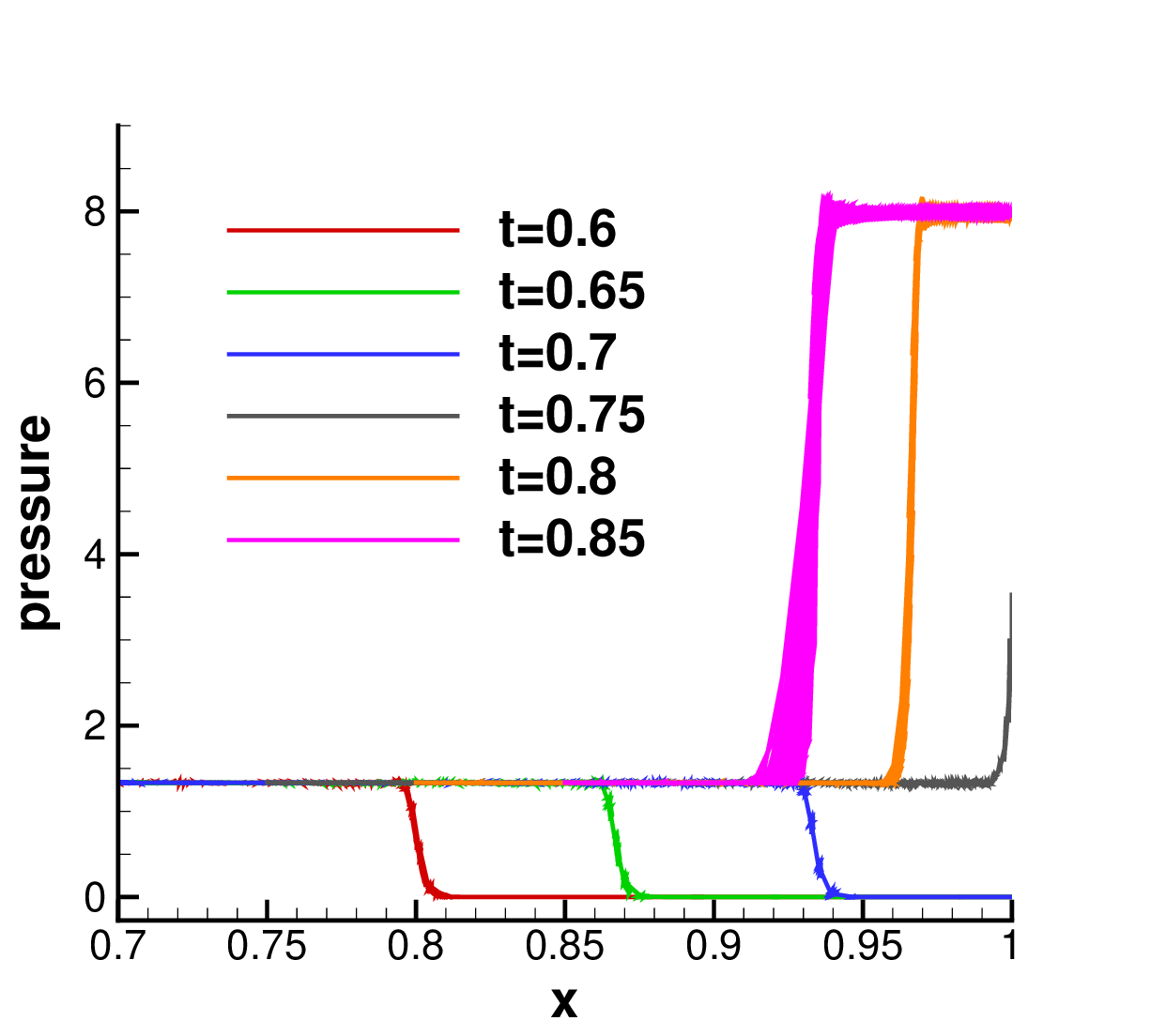}}
  \caption[]{Saltzmann problem: the density and pressure distribution at different time}\label{Saltzmann_line}
\end{figure}

\begin{figure}[hbt!]
  \centering
     \subfigure[$t=0.6$]{ \includegraphics[width=8cm]{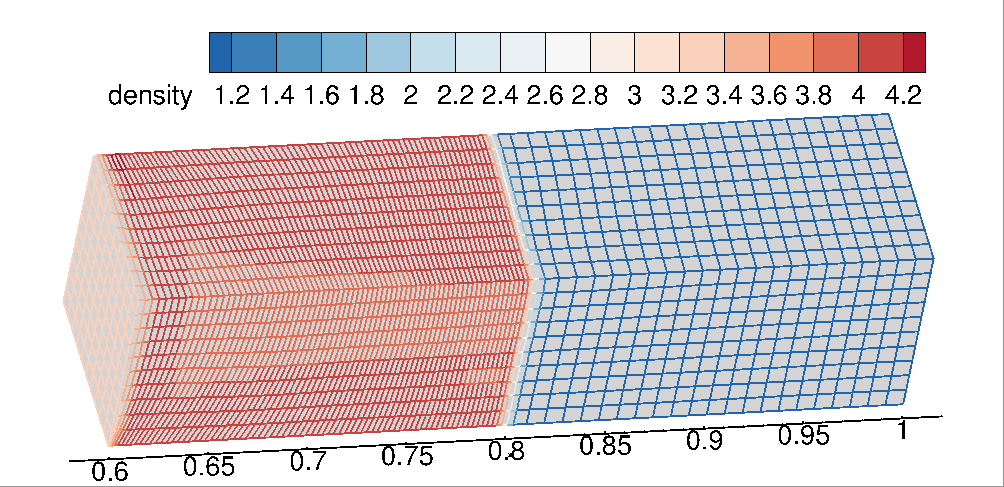}}
    \subfigure[$t=0.85$]{ \includegraphics[width=8cm]{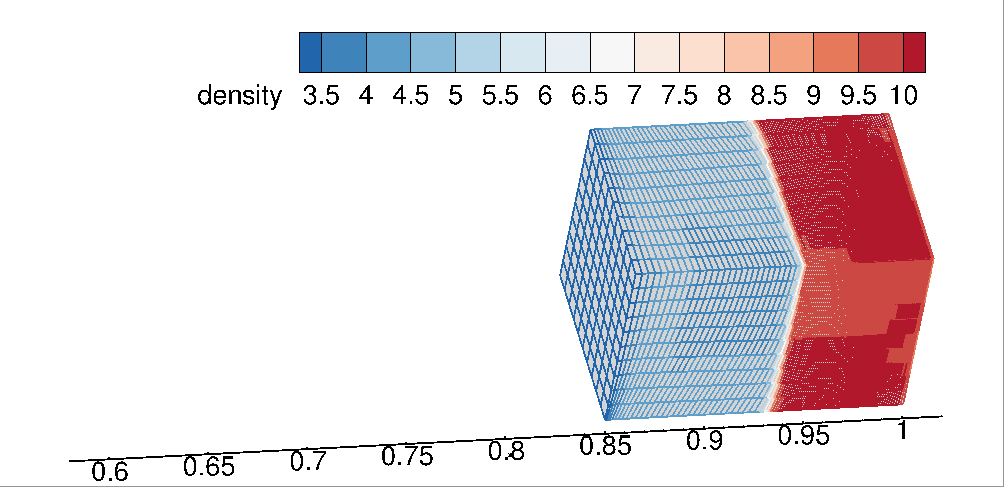}}
  \caption[]{Saltzmann problem: mesh and density distribution at different time}\label{Saltzmann_time}
\end{figure}

The Saltzmann problem is a test case that simulates fluid in a cylinder with a piston. It examines the motion of a planar shock on a skewed Cartesian grid. The computational domain is $[0,1]\times[0,0.1]\times[0,0.1]$, with uniform $h=0.01$ in three directions. A mesh distortion of $x$-coordinate is defined by
 \begin{equation*}
  \Delta x = (0.5y+z-15yz)\sin(\pi x).
 \end{equation*}
 The initial mesh is shown in Figure \ref{Saltzmann_mesh}.
  Initially, an ideal monatomic gas with $\rho=1,e=10^{-4},\gamma=5/3$ is filled in the box. The wall on the left-hand side moves at a constant speed of 1 from left to right, acting as a piston. The other boundaries act as inviscid reflecting walls. As the gas is compressed, a shock is generated and propagates to the right faster than the piston. Mesh movement is accomplished using Lagrangian velocity, and a smoothing process is applied every 20 time steps with a relaxation coefficient of $\omega=0.8$. Figure \ref{Saltzmann_line} displays the entire solution of the domain at different times, indicating that the solution tends towards a one-dimensional solution.
The drop in density may be due to the entropy increase at the wall, and further high-order boundary conditions will be considered. At $t=0.6$, the shock is located at $x=0.8$, and the post-shock solutions are $\rho=4$ and $p=1.333$. At time $t=0.75$, the shock will reach the right boundary and bounce back due to the wall. Figure \ref{Saltzmann_time} demonstrates that the Lagrangian velocity method is capable of handling shock motion issues, as it shows the mesh distribution at different time intervals.
\section{Conclusion}
 In this paper, a third-order compact gas-kinetic scheme is developed in an arbitrary Lagrangian-Eulerian (ALE) formulation for three-dimensional unstructured mesh. In order to achieve geometric consistency, the arbitrary mesh is subdivided into tetrahedrons, and the quadrilateral surface is split into two triangles. By accounting for the variation of grid velocity within the interface surface and dealing with it in a local framework, the geometric conservation law is preserved. The kinetic equation at the moving frame with grid velocity is used to construct a time-accurate gas distribution function at the Gaussian point of the cell interface, allowing for updating both cell-averaged flow variables and their gradients for the use in the high-order compact initial reconstruction. The scheme is validated using several test cases. For close boundary problems, mass conservation is well-preserved. With the help of radial basis function interpolation, the method can handle moving boundary problems and obtain high-fidelity solutions. For blast problems with strong discontinuities, the scheme retains robustness, and the Lagrangian nodal solver is adapted to track material interfaces.

 \section*{Acknowledgements}

The current research is supported by the National Key R\&D Program of China 2022YFA1004500,
National Natural Science Foundation of China (Grant No. 12172316),
and Hong Kong research grant council (16208021, 16301222).



\bibliographystyle{elsarticle-harv}
\bibliography{mybibfile}

\begin{thebibliography}{28}
\expandafter\ifx\csname natexlab\endcsname\relax\def\natexlab#1{#1}\fi
\providecommand{\url}[1]{\texttt{#1}}
\providecommand{\href}[2]{#2}
\providecommand{\path}[1]{#1}
\providecommand{\DOIprefix}{doi:}
\providecommand{\ArXivprefix}{arXiv:}
\providecommand{\URLprefix}{URL: }
\providecommand{\Pubmedprefix}{pmid:}
\providecommand{\doi}[1]{\href{http://dx.doi.org/#1}{\path{#1}}}
\providecommand{\Pubmed}[1]{\href{pmid:#1}{\path{#1}}}
\providecommand{\bibinfo}[2]{#2}
\ifx\xfnm\relax \def\xfnm[#1]{\unskip,\space#1}\fi
\bibitem[{Blazek(2015)}]{BLAZEK201573}
\bibinfo{author}{Blazek, J.}, \bibinfo{year}{2015}.
\newblock \bibinfo{title}{Chapter 4 - structured finite-volume schemes}, in:
  \bibinfo{editor}{Blazek, J.} (Ed.), \bibinfo{booktitle}{Computational Fluid
  Dynamics: Principles and Applications (Third Edition)}.
  \bibinfo{edition}{third edition} ed..
  \bibinfo{publisher}{Butterworth-Heinemann}, \bibinfo{address}{Oxford}, pp.
  \bibinfo{pages}{73--120}.
\bibitem[{Brackbill and Saltzman(1982)}]{BRACKBILL1982342}
\bibinfo{author}{Brackbill, J.}, \bibinfo{author}{Saltzman, J.},
  \bibinfo{year}{1982}.
\newblock \bibinfo{title}{Adaptive zoning for singular problems in two
  dimensions}.
\newblock \bibinfo{journal}{Journal of Computational Physics}
  \bibinfo{volume}{46}, \bibinfo{pages}{342--368}.
\newblock \DOIprefix\doi{10.1016/0021-9991(82)90020-1}.
\bibitem[{De~Boer et~al.(2007)De~Boer, Van~der Schoot and Bijl}]{RBF2007784}
\bibinfo{author}{De~Boer, A.}, \bibinfo{author}{Van~der Schoot, M.S.},
  \bibinfo{author}{Bijl, H.}, \bibinfo{year}{2007}.
\newblock \bibinfo{title}{Mesh deformation based on radial basis function
  interpolation}.
\newblock \bibinfo{journal}{Computers \& structures} \bibinfo{volume}{85},
  \bibinfo{pages}{784--795}.
\bibitem[{Gassner et~al.(2007)Gassner, Lörcher and Munz}]{dgrp}
\bibinfo{author}{Gassner, G.}, \bibinfo{author}{Lörcher, F.},
  \bibinfo{author}{Munz, C.D.}, \bibinfo{year}{2007}.
\newblock \bibinfo{title}{A contribution to the construction of diffusion
  fluxes for finite volume and discontinuous {Galerkin} schemes}.
\newblock \bibinfo{journal}{Journal of Computational Physics}
  \bibinfo{volume}{224}, \bibinfo{pages}{1049--1063}.
\bibitem[{Hirt et~al.(1974)Hirt, Amsden and Cook}]{HIRT1974227}
\bibinfo{author}{Hirt, C.}, \bibinfo{author}{Amsden, A.},
  \bibinfo{author}{Cook, J.}, \bibinfo{year}{1974}.
\newblock \bibinfo{title}{An arbitrary {Lagrangian-Eulerian} computing method
  for all flow speeds}.
\newblock \bibinfo{journal}{Journal of Computational Physics}
  \bibinfo{volume}{14}, \bibinfo{pages}{227--253}.
\newblock \DOIprefix\doi{10.1016/0021-9991(74)90051-5}.
\bibitem[{Ji et~al.(2018)Ji, Pan, Shyy and Xu}]{ji4order2018}
\bibinfo{author}{Ji, X.}, \bibinfo{author}{Pan, L.}, \bibinfo{author}{Shyy,
  W.}, \bibinfo{author}{Xu, K.}, \bibinfo{year}{2018}.
\newblock \bibinfo{title}{A compact fourth-order gas-kinetic scheme for the
  {Euler} and {Navier}–{Stokes} equations}.
\newblock \bibinfo{journal}{Journal of Computational Physics}
  \bibinfo{volume}{372}, \bibinfo{pages}{446--472}.
\bibitem[{Ji et~al.(2021a)Ji, Shyy and Xu}]{jiGC2021}
\bibinfo{author}{Ji, X.}, \bibinfo{author}{Shyy, W.}, \bibinfo{author}{Xu, K.},
  \bibinfo{year}{2021}a.
\newblock \bibinfo{title}{A gradient compression-based compact high-order
  gas-kinetic scheme on {3D} hybrid unstructured meshes}.
\newblock \bibinfo{journal}{International Journal of Computational Fluid
  Dynamics} \bibinfo{volume}{35}, \bibinfo{pages}{485--509}.
\bibitem[{Ji et~al.(2020)Ji, Zhao, Shyy and
  Xu}]{jiHWENOReconstructionBased2020}
\bibinfo{author}{Ji, X.}, \bibinfo{author}{Zhao, F.}, \bibinfo{author}{Shyy,
  W.}, \bibinfo{author}{Xu, K.}, \bibinfo{year}{2020}.
\newblock \bibinfo{title}{A {HWENO} reconstruction based high-order compact
  gas-kinetic scheme on unstructured mesh}.
\newblock \bibinfo{journal}{Journal of Computational Physics}
  \bibinfo{volume}{410}, \bibinfo{pages}{109367}.
\bibitem[{Ji et~al.(2021b)Ji, Zhao, Shyy and
  Xu}]{jiThreedimensionalCompactHighorder2020}
\bibinfo{author}{Ji, X.}, \bibinfo{author}{Zhao, F.}, \bibinfo{author}{Shyy,
  W.}, \bibinfo{author}{Xu, K.}, \bibinfo{year}{2021}b.
\newblock \bibinfo{title}{Compact high-order gas-kinetic scheme for
  three-dimensional flow simulations}.
\newblock \bibinfo{journal}{AIAA Journal} \bibinfo{volume}{59},
  \bibinfo{pages}{2979--2996}.
\bibitem[{Ji et~al.(2021c)Ji, Zhao, Shyy and Xu}]{jiMultiresolution2021}
\bibinfo{author}{Ji, X.}, \bibinfo{author}{Zhao, F.}, \bibinfo{author}{Shyy,
  W.}, \bibinfo{author}{Xu, K.}, \bibinfo{year}{2021}c.
\newblock \bibinfo{title}{Two-step multi-resolution reconstruction-based
  compact gas-kinetic scheme on tetrahedral mesh}.
\newblock \bibinfo{journal}{arXiv:2102.01366 [physics]} \bibinfo{note}{ArXiv:
  2102.01366}.
\bibitem[{Kucharik and Shashkov(2014)}]{KUCHARIK2014268}
\bibinfo{author}{Kucharik, M.}, \bibinfo{author}{Shashkov, M.},
  \bibinfo{year}{2014}.
\newblock \bibinfo{title}{Conservative multi-material remap for staggered
  multi-material arbitrary{ Lagrangian–Eulerian} methods}.
\newblock \bibinfo{journal}{Journal of Computational Physics}
  \bibinfo{volume}{258}, \bibinfo{pages}{268--304}.
\newblock \DOIprefix\doi{10.1016/j.jcp.2013.10.050}.
\bibitem[{Li(2019)}]{li2019two}
\bibinfo{author}{Li, J.}, \bibinfo{year}{2019}.
\newblock \bibinfo{title}{Two-stage fourth order: temporal-spatial coupling in
  computational fluid dynamics (cfd)}.
\newblock \bibinfo{journal}{Advances in Aerodynamics} \bibinfo{volume}{1},
  \bibinfo{pages}{1--36}.
\newblock \DOIprefix\doi{10.1186/s42774-019-0004-9}.
\bibitem[{Li and Du(2016)}]{liTwoStageFourthOrder2016}
\bibinfo{author}{Li, J.}, \bibinfo{author}{Du, Z.}, \bibinfo{year}{2016}.
\newblock \bibinfo{title}{A two-stage fourth order time-accurate discretization
  for {Lax}--{Wendroff} type flow solvers {I}. hyperbolic conservation laws}.
\newblock \bibinfo{journal}{SIAM Journal on Scientific Computing}
  \bibinfo{volume}{38}, \bibinfo{pages}{A3046--A3069}.
\bibitem[{Maire and Nkonga(2013)}]{Lagrangianvel}
\bibinfo{author}{Maire, P.H.}, \bibinfo{author}{Nkonga, B.},
  \bibinfo{year}{2013}.
\newblock \bibinfo{title}{Multi-scale {Godunov}-type method for cell-centered
  discrete {Lagrangian} hydrodynamics}.
\newblock \bibinfo{journal}{Journal of Computational Physics} ,
  \bibinfo{pages}{799--821}\DOIprefix\doi{10.1016/j.jcp.2008.10.012}.
\bibitem[{Noh(1987)}]{NOH1987}
\bibinfo{author}{Noh, W.}, \bibinfo{year}{1987}.
\newblock \bibinfo{title}{Errors for calculations of strong shocks using an
  artificial viscosity and an artificial heat flux}.
\newblock \bibinfo{journal}{Journal of Computational Physics}
  \bibinfo{volume}{72}, \bibinfo{pages}{78--120}.
\newblock \DOIprefix\doi{10.1016/0021-9991(87)90074-X}.
\bibitem[{Pan and Xu(2021)}]{PAN2021ALE3d}
\bibinfo{author}{Pan, L.}, \bibinfo{author}{Xu, K.}, \bibinfo{year}{2021}.
\newblock \bibinfo{title}{An {Arbitrary-Lagrangian-Eulerian} high-order
  gas-kinetic scheme for three-dimensional computations}.
\newblock \bibinfo{journal}{Journal of Scientific Computing}
  \bibinfo{volume}{88}, \bibinfo{pages}{1--26}.
\newblock \DOIprefix\doi{10.1007/s10915-021-01525-9}.
\bibitem[{Pan et~al.(2020)Pan, Zhao and Xu}]{PAN2020ALE2d}
\bibinfo{author}{Pan, L.}, \bibinfo{author}{Zhao, F.}, \bibinfo{author}{Xu,
  K.}, \bibinfo{year}{2020}.
\newblock \bibinfo{title}{High-order {ALE} gas-kinetic scheme with {WENO}
  reconstruction}.
\newblock \bibinfo{journal}{Journal of Computational Physics}
  \bibinfo{volume}{417}, \bibinfo{pages}{109558}.
\newblock \DOIprefix\doi{10.1016/j.jcp.2020.109558}.
\bibitem[{Ren et~al.(2016)Ren, Xu and Shyy}]{REN2016-ALE-DG}
\bibinfo{author}{Ren, X.}, \bibinfo{author}{Xu, K.}, \bibinfo{author}{Shyy,
  W.}, \bibinfo{year}{2016}.
\newblock \bibinfo{title}{A multi-dimensional high-order {DG}-{ALE} method
  based on gas-kinetic theory with application to oscillating bodies}.
\newblock \bibinfo{journal}{Journal of Computational Physics}
  \bibinfo{volume}{316}, \bibinfo{pages}{700--720}.
\newblock \DOIprefix\doi{10.1016/j.jcp.2016.04.028}.
\bibitem[{Wukie et~al.(2023)Wukie, Fidkowski, Persson and Wang}]{high_Fidelity}
\bibinfo{author}{Wukie, N.A.}, \bibinfo{author}{Fidkowski, K.},
  \bibinfo{author}{Persson, P.O.}, \bibinfo{author}{Wang, Z.J.},
  \bibinfo{year}{2023}.
\newblock \bibinfo{title}{High-fidelity {CFD} verification workshop 2024: Mesh
  motion}, in: \bibinfo{booktitle}{AIAA SCITECH 2023 Forum}, p.
  \bibinfo{pages}{1243}.
\bibitem[{Xu(1998)}]{vki1998}
\bibinfo{author}{Xu, K.}, \bibinfo{year}{1998}.
\newblock \bibinfo{title}{Gas-kinetic schemes for unsteady compressible flow
  simulations}.
\newblock \bibinfo{journal}{Computational Fluid Dynamics, Annual Lecture
  Series, 29 th, Rhode-Saint-Genese, Belgium} .
\bibitem[{Xu(2001)}]{xuGKS2001}
\bibinfo{author}{Xu, K.}, \bibinfo{year}{2001}.
\newblock \bibinfo{title}{A gas-kinetic {BGK} scheme for the
  {Navier}–{Stokes} equations and its connection with artificial dissipation
  and {Godunov} method}.
\newblock \bibinfo{journal}{Journal of Computational Physics}
  \bibinfo{volume}{171}, \bibinfo{pages}{289--335}.
\bibitem[{Zhang et~al.(2023)Zhang, Ji and Xu}]{zhang2023slidingmesh}
\bibinfo{author}{Zhang, Y.}, \bibinfo{author}{Ji, X.}, \bibinfo{author}{Xu,
  K.}, \bibinfo{year}{2023}.
\newblock \bibinfo{title}{A high-order compact gas-kinetic scheme in a rotating
  coordinate frame and on sliding mesh}.
\newblock \bibinfo{journal}{International Journal of Computational Fluid
  Dynamics} , \bibinfo{pages}{1--20}.
\bibitem[{Zhao et~al.(2019)Zhao, Ji, Shyy and
  Xu}]{zhaoCompactHigherorderGaskinetic2019}
\bibinfo{author}{Zhao, F.}, \bibinfo{author}{Ji, X.}, \bibinfo{author}{Shyy,
  W.}, \bibinfo{author}{Xu, K.}, \bibinfo{year}{2019}.
\newblock \bibinfo{title}{Compact higher-order gas-kinetic schemes with
  spectral-like resolution for compressible flow simulations}.
\newblock \bibinfo{journal}{Advances in Aerodynamics} \bibinfo{volume}{1},
  \bibinfo{pages}{13}.
\bibitem[{Zhao et~al.(2020)Zhao, Ji, Shyy and Xu}]{zhaoAcousticShockWave2020}
\bibinfo{author}{Zhao, F.}, \bibinfo{author}{Ji, X.}, \bibinfo{author}{Shyy,
  W.}, \bibinfo{author}{Xu, K.}, \bibinfo{year}{2020}.
\newblock \bibinfo{title}{An acoustic and shock wave capturing compact
  high-order gas-kinetic scheme with spectral-like resolution}.
\newblock \bibinfo{journal}{International Journal of Computational Fluid
  Dynamics} \bibinfo{volume}{34}, \bibinfo{pages}{731--756}.
\bibitem[{Zhao et~al.(2021)Zhao, Ji, Shyy and
  Xu}]{zhaoDirectModelingComputational2021}
\bibinfo{author}{Zhao, F.}, \bibinfo{author}{Ji, X.}, \bibinfo{author}{Shyy,
  W.}, \bibinfo{author}{Xu, K.}, \bibinfo{year}{2021}.
\newblock \bibinfo{title}{Direct modeling for computational fluid dynamics and
  the construction of high-order compact scheme for compressible flow
  simulations}.
\newblock \bibinfo{journal}{arXiv:2107.06555 [physics]} \bibinfo{note}{ArXiv:
  2107.06555}.
\bibitem[{Zhao et~al.(2022)Zhao, Ji, Shyy and
  Xu}]{zhaoCompactHighorderGaskinetic2022a}
\bibinfo{author}{Zhao, F.}, \bibinfo{author}{Ji, X.}, \bibinfo{author}{Shyy,
  W.}, \bibinfo{author}{Xu, K.}, \bibinfo{year}{2022}.
\newblock \bibinfo{title}{A compact high-order gas-kinetic scheme on
  unstructured mesh for acoustic and shock wave computations}.
\newblock \bibinfo{journal}{Journal of Computational Physics}
  \bibinfo{volume}{449}, \bibinfo{pages}{110812}.
\bibitem[{Zhu and Qiu(2018)}]{zhuNewFiniteVolume2018}
\bibinfo{author}{Zhu, J.}, \bibinfo{author}{Qiu, J.}, \bibinfo{year}{2018}.
\newblock \bibinfo{title}{New finite volume weighted essentially nonoscillatory
  schemes on triangular meshes}.
\newblock \bibinfo{journal}{SIAM Journal on Scientific Computing}
  \bibinfo{volume}{40}, \bibinfo{pages}{A903--A928}.
\bibitem[{Zhu and Shu(2020)}]{zhu2020}
\bibinfo{author}{Zhu, J.}, \bibinfo{author}{Shu, C.W.}, \bibinfo{year}{2020}.
\newblock \bibinfo{title}{A new type of third-order finite volume
  multi-resolution {WENO} schemes on tetrahedral meshes}.
\newblock \bibinfo{journal}{Journal of Computational Physics}
  \bibinfo{volume}{406}, \bibinfo{pages}{109212}.

\end{thebibliography}

\end{document}